\documentclass[11pt]{article}
\usepackage{epsfig}
\usepackage{axodraw}
\usepackage{graphicx}
\usepackage{rotate}
\usepackage{latexsym}
\usepackage{amssymb}
\newcommand\pubnumber{}
\newcommand\pubdate{\today}
\newcommand\hepnumber{hep-ph/0612122}

\def\csumaa{Deutsches Electronen - Synchrotron, DESY, Platanenallee 6, 15738
Zeuthen, Germany}
\def\csuma{Fakult\"at f\"ur Physik, Albert-Ludwigs Universit\"at, Freiburg,
Germany\\
Institut f\"ur Theoretische Physik, Universit\"at Z\"urich, Z\"urich,
Switzerland}
\def\csumb{Dipartimento di Fisica, Universit\`a di Padova 
and INFN, Sezione di Padova, Italy}
\def\csumc{Dipartimento di Fisica Teorica, Universit\`a di Torino, Italy\\
INFN, Sezione di Torino, Italy}
\def\support{\footnote{Work supported by MIUR under contract
2001023713$\_$006 and by the European Community's Marie Curie Research 
Training Network under contract MRTN-CT-2006-035505
`Tools and Precision Calculations for Physics Discoveries at Colliders'.}}
\def\Title#1{\begin{center} {\Large\bf #1 } \end{center}}

\def\Author#1{\begin{center}{ \sc #1} \end{center}}
\def\Address#1{\begin{center}{ \it #1} \end{center}}

\newcommand\pubblock{\rightline{\begin{tabular}{l} \pubnumber\\
         \pubdate\\ \hepnumber \\ ZU-TH 23/06 \\ DESY 06-223 \\
SFB/CPP-06-54 \end{tabular}}}
\newenvironment{Abstract}{\begin{quotation}  }{\end{quotation}}

\def\Acknowledgments{\bigskip  \bigskip \begin{center}
          \large\bf Acknowledgments\end{center}}
\def\email#1{\footnote{#1}}
\makeatletter
\def\section{\@startsection{section}{0}{\z@}{5.5ex plus .5ex minus
 1.5ex}{2.3ex plus .2ex}{\large\bf}}
\def\subsection{\@startsection{subsection}{1}{\z@}{3.5ex plus .5ex minus
 1.5ex}{1.3ex plus .2ex}{\normalsize\bf}}
\def\subsubsection{\@startsection{subsubsection}{2}{\z@}{-3.5ex plus
-1ex minus  -.2ex}{2.3ex plus .2ex}{\normalsize\sl}}

\renewcommand{\@makecaption}[2]{%
   \vskip 10pt
   \setbox\@tempboxa\hbox{\small #1: #2}
   \ifdim \wd\@tempboxa >\hsize     % IF longer than one line:
       \small #1: #2\par          %   THEN set as ordinary paragraph.
     \else                        %   ELSE  center.
       \hbox to\hsize{\hfil\box\@tempboxa\hfil}
   \fi}
 \def\citenum#1{{\def\@cite##1##2{##1}\cite{#1}}}
\def\citea#1{\@cite{#1}{}}
\newcount\@tempcntc
\def\@citex[#1]#2{\if@filesw\immediate\write\@auxout{\string\citation{#2}}\fi
  \@tempcnta\z@\@tempcntb\m@ne\def\@citea{}\@cite{\@for\@citeb:=#2\do
    {\@ifundefined
       {b@\@citeb}{\@citeo\@tempcntb\m@ne\@citea\def\@citea{,}{\bf }\@warning
       {Citation `\@citeb' on page \thepage \space undefined}}%
    {\setbox\z@\hbox{\global\@tempcntc0\csname b@\@citeb\endcsname\relax}%
     \ifnum\@tempcntc=\z@ \@citeo\@tempcntb\m@ne
       \@citea\def\@citea{,}\hbox{\csname b@\@citeb\endcsname}%
     \else
      \advance\@tempcntb\@ne
      \ifnum\@tempcntb=\@tempcntc
      \else\advance\@tempcntb\m@ne\@citeo
      \@tempcnta\@tempcntc\@tempcntb\@tempcntc\fi\fi}}\@citeo}{#1}}
\def\@citeo{\ifnum\@tempcnta>\@tempcntb\else\@citea\def\@citea{,}%
  \ifnum\@tempcnta=\@tempcntb\the\@tempcnta\else
  {\advance\@tempcnta\@ne\ifnum\@tempcnta=\@tempcntb \else\def\@citea{--}\fi
    \advance\@tempcnta\m@ne\the\@tempcnta\@citea\the\@tempcntb}\fi\fi}
\makeatother
\newcommand{\nl}{\nonumber\\}

\newcommand{\lpar}{\left(}                            % bracketing
\newcommand{\rpar}{\right)}

\newcommand{\bq}{\begin{equation}}                    % equationing
\newcommand{\eq}{\end{equation}}
\newcommand{\bqa}{\arraycolsep 0.14em\begin{eqnarray}}
\newcommand{\eqa}{\end{eqnarray}}
\newcommand{\ba}[1]{\begin{array}{#1}}
\newcommand{\ea}{\end{array}}
\newcommand{\ben}{\begin{enumerate}}
\newcommand{\een}{\end{enumerate}}
\newcommand{\bei}{\begin{itemize}}
\newcommand{\eei}{\end{itemize}}
\newcommand{\eqn}[1]{Eq.(\ref{#1})}
\newcommand{\eqns}[2]{Eqs.(\ref{#1})--(\ref{#2})}

\newcommand{\tabn}[1]{Tab.~\ref{#1}}

\newcommand{\sect}[1]{Section~\ref{#1}}

\def\Re{\mathop{\operator@font Re}\nolimits}
\def\Im{\mathop{\operator@font Im}\nolimits}
\newcommand{\ord}[1]{{\cal O}\lpar#1\rpar}

\newcommand{\ib}{i}

\newcommand{\barf}{\overline f}                
\newcommand{\barh}{\overline h}

\newcommand{\mh}{M_{_H}}

\newcommand{\gf}{G_{\ssF}}

\newcommand{\stw}{s_{\theta}}             % bare, Lagrangian parameters
\newcommand{\ctw}{c_{\theta}}
\newcommand{\stws}{s_{\theta}^2}
\newcommand{\stwc}{s_{\theta}^3}

\newcommand{\stwq}{s_{\theta}^4}
\newcommand{\stwx}{s_{\theta}^6}
\newcommand{\ctws}{c_{\theta}^2}

\newcommand{\gfd}{\gamma_5}

\newcommand{\gapu}[1]{\gamma^{#1}}

\newcommand{\tpfi}{\lpar 2\pi\rpar^4\ib}

\newcommand{\upar}[1]{u}

\newcommand{\ssA}{{\scriptscriptstyle{A}}}

\newcommand{\ssF}{{\scriptscriptstyle{F}}}

\newcommand{\ssH}{{\scriptscriptstyle{H}}}
\newcommand{\ssI}{{\scriptscriptstyle{I}}}

\newcommand{\ssL}{{\scriptscriptstyle{L}}}
\newcommand{\ssM}{{\scriptscriptstyle{M}}}

\newcommand{\ssQ}{{\scriptscriptstyle{Q}}}
\newcommand{\ssR}{{\scriptscriptstyle{R}}}
\newcommand{\ssS}{{\scriptscriptstyle{S}}}
\newcommand{\ssT}{{\scriptscriptstyle{T}}}

\newcommand{\ssV}{{\scriptscriptstyle{V}}}
\newcommand{\ssW}{{\scriptscriptstyle{W}}}

\newcommand{\ssY}{{\scriptscriptstyle{Y}}}
\newcommand{\ssZ}{{\scriptscriptstyle{Z}}}

\newcommand{\bqas}{\begin{eqnarray*}}
\newcommand{\eqas}{\end{eqnarray*}}
\def\app#1#2 {{\it Acta. Phys. Pol.} {\bf#1},#2}
\def\cpc#1#2 {{\it Computer Phys. Comm.} {\bf#1},#2}
\def\np#1#2 {{\it Nucl. Phys.} {\bf#1},#2}
\def\pl#1#2 {{\it Phys. Lett.} {\bf#1},#2}
\def\prep#1#2 {{\it Phys. Rep.} {\bf#1},#2}
\def\prev#1#2 {{\it Phys. Rev.} {\bf#1},#2}
\def\prl#1#2 {{\it Phys. Rev. Lett.} {\bf#1},#2}
\def\zp#1#2 {{\it Zeit. Phys.} {\bf#1},#2}
\def\sptp#1#2 {{\it Suppl. Prog. Theor. Phys.} {\bf#1},#2}
\def\mpl#1#2 {{\it Modern Phys. Lett.} {\bf#1},#2}
\def\jetp#1#2 {{\it Sov. Phys. JETP} {\bf#1},#2}
\def\fpj#1#2 {{\it Fortschr. Phys.} {\bf#1},#2}
\def\afp#1#2 {{\it Acta.Phys. Polon.} {\bf#1},#2}
\def\err#1#2 {{\it Erratum} {\bf#1},#2}
\def\ijmp#1#2 {{\it Int. J. Mod. Phys} {\bf#1},#2}
\def\nc#1#2 {{\it Nuovo Cimento} {\bf#1},#2}
\def\ap#1#2 {{\it Ann. Phys.} {\bf#1},#2}
\def\cmp#1#2 {{\it Comm. Math. Phys.} {\bf#1},#2}
\def\el#1#2 {{\it Europhys. Lett.} {\bf#1},#2}
\def\hpa#1#2 {{\it Helv. Phys. Acta} {\bf#1},#2}
\def\yf#1#2 {{\it Yad. Fiz.} {\bf#1},#2}
\def\nim#1#2 {{\it Nucl. Instrum. Meth.} {\bf#1},#2}
\def\spz#1#2 {{\it Sov. Pisma Zhetf} {\bf#1},#2}
\def\jetpl#1#2 {{\it JETP Lett.} {\bf#1},#2}
\def\sjnp#1#2 {{\it Sov. J. Nucl. Phys.} {\bf#1},#2}
\def\ptp#1#2 {{\it Progr. Theor. Phys. (Kyoto)} {\bf#1},#2}
\def\rmp#1#2  {{\it Rev. Mod. Phys.} {\bf#1},#2}
\def\zhetf#1#2 {{\it ZhETF} {\bf#1},#2}
\def\prs#1#2 {{\it Proc. Roy. Soc.} {\bf#1},#2}
\def\phys#1#2 {{\it Physica} {\bf#1},#2}

\def\bfi{\begin{figure}}
\def\efi{\end{figure}}

\newcommand{\LB}{{\cal L}oop{\cal B}ack}
\newcommand{\GS}{{\cal G}raph{\cal S}hot}
\newcommand{\cL}{{\cal L}}

\newcommand{\cC}{{\cal C}}
\newcommand{\ssAZ}{{\scriptscriptstyle{AZ}}}
\newcommand{\ssZA}{{\scriptscriptstyle{ZA}}}
\newcommand{\ssAA}{{\scriptscriptstyle{AA}}}
\newcommand{\ssZZ}{{\scriptscriptstyle{ZZ}}}

\newcommand{ \be}{\begin{equation}}
\newcommand{ \ee}{\end{equation}}
\newcommand{ \bea}{\begin{eqnarray}}
\newcommand{ \eea}{\end{eqnarray}}
\newcommand{ \mysmall}[1]{\scriptscriptstyle #1} % a smaller #
\newcommand{ \bm} {\boldmath}
\newcommand{ \ubm} {\unboldmath}
\newcommand{ \bb}{\beta_t}
\newcommand{ \Mgb}{\bar{g}}
\newcommand{ \MMb}{\bar{M}}
\newcommand{ \ccb}{\bar{c}_\theta}
\newcommand{ \ssb}{\bar{s}_\theta}
\newcommand{ \Ab}{\overline{A}}
\newcommand{ \Zb}{\overline{Z}}
\newcommand{ \Xb}{\overline{X}}
\newcommand{ \redcircle}[1]{\Red\circle*{#1}\Black}
\newcommand{\pww}[1]{\Pi^{#1}_{\mu \nu, \mysmall{W} \mysmall{W} }}
\newcommand{\pzz}[1]{\Pi^{#1}_{\mu \nu, \mysmall{Z} \mysmall{Z} }}
\newcommand{\paa}[1]{\Pi^{#1}_{\mu \nu, \mysmall{A} \mysmall{A} }}
\newcommand{\paz}[1]{\Pi^{#1}_{\mu \nu, \mysmall{A} \mysmall{Z} }}

\newcommand{\pwp}[1]{\Pi^{#1}_{\mu, \mysmall{W \phi} }}

\newcommand{\pzp}[1]{\Pi^{#1}_{\mu, \mysmall{Z \phi_o} }}

\newcommand{\pap}[1]{\Pi^{#1}_{\mu, \mysmall{A \phi_o}} }

\newcommand{\ppp}[1]{\Pi^{#1}_{\mysmall{\phi \phi}} }
\newcommand{\ppopo}[1]{\Pi^{#1}_{\mysmall{\phi_o \phi_o}} }
\newcommand{\ppzpz}[1]{\Pi^{#1}_{\mysmall{\phi_o \phi_o}} }
\newcommand{\dd}[2]{D^{(1)}_{\mysmall{{#1} {#2}}}}
\newcommand{\pp}[2]{P^{(1)}_{\mysmall{{#1} {#2}}}}
\newcommand{\GG}[2]{G^{(1)}_{\mysmall{{#1} {#2}}}}
\newcommand{\rr}[2]{R^{(1)}_{\mysmall{{#1} {#2}}}}
\newcommand{\WW}{\mysmall{WW}}
\newcommand{\II}{\mysmall{I}}

\def\Red        {}
\def\Black      {}
\begin{document}
\begin{titlepage}
\pubblock
\vfill
\def\thefootnote{\fnsymbol{footnote}}
\Title{Two-Loop Renormalization in the Standard Model\\[5mm]
Part I: Prolegomena\support} 
\vfill
\Author{Stefano Actis\email{Stefano.Actis@desy.de}}
\Address{\csumaa}
\Author{Andrea Ferroglia\email{andrea.ferroglia@physik.uni-freiburg.de}}
\Address{\csuma}
\Author{Massimo Passera\email{massimo.passera@pd.infn.it}}
\Address{\csumb}
\Author{Giampiero Passarino\email{giampiero@to.infn.it}}
\Address{\csumc} 
\vfill
\begin{Abstract}
\noindent 
In this paper the building blocks for the two-loop renormalization of the
Standard Model are introduced with a comprehensive discussion of the special
vertices induced in the Lagrangian by a particular diagonalization of the
neutral sector and by two alternative treatments of the Higgs tadpoles.
Dyson resummed propagators for the gauge bosons are derived, and two-loop
Ward-Slavnov-Taylor identities are discussed. In part II, the complete set
of counterterms needed for the two-loop renormalization will be derived. In
part III, a renormalization scheme will be introduced, connecting the
renormalized quantities to an input parameter set of (pseudo-)experimental
data, critically discussing renormalization of a gauge theory with unstable
particles.
\end{Abstract}
\vfill
\begin{center}
\end{center}
\end{titlepage}
\def\thefootnote{\arabic{footnote}}
\setcounter{footnote}{0}
%--
\small
\thispagestyle{empty}
\tableofcontents
\setcounter{page}{1}
\normalsize
%--
\clearpage
%--
\section{Introduction}
\label{sec-intro}

In a series of papers we developed a strategy for the algebraic-numerical
evaluation of two-loop, two-(three-)leg Feynman diagrams appearing in any
renormalizable quantum field theory. In~\cite{Passarino:2001wv} the general
strategy has been designed and in~\cite{Passarino:2001jd} a complete list of
results has been derived for two-loop functions with two external legs,
including their infrared divergent on-shell derivatives.  Results for
one-loop multi-leg diagrams have been shown in~\cite{Ferroglia:2002mz} and
additional material can be found in~\cite{Ferroglia:2002yr}. Two-loop
three-point functions for infrared convergent configurations have been
considered in~\cite{Ferroglia:2003yj}, two-loop tensor integrals
in~\cite{Actis:2004bp}, two-loop infrared divergent vertices
in~\cite{Passarino:2006gv}.  As a by-product of our general program we have
developed a set of FORTRAN/95 routines~\cite{LB} for computing everything
which is needed, from standard $A_0,\,\dots\,,D_0$
functions~\cite{Passarino:1979jh} to two-loop, two-(three-) point functions.
This new ensemble of programs, which includes the treatment of complex
poles~\cite{Argyres:1995ym}, will succeed the corresponding library of {\tt
TOPAZ0}~\cite{Montagna:1993ai}.

The next step in our project has been to introduce all those elements which
are necessary for a complete discussion of the two-loop renormalization of
the Standard Model (SM). In this paper we introduce basic aspects of
renormalization which are needed before the introduction of counterterms. In
part II we will present a detailed analysis of the counterterms with special
emphasis to the cancellation of ultraviolet poles with non-local residues
(the so-called problem of overlapping divergences), while in part III we
will deal with finite renormalization deriving renormalization equations, up
to two loops, that relate the renormalized parameters of the model to an
input parameter set, which always includes the fine structure constant
$\alpha$ and the Fermi coupling constant $\gf$. Renormalization with
unstable particles will also be addressed.

Having provided a derivation of the elements which are essential for
constructing a renormalization procedure, we will proceed in computing a
first set of pseudo-observables, including the running e.m. coupling
constant and the complex poles characterizing unstable gauge bosons.

Several authors have already contributed in developing seminal results for
the two-loop renormalization of the SM~\cite{Awramik:2004qv}. Here we want
to present our own approach, from fundamentals to applications.  The whole
set of results is completely independent from other sources; furthermore, we
wanted to collect in a single place all the formulas and algorithms that can
be used for many applications and are never there when you need them.

The code $\GS$~\cite{GraphShot} synthesizes the algebraic component of the
project (for alternative approaches see ref.~\cite{Hahn:2000kx} and
references therein) from generation of diagrams, reduction of tensor
structures, special kinematical configurations, analytical extraction of
ultraviolet/infrared poles~\cite{Passarino:2006gv} and of collinear
logarithms and check of Ward-Slavnov-Taylor identities (hereafter WST
identities)~\cite{Taylor:ff}. The corresponding output is then treated by a
FORTRAN/95 code, $\LB$~\cite{LB}, which is able to exploit the multi-scale
structure of two-loop diagrams. Future applications will include $H \to
\gamma \gamma$ and $H \to gg$, to give an example.

It is worth noticing that there are other solutions to the problems discussed 
in this paper; noticeably, one can choose to work in the background-field 
formalism~\cite{Denner:1994xt}; here we only stress that our solution has been 
extended up to the two loops and has been implemented in a complete and 
stand-alone set of procedures for two-loop renormalization.

The outline of the paper is as follows. In \sect{sec-tadpoles} we discuss
the role of tadpoles in a spontaneously broken gauge theory presenting two
alternative schemes in \sect {subsec-betah} and in \sect{subsec-betat}.
Diagonalization of the neutral sector in the SM is derived in
\sect{sec-gamma}. WST identities are discussed in \sect{sec-WSTI}. Dyson
resummation is analyzed in \sect{sec-dyson}.  Bases relevant for
renormalization are introduced in \sect{sec-LQ}.  New sets of Feynman rules,
required by our renormalization procedure, are given in the Appendices.
%--
\section{Higgs tadpoles}
\label{sec-tadpoles}

Tadpoles in a spontaneously broken gauge theory have been discussed by many
authors (see, for instance~\cite{Sirlin:1985ux}). Here we outline those
aspects which are peculiar to our approach.
%--
\subsection{The basics}
\label{subsec-basics}

Following notation and conventions of ref.~\cite{Bardin:1999ak}, the minimal
Higgs sector of the SM is provided by the Lagrangian 
\be
    \cL_S  = -(D_{\mu} K)^\dagger (D_{\mu} K) -\mu^2 K^\dagger K
                  - (\lambda/2) (K^\dagger K)^2,
\label{eq:LS}
\ee
where the covariant derivative is given by
\be
    D_{\mu} K = \left(\partial_\mu -\frac{i}{2}g B_{\mu}^a \tau^a 
                -\frac{i}{2} g' B_{\mu}^0          \right) K,
\ee
$g'/g = -\sin \theta/\cos\theta$, $\theta$ is the weak mixing angle,
$\tau^a$ are the standard Pauli matrices, $B_{\mu}^a$ is a triplet of vector
gauge bosons and $B_{\mu}^0$ a singlet.  For the theory to be stable we must
require $\lambda >0$. We choose $\mu^2<0$ in order to have spontaneous
symmetry breaking (SSB). The scalar field in the minimal realization of the
SM is
\be
    K = \frac{1}{\sqrt 2} \left( \begin{array}{c} 
                                  \zeta +i\phi_0 \\
				  -\phi_2 +i\phi_1
				  \end{array} \right),
\ee
where $\zeta$ and the Higgs-Kibble fields $\phi_0$, $\phi_1$ and
$\phi_2$ are real. For $\mu^2<0$ we have SSB, $\langle K\rangle_0 \neq 0$. In
particular, we choose $\zeta +i\phi_0$ to be the component of $K$ to develop
the non-zero vacuum expectation value (VEV), and we set 
$\langle \phi_0 \rangle_0 = 0$ and $\langle \zeta\rangle_0 \neq 0$. We then 
introduce the (physical) Higgs fields as $H =\zeta - v$.
%--
The parameter $v$ is not a new parameter of the model; its value must
be fixed by the requirement that $\langle H \rangle_0 = 0$ (i.e.~$\langle K
\rangle_0 = (1/\sqrt{2})(v,0)$), so that the vacuum doesn't absorb/create
Higgs particles. To see how this works at the lowest order, consider the
part of $\cL_S$ containing the Higgs field:
\be
    -(1/2)(\partial_\mu H)^2 -(\mu^2/2) (H+v)^2 - (\lambda/8) (H+v)^4.
\ee
These terms generate vertices that imply absorption of $H$ in the vacuum,
namely those linear in $H$,
\be
    \left[ -\mu^2 v -(\lambda/2) v^3 \right] H,
\ee
which correspond to the vertex $H$$\!\!$\put(50,3){\circle*{5}}
\put(10,3){\line(1,0){40}}~~~~~~~~~~~~~~~~. This vertex gives a
non-zero value to the diagrams with one $H$ line, and thus a
non-zero VEV. We will set it to zero, i.e. $v=(-2\mu^2/\lambda)^{1/2}$ (or
$v = 0$, but then, no SSB).

%%%%%%%%%%%%%%%%%%%%%%%%%%%%%%%%%%%%%%%%%%%%%%%%%%%%%%%%%%%%%%%%%%%%%%%%%%%%%
\subsection{The parameter \boldmath $\beta_h$ \unboldmath}
\label{subsec-betah}

\subsubsection{Definitions and Lagrangian}

More complicated diagrams contribute to $\langle H \rangle_0$ in higher
orders of perturbation theory.  The parameter $v$ must then be readjusted
such that $\langle H \rangle_0=0.$ First of all, let us introduce the new
bare parameters $M$ (the $W$ boson mass), $\mh$ (the mass of the physical
Higgs particle) and $\beta_h$ (the tadpole constant) according to the
following definitions:
\renewcommand{\arraystretch}{1.5}
\be
\left\{
  \begin{array}{lll}
    M &=& gv/2 \\
    \mh^2 &=& \lambda v^2 \\
    \beta_h &=& \mu^2  +\frac{\lambda}{2}v^2
  \end{array}
\right.
~~~~~\Longrightarrow~~~~~~~
\left\{
  \begin{array}{lll}
    v &=& 2M/g \\
    \lambda &=& \left(g \mh/2M\right)^2 \\
    \mu^2 &=&  \beta_h -\frac{1}{2}\mh^2
  \end{array}
\right.
\label{eq:betaH}
\ee
\renewcommand{\arraystretch}{1}

\noindent This parameter $\beta_h$ is the same as 
$\beta_{\ssH}$ of ~\cite{Bardin:1999ak} and $\beta_h$ 
of~\cite{Veltman:1994wz}. The new set
of (bare) parameters is therefore $g,g'$, $M,\mh$ and $\beta_h$ instead of
$g,g'$, $\mu, \lambda, v$. Remember that $\beta_h$ (like $v$) is not an
independent parameter. In terms of $g,g'$, $M,\mh$ and $\beta_h$, 
$\cL_S^I$ becomes (in ref.~\cite{Bardin:1999ak} some terms
have been dropped)
\bea
    \cL_S^I  &=& -\mu^2 K^\dagger K - (\lambda/2) (K^\dagger K)^2 
    = -\beta_h \left[\frac{2M^2}{g^2} + \frac{2M}{g}H 
      +\frac{1}{2}\left(H^2 +\phi_0^2 +2\phi_+ \phi_- \right) \right] 
              \nonumber\\
    && + \frac{\mh^2 M^2}{2g^2} -\frac{1}{2} \mh^2 H^2 
     -g \frac{\mh^2}{4M} H \left( H^2 +\phi_0^2 +2\phi_+ \phi_- \right) 
    -g^2 \frac{\mh^2}{32M^2} \left( H^2 +\phi_0^2 +2\phi_+ \phi_-
    \right)^2,
\label{eq:LSI}
\eea
with $\phi_\pm = (\phi_1 \mp i \phi_2)/\sqrt 2$. Note that $(-\mu^2
K^\dagger K)$ is the only term of $\cL_S$ containing $\beta_h$ (actually,
the only term of the whole SM Lagrangian). Let us now set $\beta_h$ such
that the VEV of the Higgs field $H$ remains zero to each order of
perturbation theory.
%%%%%%%%%%%%%%%%%%%%%%%%%%%%%%%%%%%%%%%%%%%%%%%%%%%%%%%%%%%%%%%%%%%%%%%%%%%%%
\subsubsection{$\beta_h$ fixing at the lowest order}

At the lowest order, the only diagram contributing to $\langle H \rangle_0$
is
\be
H \put(50,3){\circle*{5}}\put(10,3){\line(1,0){40}}~~~~~~~~~~~~~~~
\label{eq:betaset0}
\ee
originated by the term in $\cL_S^I$ linear in $H$, $-(2\beta_h M/g)H$.
Therefore, at the lowest order we will simply set $\beta_h =0$.
%%%%%%%%%%%%%%%%%%%%%%%%%%%%%%%%%%%%%%%%%%%%%%%%%%%%%%%%%%%%%%%%%%%%%%%%%%%%%
\subsubsection{$\beta_h$ fixing up to one loop}

Define $ \beta_h = \beta_{h_0} +\beta_{h_1} g^2 + \beta_{h_2} g^4 + \cdots
$. The lowest-order $\beta_h$ fixing of the previous section amounts to
$\beta_{h_0}=0$.  At the one-loop level, two types of diagrams contribute to
the Higgs VEV up to $\mathcal{O}(g)$: 
\be
T_0:~~~ \put(50,3){\circle*{5}}\put(10,3){\line(1,0){40}}~~~~~~~~~~~~~~~
\mbox{~~~~~~~+~~~~}
T_1:~~~ \put(50,3){\circle{20}}\put(10,3){\line(1,0){30}}~~~~~~~~~~~~~~~
\label{eq:betaset1}
\ee 
where the empty blob in the second term symbolically indicates all the
one-loop diagrams containing a scalar field ($H$, $\phi_\pm$, $\phi_0$), a
gauge field ($Z$, $W_\pm$), a Faddeev--Popov ghost field ($X_+$, $X_-$,
$X_{\ssZ}$), or a fermionic field. As an example, consider only the one-loop
diagram containing the $H$ field: $T_1^{\mysmall{(H)}}$; if this were the
only $T_1$ diagram, in order to have $\langle H \rangle_0=0$ it should
cancel with $T_0$, i.e.
\be (2\pi)^4 i \left( -\beta_h^{\mysmall{(H)}}\frac{2M}{g} \right) - g
   \frac{3\mh^2}{4M} i\pi^2 A_0(\mh) = 0, \ee
where $i\pi^2 A_0(m)={\mu}^{4-n}\int d^n q /(q^2 +m^2 -i\epsilon)$. The 
solution of this equation is $\beta_{h_0} = 0$ and
\be
    \beta_{h_1}^{\mysmall{(H)}}=
    \frac{1}{(2\pi)^4 i}\left(\frac{T_1^{\mysmall{(H)}}}{2Mg}\right) =
    -\frac{1}{16\pi^2} \left[\frac{3\mh^2}{8M^2}A_0(\mh) \right].
\ee
Of course, $\beta_{h_1}^{\mysmall{(H)}}$ is just the contribution to
$\beta_{h_1}$ arising from the one-loop tadpole diagram containing the $H$
field; the complete expression for $\beta_{h_1}$ in the $R_{\xi}$ gauge is
\bea 
     \beta_{h_1} &=&-\frac{1}{16\pi^2} \left[\!\!\!\!\!\!\!\phantom{\sum_f}
     \frac{3}{2} A_0(M) + \frac{3}{4\ctws} A_0(M_0) +M^2+\frac{M_0^2}{2\ctws}+
     \right. \nonumber \\
     && \left. + \,\,\frac{\mh^2}{8M^2} \Big(
     A_0(\xi_{\ssZ} M_0) +2A_0(\xi_{\ssW} M) \Big)+
     \frac{3\mh^2}{8M^2} A_0(\mh) 
     \,-\,\sum_f \frac{m_f^2}{M^2} A_0(m_f) \right],
\eea
where $M_0=M/\ctw$ and $m_f$ are the $Z$ and fermion masses, and $\ctw =
\cos\theta$. 

%%%%%%%%%%%%%%%%%%%%%%%%%%%%%%%%%%%%%%%%%%%%%%%%%%%%%%%%%%%%%%%%%%%%%%%%%%%%%
\subsubsection{$\beta_h$ vertices in one-loop calculations}
%--
Beyond the lowest order, $\beta_h$ is not zero and the Lagrangian $\cL_S^I$
contains the following vertices involving a $\beta_h$ factor (``$\beta_h$
vertices'', from now on):
\bea
  H & \put(60,3){\circle*{5}}\put(0,3){\line(1,0){60}}
  ~~~~~~~~~~~~~~~~~~& ~~~~~~~~~~~~\:(2\pi)^4 i\;(-2M\beta_h/g)
  \label{eq:betatad} \\
  H & \put(30,3){\circle*{5}}\put(0,3){\line(1,0){60}} 
  ~~~~~~~~~~~~~~~~~~& H_{~} ~~~~~~~~~       (2\pi)^4 i\;(-\beta_h)
  \label{eq:betavH}  \\
  \phi_0  & \put(30,3){\circle*{5}}\put(0,3){\line(1,0){60}}
  ~~~~~~~~~~~~~~~~~~& \phi_0\, ~~~~~~~~~    (2\pi)^4 i\;(-\beta_h)
  \label{eq:betavp0} \\
  \phi_+  & \put(30,3){\circle*{5}}\put(0,3){\line(1,0){60}}
  ~~~~~~~~~~~~~~~~~~& \phi_-  ~~~~~~~~~     (2\pi)^4 i\;(-\beta_h) 
\label{eq:betavppm}
\eea
(as usual, the combinatorial factors for identical fields are included; see
the Appendix D of ref.~\cite{Veltman:1994wz}).  Note that only scalar fields
appear in the $\beta_h$ vertices. These $\beta_h$ vertices must be included
in one-loop calculations. Consider, for example, the Higgs self-energy at
the one-loop level. The diagrams contributing to this ${\mathcal{O}}(g^2)$
quantity are
\be
  H~~ \put(30,3){\circle*{5}}\put(0,3){\line(1,0){60}} 
  ~~~~~~~~~~~~~~~~~~ H ~~~~~~~ + ~~~~~~~~ 
  H~~ \put(30,3){\circle{20}}\put(0,3){\line(1,0){20}}
                             \put(40,3){\line(1,0){20}}  
  ~~~~~~~~~~~~~~~~~~ H,
\label{eq:higgsSE1}
\ee 
where the empty blob in the second term represents all the one-loop
contributions (two possible topologies). The first diagram, containing a
two-leg $\beta_h$ vertex, shouldn't be forgotten, and plays an important
role in the Ward identities (see later).  One should also include diagrams
containing tadpoles: \vspace{3mm}
\be
       H~~ \put(30,30){\circle*{5}}\put(30,3){\line(0,1){25}} 
       \put(0,3){\line(1,0){60}} 
       ~~~~~~~~~~~~~~~~~~ H ~~~~~~~ + ~~~~~~~~ 
       H~~ \put(30,30){\circle{20}}\put(30,3){\line(0,1){17}} 
       \put(0,3){\line(1,0){60}} 
       ~~~~~~~~~~~~~~~~~~ H, 
\label{eq:higgsSE2}
\ee
but these diagrams add up to zero as a consequence of our choice for
$\beta_h$. 
%%%%%%%%%%%%%%%%%%%%%%%%%%%%%%%%%%%%%%%%%%%%%%%%%%%%%%%%%%%%%%%%%%%%%%%%%%%%%
\subsubsection{$\beta_h$ fixing up to two loops}

Up to terms of ${\mathcal{O}}(g^3)$, $\langle H \rangle_0$ gets
contributions from the following diagrams:
\bea
T_0:& \put(50,3){\circle*{5}}\put(10,3){\line(1,0){40}}~~~~~~~~~~~~~~~
        &\mbox{~~~~~~~~~~(1)~~~~~~~+~~} \nonumber\\ ~ \nonumber \\
T_1:& \put(50,3){\circle{20}}\put(10,3){\line(1,0){30}}~~~~~~~~~~~~~~~
        &\mbox{~~~~~~~~~~(1/2)~~~~+} \nonumber\\ ~ \nonumber \\
T_2:&\put(40,3){\circle{20}}\put(10,3){\line(1,0){40}}~~~~~~~~~~~~~~~
        & \!\!\!\!\!\ \mbox{(1/6)~~~+}
    \put(40,3){\circle{20}}\put(40,-7){\line(0,1){20}} 
        \put(10,3){\line(1,0){20}}~~~~~~~~~~~~~~~~~
	\mbox{(1/4)~~~+}
    \put(40,3){\circle{20}}\put(60,3){\circle{20}}
        \put(10,3){\line(1,0){20}}~~~~~~~~~~~~~~~
        \mbox{~~~~~~~(1/4)~~~+} \nonumber\\ ~ \nonumber \\
T_3:& \put(50,3){\circle{20}}\put(60,3){\circle*{5}}
        \put(10,3){\line(1,0){30}}~~~~~~~~~~~~~~~
        &\mbox{~~~~~~~~~~(1/2)~~~~+} \nonumber\\ ~ \nonumber \\
T_4:& \put(50,3){\circle{20}}\put(80,3){\circle{20}}
        \put(10,3){\line(1,0){30}}\put(60,3){\line(1,0){10}}~~~~~~~~~~~~~~~
        &\mbox{~~~~~~~~~~(1/4)~~~~+} 
     \put(50,3){\circle{20}}\put(80,3){\circle*{5}}
        \put(10,3){\line(1,0){30}}\put(60,3){\line(1,0){18}}~~~~~~~~~~~~~~~
        \mbox{~~~~~~~~~~~~(1/2)~~~~+} \nonumber\\ ~ \nonumber \\
T_5:& \put(50,3){\circle*{5}}\put(80,3){\circle{20}}
        \put(10,3){\line(1,0){38}}\put(52,3){\line(1,0){18}}~~~~~~~~~~~~~~~
        &\mbox{~~~~~~~~~~(1/2)~~~~+} 
     \put(50,3){\circle*{5}}\put(80,3){\circle*{5}}
        \put(10,3){\line(1,0){38}}\put(52,3){\line(1,0){27}}~~~~~~~~~~~~~~~
        \mbox{~~~~~~~~~~~~(1)~~~~~~~+} \nonumber\\ ~ \nonumber \\
T_6:& \put(50,10){\circle{20}}\put(80,0){\circle{20}}
        \put(10,0){\line(1,0){60}} ~~~~~~~~~~~~~~~
        &\mbox{~~~~~~~~~~(1/4)~~~~+} 
     \put(50,10){\circle{20}}\put(80,0){\circle*{5}}
        \put(10,0){\line(1,0){68}} ~~~~~~~~~~~~~~~
        \mbox{~~~~~~~~~~~~~(1/2)~~~+} \nonumber\\ ~ \nonumber \\
T_7:& \put(10,3){\line(1,0){20}}
     \put(30,3){\line(1,1){10}}
     \put(30,3){\line(1,-1){10}} 
     \put(44,17){\circle{10}}
     \put(44,-11){\circle{10}} ~~~~~~~~~~~~~~~ & 
     \!\!\!\!\!\mbox{(1/8)~~~~~~+~~~~~~}  
\put(0,3){\line(1,0){20}}
     \put(20,3){\line(1,1){13}}
     \put(20,3){\line(1,-1){10}} 
     \put(34,17){\circle*{5}}
     \put(34,-11){\circle{10}} \mbox{~~~~~~~~~~~~(1/2)~~~~~~+~~~~~~}  
\put(0,3){\line(1,0){20}}
     \put(20,3){\line(1,1){13}}
     \put(20,3){\line(1,-1){13}} 
     \put(34,17){\circle*{5}}
     \put(34,-11){\circle*{5}} \mbox{~~~~~~~~~~~~(1/2).}  \nonumber\\
     \nonumber
\eea
The coefficients in parentheses indicate the combinatorial factors of each
diagram when all fields are identical. Owing to our previous choice for
$\beta_{h_0}$ and $\beta_{h_1}$, all the reducible diagrams add up to
zero: $T_4=T_5=T_6=T_7=0$.  The equation
\be 
    \sum_{i=0}^3 T_i = 0 
\ee 
provides then $\beta_{h_2}$:
\be
    \beta_{h_2} = \frac{1}{(2\pi)^4 i} \left( \frac{T_2+T_3}{2Mg^3}\right).
\ee
%

%%%%%%%%%%%%%%%%%%%%%%%%%%%%%%%%%%%%%%%%%%%%%%%%%%%%%%%%%%%%%%%%%%%%%%%%%%%%%
\subsubsection{$\beta_h$ vertices in two-loop calculations}

The two-leg $\beta_h$ vertices in
Eqs.~(\ref{eq:betavH},\ref{eq:betavp0},\ref{eq:betavppm}) should be included
in all the appropriate diagrams at the two-loop level, while all graphs (up
to two loops) containing tadpoles will add up to zero as a consequence of
our choice for $\beta_{h_0}$, $\beta_{h_1}$ and $\beta_{h_2}$. Note that
two-leg $\beta_h$ vertices will also appear in the ${\mathcal{O}}(g^4)$
self-energies of fields which do not belong to the Higgs sector; for
example, in diagrams like these:
$$
Z ~~~~~~~~~~~~\put(0,3){\circle{20}}
\put(8,13){$\scriptstyle{H}$}\put(-15,13){$\scriptstyle{H}$}
\put(8,-12){$\scriptstyle{Z}$}
\put(0,13){\circle*{5}}
\put(10,3){\line(1,0){30}}
\put(-40,3){\line(1,0){30}}
~~~~~~~~~~~~Z 
~~~~~~~~~~~~~~~~
Z ~~~~~~~~~~~\put(0,3){\circle{20}}
\put(8,13){$\scriptstyle{H}$}\put(-15,13){$\scriptstyle{H}$}
\put(0,13){\circle*{5}}
\put(-40,-7){\line(1,0){80}}
~~~~~~~~~~~Z,
$$
which are representative of the only two irreducible ${\mathcal{O}}(g^4)$
$Z$ self-energy topologies containing $\beta_h$ vertices (excluding
tadpoles, of course).
%%%%%%%%%%%%%%%%%%%%%%%%%%%%%%%%%%%%%%%%%%%%%%%%%%%%%%%%%%%%%%%%%%%%%%%%%%%%%
\subsection{\bm The $\bb$ Scheme \ubm}
\label{subsec-betat}

\subsubsection{Definitions and Lagrangian}
\label{subsubsec-betatdef}

Tadpoles do not depend on any particular scale other than their internal
mass, and cancel in any renormalized self-energy. However, they play an
essential role in proving the gauge invariance of all the building blocks of
the theory. In order to exploit this option, we will now consider a slightly
different strategy to set the Higgs VEV to zero. Instead of using
Eqs.~(\ref{eq:betaH}), the ``$\beta_h$ scheme'', we will define the new bare
parameters $M'$ (the $W$ boson mass), $\mh'$ (the mass of the physical Higgs
particle) and $\bb$ (the tadpole constant) according to the following
``$\bb$ scheme'':
\renewcommand{\arraystretch}{1.5}
\be
\left\{
  \begin{array}{lll}
     M'(1+\bb) \!\! &=& g v/2 \\
     (\mh')^2 &=& \lambda \left( 2M'/g \right)^2 \\
     0 &=& \mu^2  +\frac{\lambda}{2}\left(2M'/g\right)^2
  \end{array}
\right.
\!\! \Longrightarrow~
\left\{
  \begin{array}{lll}
    v &=& 2M'(1+\bb)/g \\
    \lambda &=& \left(g \mh'/2M'\right)^2 \\
    \mu^2 &=& -\frac{1}{2}(\mh')^2
  \end{array}
\right.
\label{eq:betat}
\ee
\renewcommand{\arraystretch}{1}

\noindent 
The new set of bare parameters is therefore $g,g'$, $M',\mh'$ and $\bb$ instead
of $g,g'$ and $\mu,\lambda,v$ or $g,g'$ $M,\mh$ and $\beta_h$. Remember 
that $\bb$
(like $v$ and $\beta_h$) is not an independent parameter.  Note that,
contrary to $\beta_h$, the parameter $\bb$ appears in the Higgs doublet $K$
via $\zeta=H+v$, with $v=2M'(1+\bb)/g$. As a consequence, all three terms of
the Lagrangian $\cL_S$ in \eqn{eq:LS} depend on this parameter. In
particular, the interaction part of $\cL_S$ becomes
%--
\bea
    \cL_S^I  &=& -\mu^2 K^\dagger K - (\lambda/2) (K^\dagger K)^2 \\
    &=& \left(1+\bb\right)^2 \!
        \Big( 1-\bb \left(2+\bb\right)\Big)\frac{\mh'^2 M'^2}{2g^2}   
      -\bb \left(\bb+1 \right)  \left(\bb+2 \right)
	\frac{\mh'^2 M'}{g}H \nonumber\\
    &&  -\frac{1}{2} \mh'^2 H^2  
        -\frac{1}{4} \mh'^2 \bb \left( \bb+2\right)
	      \left(3H^2+\phi_0^2+2\phi_+\phi_- \right) 
		\nonumber\\
    && -g \left(1+\bb\right) \frac{\mh'^2}{4M'} H 
		\left( H^2 +\phi_0^2 +2\phi_+ \phi_- \right) 
     -g^2 \frac{\mh'^2}{32M'^2} \left( H^2 +\phi_0^2 +2\phi_+ \phi_-
    \right)^2,
\label{eq:LSIt}
\eea
while the term of $\cL_S$ involving $-(D_{\mu} K)^\dagger (D_{\mu} K)$, 
yields a (lengthy) {\em $\bb$-independent}
expression (see refs.~\cite{Bardin:1999ak} and \cite{Veltman:1994wz}), {\em
plus} the following terms containing $\bb$:
\bea
     \bb &\times& \left[ ig \stw M'
                 \left(\phi^- W^+_\mu - \phi^+ W^-_\mu\right)
		 \left(A_\mu -\frac{\stw}{\ctw}Z_\mu\right) 
		 \phantom{\frac{Z_\mu^2}{\ctws}}\!\!\!\!\!\!
	        -\frac{gM'}{2} H \left(2W^+_\mu W^-_\mu  +
		 \frac{Z_\mu Z_\mu}{\ctws}\right) \right.
		 \nonumber \\
	      && \left. -\frac{M'^2}{2} \left(\bb+2\right)
		 \left(2W^+_\mu W^-_\mu  +
		 \frac{Z_\mu Z_\mu}{\ctws}\right) \right.
	       \left. + \frac{M'}{\ctw} Z_\mu \partial_\mu \phi_0
		 + M' W^+_\mu \partial_\mu \phi_-
		 + M' W^-_\mu \partial_\mu \phi_+ 
		 \phantom{\frac{Z_\mu^2}{\ctws}} \!\!\!\!\!\!\!\!\right], 
\label{eq:LSDt}
\eea
where, as usual, $W_{\mu}^{\pm} = (B_{\mu}^1 \mp iB_{\mu}^2)/\sqrt{2}$, and
\be
        \left(\begin{array}{c} Z_\mu \\ A_\mu \end{array} \right) =
	\left(\begin{array}{cc} \ctw& -\stw \\ \stw& \ctw\end{array} \right)
	\left(\begin{array}{c} B^3_\mu \\ B^0_\mu \end{array} \right).
\ee

Where else, in the SM Lagrangian, does the parameter $\bb$ appear?  Wherever
$v$ does --- as it can be readily seen from \eqn{eq:betat}. Let us now
quickly discuss the other sectors of the SM: Yang--Mills, fermionic,
Faddeev--Popov (FP) and gauge-fixing.
The pure Yang--Mills Lagrangian obviously contains no $\bb$ terms.

The gauge-fixing part of the Lagrangian, $\cL_{gf}$, cancels in the $R_\xi$
gauges the gauge--scalar mixing terms $Z$--$\phi_0$ and $W^\pm$--$\phi^\pm$
contained in the scalar Lagrangian $\cL_S$.  These terms are proportional to
$gv/2$, i.e., to $M'(1+\bb)$ in the $\beta_t$ scheme, and to $M$ in the
$\beta_h$ scheme. The gauge-fixing Lagrangian $\cL_{gf}$ is a matter of
choice: we adopt the usual definition
\be
       \cL_{gf} = -\cC_{+} \cC_{-} 
                       -\frac{1}{2}\cC_{\ssZ}^2
                       -\frac{1}{2}\cC_{\ssA}^2,
\label{eq:Lgf}
\ee
with
%--
\be
       \cC_{\ssA} = -\frac{1}{\xi_{\ssA}}
                   \partial_{\mu} A_{\mu},
\quad
       \cC_{\ssZ} = -\frac{1}{\xi_{\ssZ}}
                   \partial_{\mu} Z_{\mu} +{\xi_{\ssZ}} 
		   \frac{M'}{\ctw} \phi_0,   
\quad
       \cC_{\pm} = -\frac{1}{\xi_{\ssW}}
              \partial_{\mu} W_{\mu}^\pm +{\xi_{\ssW}}M' \phi_{\pm} 
\label{eq:CBgf}
\ee
%--
(note: no $\bb$ terms), thus canceling the $\cL_S$ $g$-independent
gauge--scalar mixing terms proportional to $M'$, but not those proportional
to $M'\bb$ (appearing at the end of \eqn{eq:LSDt}), which are of
${\mathcal{O}}(g^2)$.  Alternatively, one could choose $M'(1+\bb)$ instead
of $M'$ in \eqn{eq:CBgf}, thus canceling all $\cL_S$ gauge--scalar mixing
terms, both proportional to $M'$ and $M'\bb$, but introducing then new
two-leg $\bb$ vertices. In this latter case, as $M=M'(1+\bb)$, the gauge
fixing Lagrangian would be identical to the one of the $\beta_h$ scheme. We
will not follow this latter approach. Of course it is only a matter of
choice, but the explicit form of $\cL_{gf}$ determines the FP ghost
Lagrangian.

The parameter $\bb$ shows up also in the FP ghost sector. The FP
Lagrangian depends on the gauge variations of the chosen gauge-fixing
functions $\cC_{\ssA}$, $\cC_{\ssZ}$ and ${\cal
C}_{\pm}$. If, under gauge transformations, the functions $\cC_{i}$
transform as
\be
\cC_{i}\rightarrow \cC_{i}+\left(M_{ij}+gL_{ij}\right)\Lambda_j,
\label{eq:Ctransf}
\ee
with $i= (A,Z,\pm)$, then the FP ghost Lagrangian is given by
\be
    \cL_{FP} = \Xb_i \left(M_{ij}+gL_{ij}\right) X_j.
\label{eq:LFPdef}
\ee
With the choice for $\cL_{gf}$ given in \eqn{eq:Lgf} (and the relation 
$gv/2=M'(1+\bb)$) it is easy to check that the FP ghost Lagrangian contains 
the $\bb$ terms
\be
        \cL_{\mysmall{FP}} = -\left(M'\right)^2 \bb \left( 
	\xi_{\ssW}\Xb^+ X^+ + 
	\xi_{\ssW}\Xb^- X^- + 
	\xi_{\ssZ}\Xb_{\ssZ} X_{\ssZ} /\ctws
	\right) + \cdots,
\label{eq:LFP}
\ee
where the dots indicate the usual $\bb$--independent terms. Had we chosen
$\cL_{gf}$ with $M'(1+\bb)$ instead of $M'$ in \eqn{eq:CBgf},
additional $\bb$ terms would now arise in the FP Lagrangian.

In the fermionic sector, the tadpole constant $\bb$ appears in the mass
terms:
\be
        \frac{v}{\sqrt2}\left(-\alpha \bar{u} u+\beta \bar{d} 
	                           d \right)= 
	                  -\left(1+\bb\right) 
			  \left(m_u \bar{u} u + m_d \bar{d} d \right)
\label{eq:Lfmass}
\ee
$(v=2M'(1+\bb)/g)$, where $\alpha$ and $\beta$ are the Yukawa couplings, and
$m_u$, $m_d$ are the masses of the fermions. The rest of the fermion
Lagrangian does not contain $\bb$, as it doesn't depend on $v$.

The Feynman rules for vertices involving a $\bb$ factor (``$\bb$ vertices'')
are listed in Appendix B, dropping the primes over $M'$ and $\mh'$. In the
$\bb$ scheme, contrary to the $\beta_h$ one, we have (many) two- and
three-leg $\bb$ vertices containing also non-scalar fields. Note that
three-leg $\bb$ vertices introduce a fourth irreducible topology for
${\mathcal{O}}(g^4)$ self-energy diagrams containing $\beta_t$ vertices,
namely:
$$
~~~~~~~~~~~~~
\put(-30,3){\line(1,0){60}}
\put(0,3){\circle*{5}}
~~~~~~~~~~~~~~~~~~~~~~~~~~~
\put(0,3){\circle{20}}
\put(0,13){\circle*{5}}
\put(10,3){\line(1,0){20}}
\put(-30,3){\line(1,0){20}}
~~~~~~~~~~~~~~~~~~~~~~~~~~~
\put(0,3){\circle{20}}
\put(0,13){\circle*{5}}
\put(-30,-7){\line(1,0){60}}
~~~~~~~~~~~~~~~~~~~~~~~~~~~
\put(0,3){\circle{20}}
\put(-10,3){\circle*{5}}
\put(10,3){\line(1,0){20}}
\put(-30,3){\line(1,0){20}}
~~~~~~~~~~~~~.
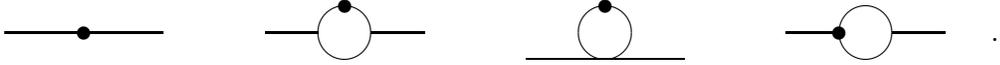
$$
%%%%%%%%%%%%%%%%%%%%%%%%%%%%%%%%%%%%%%%%%%%%%%%%%%%%%%%%%%%%%%%%%%%%%%%%%%%%%
\subsubsection{$\bb$ up to one loop}

Define $\bb = \beta_{t_0} +\beta_{t_1} g^2 + \beta_{t_2} g^4 + \cdots $.  As
we did for $\beta_h$, we will now fix the parameter $\bb$ such that the VEV
of the Higgs field $H$ remains zero order by order in perturbation theory.
At the lowest order, the only diagram contributing to $\langle H \rangle_0$
is the same one depicted in \eqn{eq:betaset0}, which origins from the term
in $\cL_S^I$ linear in $H$, $-\bb(\bb+1)(\bb+2)(\mh'^2 M'/g)H$. Therefore,
at the lowest order we can simply set $\bb=0$, i.e. $\beta_{t_0} = 0$.

Up to one loop, the diagrams $T'_0$ and $T'_1$ contributing to the Higgs VEV
are analogous to $T_0$ and $T_1$ appearing in \eqn{eq:betaset1}, so that
$\beta_{t_1}$ can be set in analogy with $\beta_{h_1}$:
\be
    \beta_{t_1} = \frac{1}{(2\pi)^4 i} \left( \frac{T'_1}{2M'g\mh'^2}\right).
\label{eq:betat1set}
\ee
Note that $T'_1$ and $T_1$ have the same functional form, but depend on
different mass parameters.
%moreover, one gets $\beta_{t_1} =
%\beta_{h_1}/\mh^2 +{\mathcal{O}}(g^2)$.

%%%%%%%%%%%%%%%%%%%%%%%%%%%%%%%%%%%%%%%%%%%%%%%%%%%%%%%%%%%%%%%%%%%%%%%%%%%%%
\subsubsection{$\bb$ up to two loops}

The two-loop $\bb$ fixing slightly differs from the $\beta_h$ one.  Up to
terms of ${\mathcal{O}}(g^3)$, $\langle H \rangle_0$ gets contributions from
the following diagrams:
\bea
T'_0:& \put(50,3){\circle*{5}}\put(10,3){\line(1,0){40}}~~~~~~~~~~~~~~~
        &\mbox{~~~~~~~~~~(1)~~~~~~~+~~} \nonumber\\ ~ \nonumber \\
T'_1:& \put(50,3){\circle{20}}\put(10,3){\line(1,0){30}}~~~~~~~~~~~~~~~
        &\mbox{~~~~~~~~~~(1/2)~~~~+} \nonumber\\ ~ \nonumber \\
T'_2:&\put(40,3){\circle{20}}\put(10,3){\line(1,0){40}}~~~~~~~~~~~~~~~
        & \!\!\!\!\!\ \mbox{(1/6)~~~+}
    \put(40,3){\circle{20}}\put(40,-7){\line(0,1){20}} 
        \put(10,3){\line(1,0){20}}~~~~~~~~~~~~~~~~~
	\mbox{(1/4)~~~+}
    \put(40,3){\circle{20}}\put(60,3){\circle{20}}
        \put(10,3){\line(1,0){20}}~~~~~~~~~~~~~~~
        \mbox{~~~~~~~(1/4)~~~+} \nonumber\\ ~ \nonumber \\
T'_3:& \put(50,3){\circle{20}}\put(60,3){\circle*{5}}
        \put(10,3){\line(1,0){30}}~~~~~~~~~~~~~~~
        &\mbox{~~~~~~~~~~(1/2)~~~~+~~~~~~} 
\put(50,3){\circle{20}}\put(40,3){\circle*{5}}
        \put(10,3){\line(1,0){30}}~~~~~~~~~~~~~~~
        \mbox{~~~~~~~~~~(1/2),}
\nonumber
\eea
plus reducible diagrams (analogous to those appearing in $T_4$--$T_7$
of section 2.4) which add up to zero because of our choice for $\beta_{t_0}$
and $\beta_{t_1}$. Note the new diagrams in $T'_3$, with three-leg $\bb$
vertices, not present in the $\beta_h$ case ($T_3$). The parameter
$\beta_{t_2}$ can be set in the usual manner, requiring
\be 
    \sum_{i=0}^3 T'_i = 0, 
    ~~~~~\Longrightarrow~~~~~
    \beta_{t_2} = \frac{1}{(2\pi)^4 i} \left(
    \frac{T'_2+T'_3}{2M'g^3\mh'^2}\right) -\frac{3}{2}\beta_{t_1}^2.  
\ee
Note that $T'_{1,2}$ and $T_{1,2}$ have the same functional form (but depend
on different mass parameters) while $T'_3$ and $T_3$ are different.

%%%%%%%%%%%%%%%%%%%%%%%%%%%%%%%%%%%%%%%%%%%%%%%%%%%%%%%%%%%%%%%%%%%%%%%%%%%%%
\subsection{\bm $\beta_h$ and $\bb$: two comments \ubm}

Consider the (doubly-contracted) WST identity relating the $Z$ self-energy
$\Pi_{\mu\nu,\ssZZ}(p)$, the $\phi_0$ self-energy $\Pi_{\mysmall{\phi_o
\phi_o}}(p)$, and the $Z$--$\phi_0$ transition $\Pi_{\mu,\mysmall{Z
\phi_o}}(p)$ (see \sect{sec-WSTI}):
\be
     p_{\mu}p_{\nu} \Pi_{\mu\nu,\ssZZ}(p) ~+~
     M_0^2 \Pi_{\mysmall{\phi_o \phi_o}}(p)                 ~+~
     2ip_{\mu} M_0 \Pi_{\mu,\mysmall{Z \phi_o}}(p)     ~=~ 0.
\label{eq:WI}
\ee
Both in the $\beta_h$ and $\beta_t$ schemes, each of the three terms in
\eqn{eq:WI} contains contributions from the tadpole diagrams, but they add
up to zero, within each term.  For example, at the one-loop level, the first
term in \eqn{eq:WI} contains the tadpole diagrams \vspace{3mm}
\be
       Z~~ \put(30,30){\circle*{5}}\put(30,3){\line(0,1){25}} 
       \put(0,3){\line(1,0){60}} 
       ~~~~~~~~~~~~~~~~~~ Z ~~~~~~~ \put(0,15){\mbox{and}} ~~~~~~~~ 
       Z~~ \put(30,30){\circle{20}}\put(30,3){\line(0,1){17}} 
       \put(0,3){\line(1,0){60}} 
       ~~~~~~~~~~~~~~~~~~ Z 
\label{eq:WItadpoles}
\ee
which cancel each other. In the $\beta_h$ scheme at the one-loop level, only
the second term of the l.h.s.\ of \eqn{eq:WI} includes a diagram with a
two-leg $\beta_h$ vertex (\eqn{eq:betavp0}), while in higher orders, two-leg
$\beta_h$ vertices appear in all three terms. In the $\beta_t$ scheme, all
three terms of \eqn{eq:WI} contain the two-leg $\beta_t$ vertices already at
the one-loop level. Similar comments are valid for the WST identity
involving the $W$ self-energy.

Concerning renormalization, the constraints imposed on $\beta_h$ and $\bb$
in the previous sections are the renormalization conditions to insure that
$\langle 0|H|0 \rangle=0$, also in the presence of radiative corrections. In
particular, the renormalized $\beta_{h,t}$ parameters are
$\beta_{h,t}^{\mysmall{(R)}} = \beta_{h,t} +\delta \beta_{h,t} =0$.  The
equivalent of Eqs.~(\ref{eq:betaH}) and (\ref{eq:betat}) for the
renormalized parameters are just the same equations with the tadpole
constants set to zero.  In the $\beta_h$ scheme, the one-loop
renormalization of the $W$ and $Z$ masses involves the diagrams
\vspace{2mm}
\be
       (a)~~\put(15,10){\circle{10}}\put(20,10){\line(1,0){10}} 
       \put(0,10){\line(1,0){10}} 
       ~~~~~~~~~~~~~~~~~~ 
       (b)~~\put(15,17){\circle{10}}\put(15,3){\line(0,1){9}} 
       \put(0,3){\line(1,0){30}} 
       ~~~~~~~~~~~~~~~~~~
       (c)~~\put(15,17){\circle*{4}}\put(15,3){\line(0,1){14}} 
       \put(0,3){\line(1,0){30}} 
       ~~~~~~~~~~~. 
\label{eq:MassRenH}
\ee
(Diagrams $(a)$ have two possible loop topologies.)  Both $(a)$ and $(b)$
are gauge-dependent, but their sum is gauge-independent on-shell. However,
as we choose the $\beta_h$ tadpole $(c)$ to cancel $(b)$, the mass
counterterm contains only $(a)$ and is therefore gauge-dependent. On the
contrary, in the $\beta_t$ scheme, the one-loop renormalization of the $W$
and $Z$ masses involves the diagrams
\vspace{2mm}
\be
       (a)~~\put(15,10){\circle{10}}\put(20,10){\line(1,0){10}} 
       \put(0,10){\line(1,0){10}} 
       ~~~~~~~~~~~~~~~~~~ 
       (c)~~\put(15,10){\circle*{4}}\put(0,10){\line(1,0){30}} 
       ~~~~~~~~~~~~~~~~~~ 
       (b)~~\put(15,17){\circle{10}}\put(15,3){\line(0,1){9}} 
       \put(0,3){\line(1,0){30}} 
       ~~~~~~~~~~~~~~~~~~
       (d)~~\put(15,17){\circle*{4}}\put(15,3){\line(0,1){14}} 
       \put(0,3){\line(1,0){30}} 
       ~~~~~~~~~~~. 
\label{eq:MassRenT}
\ee
Once again, both $(a)$ and $(b)$ diagrams are gauge-dependent, their sum is
gauge-independent on-shell, and the $\beta_t$ tadpole $(d)$ is chosen to
cancel $(b)$. But, the mass counterterm is now gauge-independent, as it
contains both $(a)$ and the two-leg $\beta_t$ vertex diagram $(c)$ (which is
missing in the $\beta_h$ case).
%%%%%%%%%%%%%%%%%%%%%%%%%%%%%%%%%%%%%%%%%%%%%%%%%%%%%%%%%%%%%%%%%%%%%%%%%%%%%
\section{Diagonalization of the neutral sector}
\label{sec-gamma}

\subsection{New coupling constant in the \bm $\beta_h$ \ubm scheme} 
\label{subsec-gammah}

The $Z$--$\gamma$ transition in the SM does not vanish at zero squared
momentum transfer. Although this fact does not pose any serious problem, not
even for the renormalization of the electric charge, it is preferable to use
an alternative strategy. We will follow the treatment of
Ref.~\cite{Passarino:1990xx}. Consider the new $SU(2)$ coupling
constant $\Mgb$, the new mixing angle $\bar{\theta}$ and the new $W$ mass
$\bar{M}$ in the $\beta_h$ scheme:
\renewcommand{\arraystretch}{1.5}
\be
  \begin{array}{lll}
    g &=& \Mgb \left(1+\Gamma\right)  \qquad
    g'= -(\sin\bar{\theta}/\cos\bar{\theta}) \, \Mgb \\
    v &=& 2\bar{M}/\Mgb \quad
    \lambda = \left(\Mgb \mh/2\bar{M}\right)^2 \quad
    \mu^2 =  \beta_h -\frac{1}{2}\mh^2
  \end{array}
\label{eq:gbinbetah}
\ee
(note: $g\sin\theta/\cos\theta=\Mgb\sin\bar{\theta}/\cos\bar{\theta}$), where
$\Gamma = \Gamma_1 \, \Mgb^2 + \Gamma_2 \,\Mgb^4 + \cdots~$ is a new parameter
yet to be specified.  This change of parameters entails new $\Ab_{\mu}$
and $\Zb_{\mu}$ fields related to $B^3_\mu$ and $B^0_\mu$ by
\renewcommand{\arraystretch}{1}
\be
    \left(\begin{array}{c} 
      \Zb_\mu \\ \Ab_\mu \end{array} \right) =
    \left(\begin{array}{cc} 
      \cos\bar{\theta}& -\sin\bar{\theta} \\ 
      \sin\bar{\theta}& \cos\bar{\theta}\end{array} \right)
    \left(\begin{array}{c} B^3_\mu \\ B^0_\mu \end{array} \right).
\label{eq:newAZ}
\ee
The replacement $g\rightarrow\Mgb(1+\Gamma)$ introduces in the SM Lagrangian
several terms containing the new parameter $\Gamma$. 
In our approach $\Gamma$ is fixed, order-by-order, by requiring that
the $Z$--$\gamma$ transition is zero at $p^2 = 0$ in the $\xi = 1$ gauge.
Let us take a close look at these `$\Gamma$ terms' in each sector of the SM.

\vspace{2mm}
\noindent $\bullet$ The pure Yang--Mills Lagrangian 
\be
    \cL_{\ssY\ssM}  = 
                    -\frac{1}{4}F_{\mu\nu}^a F_{\mu\nu}^a
                    -\frac{1}{4}F_{\mu\nu}^0 F_{\mu\nu}^0,
\label{eq:YM}
\ee
with $~F_{\mu\nu}^a = \partial_{\mu} B_{\nu}^a - \partial_{\nu} B_{\mu}^a +g
\epsilon^{abc} B_{\mu}^b B_{\nu}^c~$ and $~F_{\mu\nu}^0 = \partial_{\mu}
B_{\nu}^0 - \partial_{\nu} B_{\mu}^0$, contains the following new $\Gamma$
terms when we replace $g$ by $\Mgb(1+\Gamma)$:
\bea
     \Delta \cL_{\ssY\ssM}  \!\!\!\!&=&
      \!\!\!\!-i \Mgb \Gamma \ccb \left[ \partial_{\nu} \Zb_{\mu} 
      \left(W^+_{\mu} W^-_{\nu} - W^+_{\nu} W^-_{\mu} \right) \right.
     \left. - \Zb_{\nu} \left(W^+_{\mu} \partial_{\nu} W^-_{\mu} - 
      W^-_{\mu} \partial_{\nu} W^+_{\mu} \right) + \right.
      \nonumber \\ 
      && \left. ~+ \Zb_{\mu} \left(W^+_{\nu} \partial_{\nu} W^-_{\mu} - 
      W^-_{\nu} \partial_{\nu} W^+_{\mu} \right) \right]
     -i \Mgb \Gamma \ssb \left[ \partial_{\nu} \Ab_{\mu} 
      \left(W^+_{\mu} W^-_{\nu} - W^+_{\nu} W^-_{\mu} \right) \right.
      \nonumber \\
     &&\left. ~- \Ab_{\nu} \left(W^+_{\mu} \partial_{\nu} W^-_{\mu} - 
      W^-_{\mu} \partial_{\nu} W^+_{\mu} \right) 
      + \Ab_{\mu} \left(W^+_{\nu} \partial_{\nu} W^-_{\mu} - 
      W^-_{\nu} \partial_{\nu} W^+_{\mu} \right) \right]
      \nonumber \\
     && ~\!\!\!\!\!\!\!+\,\Mgb^2 \Gamma \left(2+\Gamma \right) 
      \left[\frac{1}{2}\left(W^+_{\mu}W^-_{\nu}W^+_{\mu}W^-_{\nu}
           -W^+_{\mu}W^-_{\mu}W^+_{\nu}W^-_{\nu}\right) \right.
      \nonumber \\
     && \left. ~
      +\ccb^2\left(\Zb_{\mu}W^+_{\mu}\Zb_{\nu}W^-_{\nu}
      -\Zb_{\mu}\Zb_{\mu}W^+_{\nu}W^-_{\nu} \right) \right.
      \left. ~
      +\ssb^2\left(\Ab_{\mu}W^+_{\mu}\Ab_{\nu}W^-_{\nu} 
      -\Ab_{\mu}\Ab_{\mu}W^+_{\nu}W^-_{\nu} \right) \right.  
      \nonumber \\
     && \left. ~
      +\ssb \ccb \left(\Ab_{\mu}\Zb_{\nu}
      (W^+_{\mu} W^-_{\nu} + W^+_{\nu} W^-_{\mu}) 
      -2\Ab_{\mu}\Zb_{\mu}W^+_{\nu}W^-_{\nu} \right) 
      \phantom{\frac{1}{2}} \!\!\!\!\!\right],
\label{eq:SpecialYMh}
\eea
where $\ssb = \sin \bar{\theta}$ and $\ccb = \cos \bar{\theta}$. As these
terms are of ${\mathcal{O}}(\Mgb^3)$ or ${\mathcal{O}}(\Mgb^4)$, they do not
contribute to the calculation of self-energies at the one-loop level, but
they do beyond it.

\vspace{2mm}
\noindent $\bullet$ The Lagrangian $\cL_S$, \eqn{eq:LS}, contains several
new $\Gamma$ terms when we employ the relation $g=\Mgb(1+\Gamma)$ and the
$\beta_h$ scheme of Eqs.~(\ref{eq:gbinbetah}).  They can be arranged in the
following three classes
\be \Delta \cL_{S,\,h} = \Delta \cL_{S,\,h}^{(n_f=2)} + 
                              \Delta \cL_{S,\,h}^{(n_f=3)} +
			      \Delta \cL_{S,\,h}^{(n_f=4)},
\label{eq:SpecialLSh}
\ee
according to the number of fields ($n_f$) appearing in each interaction term
(indicated by the superscript in parentheses). The explicit expressions, up
to terms of ${\mathcal{O}}(\Mgb^4)$, are
\bea
     \Delta \cL_{S,\,h}^{(n_f=2)} &=& \MMb\Gamma \left[ 
       -\frac{1}{2} \MMb \ssb^2 \Gamma \Ab_{\mu} \Ab_{\mu}
       -\frac{1}{2} \MMb \left(2+\Gamma \ccb^2\right) \Zb_{\mu} \Zb_{\mu}
        \right. \nonumber \\  & & \left.
      - \MMb \,\frac{\ssb}{\ccb} \left(1+\Gamma \ccb^2\right) 
        \Ab_{\mu} \Zb_{\mu} 
       +\partial_{\mu} \phi_0 \left( \ssb \Ab_{\mu}+\ccb \Zb_{\mu} \right) 
        \right. \nonumber \\ & & \left.
       -\MMb \left(2+\Gamma\right) W^+_{\mu} W^-_{\mu}
	+ W^-_{\mu} \partial_{\mu} \phi^+ 
	+ W^+_{\mu} \partial_{\mu} \phi^- 
	\phantom{\frac{1}{2}} \!\!\!\!\!\right],
\label{eq:SpecialLSh2}
     \\ ~ \nonumber \\
     \Delta \cL_{S,\,h}^{(n_f=3)} &=& \Mgb\Gamma \left[ 
       -\MMb H \left( \Zb_{\mu} \Zb_{\mu} + 
         \frac{\ssb}{\ccb} \Ab_{\mu} \Zb_{\mu}  
       +2 W^+_{\mu} W^-_{\mu} \right) 
        \right. \nonumber \\  & & \left.
	+\frac{1}{2} \left( \ssb \Ab_{\mu} + \ccb \Zb_{\mu} \right) 
	\left( H \partial_{\mu} \phi^0 - \phi^0 \partial_{\mu} H +
	i \phi^+ \partial_{\mu} \phi^- 
	- i \phi^- \partial_{\mu} \phi^+\right)
        \right. \nonumber \\  & & \left.
	+ i \left( \phi^- W^+_{\mu} - \phi^+ W^-_{\mu}  \right)
	\left(\ssb \MMb \Ab_{\mu} - (\ssb^2/\ccb) \MMb \Zb_{\mu}  
	+\frac{1}{2} \partial_{\mu} \phi^0 \right)
        \right. \nonumber \\  & & \left.
	+\frac{1}{2} W^-_{\mu}\partial_{\mu}\phi^+ \left(H+i\phi^0\right)
	+\frac{1}{2} W^+_{\mu}\partial_{\mu}\phi^- \left(H-i\phi^0\right)
        \right. \left.
	-\frac{1}{2} \partial_{\mu} H 
	\left(\phi^+ W^-_{\mu}+\phi^- W^+_{\mu} \right) \right],
\label{eq:SpecialLSh3}
     \\ ~ \nonumber \\
     \Delta \cL_{S,\,h}^{(n_f=4)} &=& \frac{\Mgb^2}{2}\Gamma \left\{ 
       -\frac{1}{2}\left( H^2 + \phi_0^2\right) 
       \left( \Zb_{\mu} \Zb_{\mu} 
       + \frac{\ssb}{\ccb} \Ab_{\mu} \Zb_{\mu} + 
          2 W^+_{\mu} W^-_{\mu} \right) 
       \right. \nonumber \\  & & \left.
       + \phi^ + \phi^- \left(-2\ssb^2 \Ab_{\mu} \Ab_{\mu} + 
       (1-2\ccb^2)\Zb_{\mu} \Zb_{\mu} 
       +\left(\ssb/\ccb -4\ssb\ccb \right) \Ab_{\mu} \Zb_{\mu} \right)
       \right. \nonumber \\  & & \left.
       -2 W^+_{\mu} W^-_{\mu} \phi^+ \phi^- 
       + \left(\ssb \Ab_{\mu} -(\ssb^2/\ccb) \Zb_{\mu} \right) \times
       \right. \nonumber \\  & & \left.
       \,\,\,\,\, \times \left[
	 \phi_0 \left( \phi^+ W^-_{\mu} + \phi^- W^+_{\mu}\right)
       -i H \left( \phi^+ W^-_{\mu} - \phi^- W^+_{\mu}\right)
       \phantom{\frac{1}{2}} \!\!\!\!\! \right] \right\}.
\label{eq:SpecialLSh4}
\eea
The interaction part of the scalar Lagrangian, $\cL_S^I = -\mu^2 K^\dagger K
- (\lambda/2) (K^\dagger K)^2$, does not induce $\Gamma$ terms; these are
only originated by the term involving the covariant derivatives, $-(D_{\mu}
K)^\dagger (D_{\mu} K)$. On the other hand, as $M/g=\bar{M}/\Mgb$, the
$\beta_h$ terms induced by $\cL_S^I$ are given by \eqn{eq:LSI} expressed in
terms of the ratio $\bar{M}/\Mgb$ of the barred parameters.

\vspace{2mm}
\noindent $\bullet$ We choose the gauge-fixing Lagrangian $\cL_{gf}$
of \eqn{eq:Lgf} with the following gauge functions:

\be
       \cC_{\ssA} = -\frac{1}{\xi_{\ssA}}
                   \partial_{\mu} \Ab_{\mu},
\quad
       \cC_{\ssZ} = -\frac{1}{\xi_{\ssZ}}
                   \partial_{\mu} \Zb_{\mu} +{\xi_{\ssZ}} 
		   \frac{\bar{M}}{\ccb} \phi_0,   
\quad
       \cC_{\pm} = -\frac{1}{\xi_{\ssW}}
              \partial_{\mu} W_{\mu}^\pm +{\xi_{\ssW}}\bar{M} 
	      \phi_{\pm}. 
\label{eq:CBgfnewgH}
\ee
%--
This $R_\xi$ gauge $\Gamma$-independent $\cL_{gf}$ cancels the zeroth
order (in $\Mgb$) gauge--scalar mixing terms introduced by $\cL_{S}$,
but not those proportional to $\Gamma$. Had one chosen gauge-fixing
functions Eqs.~(\ref{eq:CBgfnewgH}) with unbarred
quantities, all the gauge--scalar mixing terms of $\cL_{S}$ would be
canceled, including those proportional to $\Gamma$, but additional new
$\Gamma$ vertices would also be introduced.

\vspace{2mm}
\noindent $\bullet$ New $\Gamma$ terms are also originated in the
Faddeev--Popov ghost sector. Studying the gauge transformations
(\eqn{eq:Ctransf}) of the gauge-fixing functions $\cC_{\ssA}$,
$\cC_{\ssZ}$ and $\cC_{\pm}$ defined in Eqs.~(\ref{eq:CBgfnewgH}), the 
additional new $\Gamma$ terms of the FP Lagrangian (which is defined in 
\eqn{eq:LFPdef}) in the $\beta_h$ scheme are:
\be \Delta \cL_{FP,\,h} = \Delta \cL_{FP,\,h}^{(n_f=2)} + 
                               \Delta \cL_{FP,\,h}^{(n_f=3)},
\label{eq:SpecialLFPh}
\ee
where the two-field terms are,
\be
     \Delta \cL_{FP,\,h}^{(n_f=2)} = 
      -\Gamma \MMb^2 \left[ \xi_{\ssZ} \Xb_{\ssZ}
	\left(X_{\ssZ} +\frac{\ssb}{\ccb} X_{\ssA}
          \right) + \xi_{\ssW} \left( \Xb_{+} X_{+}+\Xb_{-} X_{-}
	  \right) \right]\, ,
\label{eq:SpecialLFPh2}
\ee
and the three-field terms are
\bea
\label{eq:SpecialLFPh3}
     \Delta \cL_{FP,\,h}^{(n_f=3)} &=&  \Gamma \Mgb \left\{
       \phantom{\frac{1}{2}} \!\!\!\!\!
         \, i \ccb W^+_{\mu} \Big((\partial_{\mu}\Xb_{\ssZ}
	 /\xi_{\ssZ})X_{-}
                              -(\partial_{\mu}\Xb_{+}
         /\xi_{\ssW})X_{\ssZ}\Big)
			   \right.            \\ &
      +& \left.\,i \ssb W^+_{\mu} \Big((\partial_{\mu}\Xb_{\ssA}
         /\xi_{\ssA})X_{-}
                              -(\partial_{\mu}\Xb_{+}
	 /\xi_{\ssW})X_{\ssA}\Big)
			   \right.  \nonumber \\ &
      +& \left.\,i \ccb W^-_{\mu} \Big((\partial_{\mu}\Xb_{-}
         /\xi_{\ssW})X_{\ssZ}
                              -(\partial_{\mu}\Xb_{\ssZ}
	 /\xi_{\ssZ})X_{+}\Big)
			   \right.  \nonumber \\ &
      +& \left.\,i \ssb W^-_{\mu} \Big((\partial_{\mu}\Xb_{-}
         /\xi_{\ssW})X_{\ssA}
                              -(\partial_{\mu}\Xb_{\ssA}
	 /\xi_{\ssA})X_{+}\Big)
			   \right.  \nonumber \\ &
      +& \left.\,i \ccb \Zb_{\mu}\,\,\, \Big((\partial_{\mu}\Xb_{+}
         /\xi_{\ssW})X_{+}
                              -(\partial_{\mu}\Xb_{-}
	 /\xi_{\ssW})X_{-}\Big)
			   \right.  \nonumber \\ &
      +& \left.\,i \ssb \Ab_{\mu}\,\,   \Big((\partial_{\mu}\Xb_{+}
         /\xi_{\ssW})X_{+}
                              -(\partial_{\mu}\Xb_{-}
	 /\xi_{\ssW})X_{-}\Big)
			   \right.  \nonumber \\ &
      +& \left.\frac{1}{2}\xi_{\ssW} \MMb \Big[
        i\phi_0 \left(\Xb_{+} X_{+} 
                           - \Xb_{-} X_{-} \right)
        -H \left(\Xb_{+} X_{+} 
                           + \Xb_{-} X_{-} \right)\!\Big]
   			   \right.  \nonumber \\ &
      +& \left.\frac{1}{2\ccb} \xi_{\ssZ} \MMb \Xb_{\ssZ}\Big[
                               i X_{-}\phi_{+} - i X_{+}\phi_{-}
                      -\ssb H X_{\ssA} -\ccb H X_{\ssZ} \Big]
   			   \right.  \nonumber \\ &
      +& \left.\frac{i}{2} \xi_{\ssW} \MMb \Big[
                           \Xb_{-} \phi_{-} \left(\ccb X_{\ssZ} 
			   +\ssb X_{\ssA}\right)
			   -\Xb_{+} \phi_{+} \left(\ccb X_{\ssZ} 
			   +\ssb X_{\ssA}\right) \Big]
			   \right\}. \nonumber
\eea
(The bars over the FP ghost fields indicate conjugation. Obviously, the new
FP fields $X_{\ssA}$ and $X_{\ssZ}$ should also be denoted with the bar
indicating the field rediagonalization, just like the new fields $\Ab_{\mu}$
and $\Zb_{\mu}$. However, this notation would be confusing and we will leave
this point understood.) Note that the FP ghost -- gauge boson vertices are
simply the usual ones with $g$ replaced by $\Mgb \Gamma$. This is not the
case, in general, for the FP ghost -- scalar terms.

\vspace{2mm}
\noindent $\bullet$ Finally, the fermionic sector. The fermion -- gauge
boson Lagrangian,
\bea
     \cL_{fG}&=& \frac{i}{2\sqrt2}\,g \!\left[
       W^+_{\mu} \bar{u} \gamma_{\mu} \left(1+\gamma_5\right) d \,\,+
       W^-_{\mu} \bar{d} \gamma_{\mu} \left(1+\gamma_5\right) u\,
       \right] \nonumber \\
     && +\, \frac{i}{2c}\,g \, Z_{\mu} \,\bar{f} \gamma_{\mu}
       \!\left(I_3 -2Q_fs^2 +I_3 \gamma_5 \right)\!f 
        \,+i\, g s \,Q_f A_{\mu} \bar{f} \gamma_{\mu} f,
\label{eq:LfGbefore}
\eea
(where $I_3=\pm 1/2$ is the third component of the weak isospin of the
fermion $f$, and $Q_f$ its charge in units of $|e|$) becomes, under the
replacement $g\rightarrow\Mgb(1+\Gamma)$ and the $\theta$, $A_{\mu}$ and
$Z_{\mu}$ redefinitions,
\bea
     \cL_{fG}&=& \frac{i}{2\sqrt2}\,\Mgb \left(1+\Gamma\right)
       \left[
       W^+_{\mu}\bar{u} \gamma_{\mu} \left(1+\gamma_5\right) d \,+
       W^-_{\mu}\bar{d} \gamma_{\mu} \left(1+\gamma_5\right) u \,
       \right] \nonumber \\
     && +\, \frac{i}{2\ccb}\,\Mgb\, \Zb_{\mu}\, \bar{f} \gamma_{\mu}
       \left(I_3 -2Q_f\ssb^2 +I_3 \gamma_5 \right) \!f 
        \,+\,i \,\Mgb \ssb \,Q_f \Ab_{\mu} \bar{f} \gamma_{\mu} f
	\nonumber \\
     &&	+\, \frac{i}{2}\, \Mgb\, \Gamma 
	\left(\ssb \Ab_{\mu} +\ccb \Zb_{\mu} \right) I_3\,
	\bar{f} \gamma_{\mu} \left(1+\gamma_5\right) f.
\label{eq:LfGafter}
\eea
The new neutral and charged current $\Gamma$ vertices are immediately
recognizable. The CKM matrix has been set to unity.

The fermion--scalar Lagrangian does not induce $\Gamma$ terms. Indeed, the
Yukawa couplings $\alpha$ and $\beta$ in
\be
   \cL_{fS} =  -\alpha \bar{\psi}_{\ssL} K u_{\ssR} 
               \,-\, \beta \bar{\psi}_{\ssL} K^{c} d_{\ssR} 
	       \,+\, \mbox{h.c.}
\label{eq:LfS}
\ee
(where $K^{c}=i \tau_2 K^{\star}$ is the conjugate Higgs doublet) are set by
$\alpha v/\!\sqrt2 = m_u$ and $\beta v/\!\sqrt2 = -m_d$. As $v=2\MMb/\Mgb$,
it is $\alpha = \Mgb m_u/\!\sqrt2~\MMb$ and $\beta = -\Mgb
m_d/\!\sqrt2~\MMb$, and no $\Gamma$ appears in \eqn{eq:LfS}.

%Note that the relation between the Yukawa coupling $\alpha$ ($\beta$) and 
%$m_u$ ($m_d$) is not identical to that of \eqn{eq:Lfmass}.

\vspace{3mm} The Feynman rules for all these new $\Gamma$ vertices are
listed in Appendix C, up to terms of ${\mathcal{O}}(\Mgb^4)$.  Those
corresponding to the pure Yang--Mills Lagrangian (\eqn{eq:SpecialYMh}) are
not listed, as they are identical to the usual Yang--Mills ones, except for
the replacement $g \rightarrow \Mgb \Gamma$ in the three-leg vertices, and
$g^2 \rightarrow \Mgb^2 \Gamma(2+\Gamma)$ in the four-leg ones. In Appendix
C, all bars over the symbols (indicating rediagonalization) have been
dropped, except over $\Mgb$.
%%%%%%%%%%%%%%%%%%%%%%%%%%%%%%%%%%%%%%%%%%%%%%%%%%%%%%%%%%%%%%%%%%%%%%%%%%%%%
\subsection{New coupling constant in the \bm $\beta_t$ \ubm scheme} 
\label{subsec-gammat}
The $\beta_t$ scheme equations corresponding to Eqs.~(\ref{eq:gbinbetah})
are the following
\renewcommand{\arraystretch}{1.5}
\be
  \begin{array}{lll}
    g &=& \Mgb \left(1+\Gamma\right)  \qquad
    g'= -(\sin\bar{\theta}/\cos\bar{\theta})\, \Mgb \\
    v &=& 2\bar{M'}(1+\bb)/\Mgb \quad
    \lambda = \left(\Mgb \mh'/2\bar{M'}\right)^2 \quad
    \mu^2 = -\frac{1}{2}(\mh')^2.
  \end{array}
\label{eq:gbinbetat}
\ee
(Note: $g\sin\theta/\cos\theta=\Mgb\sin\bar{\theta}/\cos\bar{\theta}$.) The
analysis of the $\Gamma$ terms presented in the previous section for the
$\beta_h$ scheme can be repeated for the $\beta_t$ scheme using
Eqs.~(\ref{eq:gbinbetat}) instead of Eqs.~(\ref{eq:gbinbetah}).  The new
fields $\Ab_{\mu}$ and $\Zb_{\mu}$ are related to $B^3_\mu$ and
$B^0_\mu$ by \eqn{eq:newAZ}. Thus, we obtain the following results:

\vspace{2mm}
\noindent $\bullet$ The replacement $g\rightarrow\Mgb(1+\Gamma)$ in the pure
Yang--Mills sector introduces new $\Gamma$ vertices collected in $\Delta
\cL_{\ssY\ssM}$, which does not depend on the parameters
of the $\beta_{h,t}$ schemes. $\Delta \cL_{\ssY\ssM}$ has
already been given in \eqn{eq:SpecialYMh}.

\vspace{2mm}
\noindent $\bullet$ The new $\Gamma$ terms introduced in $\cL_S$ by
Eqs.~(\ref{eq:gbinbetat}) can be arranged once again in the three classes
\be \Delta \cL_{S,\,t} = \Delta \cL_{S,\,t}^{(n_f=2)} + 
                              \Delta \cL_{S,\,t}^{(n_f=3)} +
			      \Delta \cL_{S,\,t}^{(n_f=4)},
\label{eq:SpecialLSt}
\ee
according to the number of fields appearing in the $\Gamma$ terms. The
explicit expression for $\Delta \cL_{S,\,t}^{(2)}$ is, up to terms of
${\mathcal{O}}(\Mgb^4)$,
\bea
\label {eq:SpecialLSt2}
     \Delta \cL_{S,\,t}^{(n_f=2)} &=& \MMb'\Gamma \left[ 
       -\frac{1}{2} \MMb' \ssb^2 \Gamma \Ab_{\mu} \Ab_{\mu}
       -\frac{1}{2} \MMb' \left(2+\Gamma \ccb^2 +4\bb\right) 
        \Zb_{\mu} \Zb_{\mu}
        \right.  \\  & & \left.
      - \MMb' \,\frac{\ssb}{\ccb} \left(1+\Gamma \ccb^2+2\bb \right) 
        \Ab_{\mu} \Zb_{\mu} 
       +\partial_{\mu} \phi_0 \left( \ssb \Ab_{\mu} + 
        \ccb \Zb_{\mu} \right) 
       \left( 1+ \bb\right)
        \right. \nonumber \\  & & \left.
       -\MMb' \left(2+\Gamma+4\bb\right) W^+_{\mu} W^-_{\mu}
	+ \left( W^-_{\mu} \partial_{\mu} \phi^+ 
	+ W^+_{\mu} \partial_{\mu} \phi^- \right) \left( 1+ \bb\right)
	\phantom{\frac{1}{2}} \!\!\!\!\!\right] \nonumber
\eea
with $\ssb = \sin \bar{\theta}$ and $\ccb = \cos \bar{\theta}$, while, up to
the same ${\mathcal{O}}(\Mgb^4)$,
\be
     \Delta \cL_{S,\,t}^{(n_f=3,4)} = 
     \Delta \cL_{S,\,h}^{(n_f=3,4)}
     \left(\MMb\rightarrow \MMb'\right)
\ee
($\Delta \cL_{S,\,h}^{(n_f=3)}$ and $\Delta \cL_{S,\,h}^{(n_f=4)}$
are given in \eqn{eq:SpecialLSh3} and \eqn{eq:SpecialLSh4}).  The
subscripts $t$ and $h$ indicate the $\beta_t$ and $\beta_h$ schemes.  Note
the presence of $\bb$ factors in the new $\Gamma$ terms of
\eqn{eq:SpecialLSt2}. We will comment on this in \sect{subsec-mixing}.

\vspace{2mm}
\noindent $\bullet$ Our recipe for gauge-fixing is the same as in the
previous sections: we choose the $R_\xi$ gauge $\cL_{gf}$ to cancel the
zeroth order (in $\Mgb$) gauge--scalar mixing terms introduced by ${\cal
L}_{S}$, but not those of higher orders (see discussions in Sections
\ref{subsubsec-betatdef} and \ref{subsec-gammah}). Here, this prescription
is realized by $\cL_{gf}$ (\eqn{eq:Lgf}) with
\be
       \cC_{\ssA} = -\frac{1}{\xi_{\ssA}}
                   \partial_{\mu} \Ab_{\mu},
\quad
       \cC_{\ssZ} = -\frac{1}{\xi_{\ssZ}}
                   \partial_{\mu} \Zb_{\mu} +{\xi_{\ssZ}} 
		   \frac{\bar{M'}}{\ccb} \phi_0,   
\quad
       \cC_{\pm} = -\frac{1}{\xi_{\ssW}}
              \partial_{\mu} W_{\mu}^\pm +{\xi_{\ssW}}\bar{M'} 
	      \phi_{\pm},
\label{eq:CBgfnewgT}
\ee
clearly $\Gamma$-independent. The new $\Gamma$ terms of the FP ghost
Lagrangian in the $\bb$ scheme are:
\be \Delta \cL_{FP,\,t} = \Delta \cL_{FP,\,t}^{(n_f=2)} + 
                               \Delta \cL_{FP,\,t}^{(n_f=3)}\,\,,
\label{eq:SpecialLFPt}
\ee
where the two-field terms are 
%(see footnote before  \eqn{eq:SpecialLFPh2}):
%
\be
     \Delta \cL_{FP,\,t}^{(n_f=2)} = 
      -\left(1+\bb\right)
      \Gamma \MMb'^2 \left[ \xi_{\ssZ} \Xb_{\ssZ}
	\left(X_{\ssZ} +\frac{\ssb}{\ccb} X_{\ssA}
          \right) + \xi_{\ssW} \left( \Xb_{+} X_{+}+\Xb_{-} X_{-}
	  \right) \right],
\ee
and the three-field terms are the same as in the $\beta_h$ scheme, with
$\MMb$ replaced by $\MMb'$: $\Delta \cL_{FP,\,t}^{(n_f=3)}=\Delta {\cal
L}_{FP,\,h}^{(n_f=3)}(\MMb\rightarrow \MMb')$ (\eqn{eq:SpecialLFPh3}). 
Like in the scalar sector, the $\Gamma$ and $\bb$ factors are entangled.

\vspace{2mm}
\noindent $\bullet$ We conclude this analysis with the fermionic sector.  As
in the Yang--Mills case, the fermion -- gauge boson Lagrangian ${\cal
L}_{fG}$ does not depend on the parameters of the $\beta_h$ or $\bb$
schemes. Its expression in terms of the new coupling constant $\Mgb$ contains
new $\Gamma$ terms and is given in \eqn{eq:LfGafter}. The neutral sector
rediagonalization induces no $\Gamma$ terms in the fermion--scalar
Lagrangian $\cL_{fS}$ (\eqn{eq:LfS}), which contains, however, the $\bb$
vertices discussed in \sect{subsec-betat} (\eqn{eq:Lfmass}) (the ratio
$M'/g$ is now replaced by the identical ratio $\MMb'/\Mgb$).

\vspace{3mm} The Feynman rules for all $\Gamma$ vertices are listed in
Appendix C, up to terms of ${\mathcal{O}}(\Mgb^4)$. All primes and bars over
$A_{\mu}$, $Z_{\mu}$, $M$, $\mh$ and $\theta$ have been dropped (but not
over $\Mgb$).  As we mentioned at the end of the previous section, the
$\Gamma$ vertices of the pure Yang--Mills sector need not be listed.

%%%%%%%%%%%%%%%%%%%%%%%%%%%%%%%%%%%%%%%%%%%%%%%%%%%%%%%%%%%%%%%%%%%%%%%%%%%%%
\subsection{\bm The $\Gamma$--$\bb$ mixing \ubm}
\label{subsec-mixing}

A comment on the presence of $\bb$ factors in the new $\Gamma$ vertices is
now appropriate. Consider the Lagrangian $\cL_S$.  As we already pointed out
in \sect{subsec-gammah}, the interaction part $\cL_S^I = -\mu^2 K^\dagger K
- (\lambda/2) (K^\dagger K)^2$ does not induce $\Gamma$ terms, but gives
rise to $\bb$ terms: as $M'/g=\bar{M'}/\Mgb$, these $\bb$ terms are simply
given by \eqn{eq:LSIt} expressed in terms of $\bar{M'}/\Mgb$ instead of
$M'/g$. On the other hand, the derivative part of $\cL_S$, $-(D_{\mu}
K)^\dagger (D_{\mu} K)$, induces both $\Gamma$ and $\bb$ vertices, plus
mixed ones which we still call $\Gamma$ vertices (see the $\bb$ factors in
the two-leg $\Gamma$ terms of $\Delta \cL_{S,\,t}^{(n_f=2)}$).  It works
like this: first, we replace $g \rightarrow \Mgb (1+\Gamma)$ and $g'
\rightarrow -\Mgb (\ssb/\ccb)$ in $-(D_{\mu} K)^\dagger (D_{\mu} K)$,
splitting the result in two classes of terms, both written in terms of
$\Mgb$, with or without $\Gamma$.  Then we substitute in both classes $v
\rightarrow 2\bar{M'}(1+\bb)/\Mgb$: the class containing $\Gamma$ is, up to
terms of ${\mathcal{O}}(\Mgb^4)$, $\Delta {\cal L}_{S,\,t}$
(\eqn{eq:SpecialLSt}), and includes also $\bb$ factors, while the class free
of $\Gamma$ has the same $\bb$ vertices as \eqn{eq:LSDt} with $g$, $\theta$,
$M'$, $A_{\mu}$ and $Z_{\mu}$ replaced by $\bar{g}$, $\bar{\theta}$,
$\bar{M'}$, $\Ab_{\mu}$ and $\Zb_{\mu}$.
%
%The upshot is that you need both the results for the new $\Gamma$ vertices
%derived in the previous section \ref{subsec-gammat} (containing $\bb$), and
%the expressions for the $\bb$ terms obtained in \sect{subsec-betat}.
%
The $\Gamma$ and $\bb$ terms of the Faddeev--Popov sector are intertwined
just as in the case of the scalar Lagrangian.

%%%%%%%%%%%%%%%%%%%%%%%%%%%%%%%%%%%%%%%%%%%%%%%%%%%%%%%%%%%%%%%%%%%%%%%%%%%%%
\subsection{Summary of the special vertices}
\label{subsec-specialsummary}

The upshot of this first part of the paper lies in the Appendices. There the
readers find the full set of SM $\Gamma$ (up to ${\mathcal{O}}(\Mgb^4)$) and
$\beta_{h,t}$ special vertices in the $R_\xi$ gauges. All primes and bars
over $A_{\mu}$, $Z_{\mu}$, $M$, $\mh$ and $\theta$ have been dropped, but
not over $\Mgb$, the $SU(2)$ coupling constant of the rediagonalized neutral
sector. The readers can pick their preferred tadpole scheme, $\beta_h$ or
$\bb$, and compute their Feynman diagrams including the $\beta_{h,t}$
vertices of Appendix A or B, respectively.  If they prefer to work with the
rediagonalized neutral sector, they should simply replace $g$ by $\Mgb$ in
the $\beta_{h,t}$ vertices, and add to them the $\Gamma$ ones of Appendix
C. There, $\Gamma$ vertices are listed for the $\bb$ scheme (note that
$\Gamma$ and $\bb$ terms are intertwined --- see \sect{subsec-mixing}); just
set $\bb=0$ to use the $\beta_h$ scheme instead.

Finally, \tabn{summa} graphically summarizes which of the SM sectors
provides each type of special vertex. Note the overlap of $\Gamma$ and $\bb$
terms in the scalar and Faddeev--Popov sectors.
\renewcommand{\arraystretch}{1.3}
\renewcommand{\tabcolsep}{15pt}
%\vspace{5mm}
%\begin{center}
\begin{table}[ht]\centering
\begin{tabular}[t]{||l||c|c|c||}                            \hline
  SECTOR           & $\beta_h$ & $\bb$      &  $\Gamma$     \\ \hline \hline  
  Scalar: $(D_{\mu} K)^\dagger (D_{\mu} K)$
                   &           & $\bullet$  &  $\bullet$    \\ \hline 
  Scalar: $\mu^2 K^\dagger K+(\lambda/2) (K^\dagger K)^2$          
                   & $\bullet$ & $\bullet$  &               \\ \hline 
  Yang--Mills      &           &            &  $\bullet$    \\ \hline 
  Gauge-Fixing     &           &            &               \\ \hline 
  Faddeev--Popov   &           & $\bullet$  &  $\bullet$    \\ \hline 
  Fermion -- gauge boson
                   &           &            &  $\bullet$    \\ \hline 
  Fermion -- Higgs       
                   &           & $\bullet$  &               \\ \hline 

\end{tabular}
\caption[]{Special vertices in the Standard Model.}
\label{summa}
%\end{center} 
\end{table}
\renewcommand{\tabcolsep}{6pt}
\renewcommand{\arraystretch}{1}
\vspace{5mm}

%%%%%%%%%%%%%%%%%%%%%%%%%%%%%%%%%%%%%%%%%%%%%%%%%%%%%%%%%%%%%%%%%%%%%%%%%%%%%
\section{WST identities for two-loop gauge boson self-energies}
\label{sec-WSTI}

The purpose of this section is to discuss in detail the structure of the
(doubly-contracted) Ward-Slavnov-Taylor identities (WSTI) for the two-loop
gauge boson self-energies in the SM, focusing in particular on
the role played by the reducible diagrams. This analysis is performed in the
't~Hooft--Feynman gauge.
%%%%%%%%%%%%%%%%%%%%%%%%%%%%%%%%%%%%%%%%%%%%%%%%%%%%%%%%%%%%%%%%%%%%%%%%%%%%%
\subsection{Definitions and WST identities}
\label{subsec-WSTIdef}

Let $\Pi_{ij}$ be the sum of all diagrams (both one-particle reducible and
irreducible) with two external boson fields, $i$ and $j$, to all orders in
perturbation theory (as usual, the external Born propagators are not to be
included in the expression for ${\Pi}_{ij}$)
\be 
\Pi_{ij} = \sum_{n = 1}^{\infty} 
               \frac{g^{2 n}}{(16\pi^2)^{n}}\,
	       \Pi^{(n)}_{ij} \, . 
\label{eq:Piij}
\ee 
In the subscripts of the quantities $\Pi_{ij}^{(n)}$ we will also explicitly
indicate, when necessary, the appropriate Lorentz indices with Greek
letters.  At each order in the perturbative expansion it is convenient to
make explicit the tensor structure of these functions by employing the
following definitions:
\be 
\Pi^{(n)}_{\mu \nu, \ssV \ssV} = D^{(n)}_{\ssV
\ssV}\,\delta_{\mu \nu} + P^{(n)}_{\ssV
\ssV}\,p_{\mu}\,p_{\nu} 
        \quad\quad 
\Pi^{(n)}_{\mu, \ssV\ssS} = 
- i p_{\mu}\,M_{\ssS}\,G^{(n)}_{\ssV\ssS} 
         \quad\quad \Pi^{(n)}_{\ssS \ssS} =
R^{(n)}_{\ssS \ssS}\,, 
\label{eq:WSTIl}
\ee
where the subscripts $V$ and $S$ indicate vector and scalar fields,
$M_{\ssS}$ is the mass of the Higgs-Kibble scalar $S$, and $p$ is
the incoming momentum of the vector boson (note: 
$\Pi^{(n)}_{\mu, \ssS\ssV} =-
\Pi^{(n)}_{\mu, \ssV\ssS}$). The quantities $D_{i j}$, $P_{i
j}$, $G_{i j}$, and $R_{i j}$ depend only on the squared four-momentum and
are symmetric in $i$ and $j$. Furthermore, $D$ and $R$ have the
dimensions of a mass squared, while $G$ and $P$ are dimensionless.

The WST identities require that, at each perturbative order, the gauge-boson
self-energies satisfy the equations
\bea
&p_{\mu}\,p_{\nu}\,\paa{(n)} = 0& 
                                    \nonumber \\
&p_{\mu}\,p_{\nu}\,\paz{(n)} +i p_{\mu}\,M_0\,\pap{(n)}  = 0& 
				    \nonumber \\
&p_{\mu}\,p_{\nu}\,\pzz{(n)} + M_0^2\,\ppopo{(n)} + 
				    2\,i p_{\mu}\,M_0\,\pzp{(n)} = 0&
				    \nonumber \\
&p_{\mu}\,p_{\nu}\,\pww{(n)} + M^2\,\ppp{(n)} + 
				    2\,ip_{\mu}\,M\,\pwp{(n)} = 0\,,& 
\label{aWSTIl}
\eea
which imply the following relations among the form factors $D$, $P$, $G$, and
$R$
\bea
     D^{(n)}_{\ssAA} + p^2\,P^{(n)}_{\ssAA} &=& 0  
\label{WSTIexp1}                                   \\
     D^{(n)}_{\ssAZ} + p^2\,P^{(n)}_{\ssAZ}+ 
             M_0^2\,G^{(n)}_{\mysmall{A \phi_o}} &=& 0  
\label{WSTIexp2}	                           \\
     p^2\,D^{(n)}_{\ssZZ} + p^4\,P^{(n)}_{\ssZZ} +  
	     M_0^2\,R^{(n)}_{\mysmall{\phi_o \phi_o}} &=&
	     -2\,M_0^2\,p^2 G^{(n)}_{\mysmall{Z \phi_o}} 
\label{WSTIexp3}	                           \\
     p^2\,D^{(n)}_{\mysmall{W W}} + p^4\,P^{(n)}_{\mysmall{W W}} +  
             M^2\,R^{(n)}_{\mysmall{\phi \phi}} &=&
	     -2\,M^2\,p^2 G^{(n)}_{\mysmall{W \phi}}  \, . 
\label{WSTIexp4}
\eea 
The subscripts $A$, $Z$, $W$, $\phi$ and $\phi_0$ clearly indicate the SM
fields. We have verified these WST Identities at the two-loop level
(i.e.~$n=2$) with our code $\GS$~\cite{GraphShot}.

%%%%%%%%%%%%%%%%%%%%%%%%%%%%%%%%%%%%%%%%%%%%%%%%%%%%%%%%%%%%%%%%%%%%%%%%%%%%%
\subsection{WST identities at two loops: the role of reducible diagrams} 
\label{subsec-WSTIred}

At any given order in the coupling constant expansion, the SM gauge boson
self-energies satisfy the WSTI (\ref{aWSTIl}). For $n \geq 2$, the
quantities $\Pi^{(n)}_ {ij}$ contain both one-particle irreducible (1PI) and
reducible (1PR) contributions. At $\mathcal{O}(g^4)$, the SM $\Pi^{(n)}_
{ij}$ functions contain the following irreducible topologies: eight
two-loop topologies, three one-loop topologies with a $\beta_{t_1}$ vertex,
four one-loop topologies with a $\Gamma_1$ vertex, and one tree-level
diagram with a two-leg $\mathcal{O}(g^4)$ $\beta_{t}$ or $\Gamma$ vertex
(see figure at the end of \sect{subsubsec-betatdef}). Reducible
$\mathcal{O}(g^4)$ graphs involve the product of two $\mathcal{O}(g^2)$
ones: two one-loop diagrams, one one-loop diagram and a tree-level diagram
with a ${\mathcal O}(g^2)$ two-leg vertex insertion, or two tree-level
diagrams, each with a ${\mathcal O}(g^2)$ two-leg vertex insertion. There
are also $\mathcal{O}(g^4)$ topologies containing tadpoles but, as we
discussed in previous sections, their contributions add up to zero as a
consequence of our choice for $\beta_t$.

In the following we analyze the structure of the $\mathcal{O}(g^4)$ WSTI for
photon, $Z$, and $W$ self-energies, as well as for the photon--$Z$
mixing, emphasizing the role played by the reducible diagrams.
%%%%%%%%%%%%%%%%%%%%%%%%%%%%%%%%%%%%%%%%%%%%%%%%%%%%%%%%%%%%%%%%%%%%%%%%%%%%%
\subsubsection{The photon self-energy}

The contribution of the 1PR diagrams to the photon self-energy at ${\mathcal
O}(g^4)$ is given, in the 't~Hooft--Feynman gauge, by (with obvious
notation)
\be
\paa{\mysmall{(2)R}} = \frac{1}{(2\pi)^4 i} \left[ 
      \frac{1}{p^2}\,\tilde{\Pi}^{\mysmall{(2)R}}_{\mu \nu,\ssAA}\, 
      + \frac{1}{p^2 + M_0^2}\,
      \hat{\Pi}^{\mysmall{(2)R}}_{\mu\nu, \ssAA}\right], 
\label{eq:photon2R}
\ee
where
\bea
\tilde{\Pi}^{\mysmall{(2)R}}_{\mu \nu,
\ssAA} &=& \Pi^{(1)}_{\mu \alpha, \ssAA}\,\Pi^{(1)}_{\alpha
\nu, \ssAA} 
                                      \qquad
\hat{\Pi}^{\mysmall{(2)R}}_{\mu \nu,
\ssAA} = \Pi^{(1)}_{\mu \alpha, \ssAZ}\,\Pi^{(1)}_{\alpha
\nu, \ssZA} +\Pi^{(1)}_{\mu, \mysmall{A
\phi_o}}\,\Pi^{(1)}_{\nu,\mysmall{\phi_o A}} \nonumber \,.
\eea
It is interesting to consider separately the reducible diagrams that involve
an intermediate photon propagator ($\tilde{\Pi}^{\mysmall{(2)R}}_{\mu \nu,
\ssAA}$) and those including an intermediate $Z$ or $\phi_0$
propagator ($\hat{\Pi}^{\mysmall{(2)R}}_{\mu \nu, \ssAA}$).  By
employing the definitions given in the previous subsection and
\eqn{WSTIexp1} with $n=1$, one verifies that $\tilde{\Pi}^{\mysmall{2
R}}_{\mu \nu, \ssAA}$ obeys the photon WSTI by itself,
\be p_{\mu}\,p_{\nu}\,\tilde{\Pi}^{\mysmall{\mysmall{(2)R}}}_{\mu \nu,
\ssAA} = p^2\,\left[\dd{A}{A} + p^2\,\pp{A}{A}\right]^2 = 0\,.
\label{aa51}
\ee
This is not the case for $\hat{\Pi}^{\mysmall{(2)R}}_{\mu\nu,\ssAA}$,
although most of its contributions cancel when contracted by $p_{\mu}
p_{\nu}$ as a consequence of \eqn{WSTIexp2} ($n=1$),
\be 
   p_{\mu}\,p_{\nu}\hat{\Pi}^{\mysmall{(2)R}}_{\mu \nu,\ssAA} = 
   p^2\,M_0^2\, \left(p^2 +M_0^2 \right) \left[\GG{A}{\phi_o} \right]^2.
\label{aar}
\ee
The only diagrams contributing to the $A$--$\phi_0$ mixing up to ${\mathcal
O}(g^2)$ are those with a $W$--$\phi$ or FP ghosts loop, and the tree-level
diagram with a $\Gamma$ insertion. Their contribution, in the 't
Hooft--Feynman gauge, is
\be \GG{A}{\phi_0} = (2\pi)^4 i\, \stw \ctw \,\left[
                 2B_0(p^2,M,M) + 16 \pi^2\Gamma_1\right].
\ee
A direct calculation (e.g.~with $\GS$) shows that this residual contribution
of the reducible diagrams to the ${\mathcal O}(g^4)$ photon WSTI, \eqn{aar},
is exactly canceled by the contribution of the $\mathcal{O}(g^4)$
irreducible diagrams, which include two-loop diagrams as well as one-loop
graphs with a two-leg vertex insertion.
%%%%%%%%%%%%%%%%%%%%%%%%%%%%%%%%%%%%%%%%%%%%%%%%%%%%%%%%%%%%%%%%%%%%%%%%%%%%%
\subsubsection{The photon--$Z$ mixing}

We now consider the second of Eqs.~(\ref{aWSTIl}) for $n=2$. Reducible
diagrams contribute to both $A$--$Z$ and $A$--$\phi_0$
transitions. Following the example of \eqn{eq:photon2R}, we divide these
contributions in two classes: the diagrams that include an intermediate
photon propagator and those mediated by a $Z$ or a $\phi_0$, namely,
for the photon--$Z$ transition in the 't~Hooft--Feynman gauge,
\bea 
\paz{\mysmall{(2)R}} 
&=& \frac{1}{(2\pi)^4 i} \left[ 
\frac{1}{p^2}\,\tilde{\Pi}^{\mysmall{(2)R}}_{\mu
\nu, \ssAZ}\, + \frac{1}{p^2 + 
M_0^2}\,\hat{\Pi}^{\mysmall{(2)R}}_{\mu \nu, \ssAZ} \right] 
                           \nonumber \\ 
\tilde{\Pi}^{\mysmall{(2)R}}_{\mu \nu, \ssAZ} 
&=& 
\Pi^{(1)}_{\mu \alpha, \ssAA}\,\Pi^{(1)}_{\alpha \nu, \ssAZ} 
                           \nonumber \\
\hat{\Pi}^{\mysmall{(2)R}}_{\mu \nu, \ssAZ}
&=& 
\Pi^{(1)}_{\mu \alpha, \ssAZ}\,\Pi^{(1)}_{\alpha \nu, \ssZZ} 
+\Pi^{(1)}_{\mu,\mysmall{A \phi_o}}\,\Pi^{(1)}_{\nu,\mysmall{\phi_o Z}} \,,  
\eea
and, for the photon--$\phi_0$ transition in the same gauge,
\bea 
\pap{\mysmall{(2)R}} &=& \frac{1}{(2\pi)^4 i} \left[ 
\frac{1}{p^2}\,\tilde{\Pi}^{\mysmall{(2)R}}_{\mu, 
\mysmall{A \phi_o}}\, + \frac{1}{p^2 + 
M_0^2}\,\hat{\Pi}^{\mysmall{(2)R}}_{\mu, \mysmall{A \phi_o}}\right]
                               \nonumber \\
\tilde{\Pi}^{\mysmall{(2)R}}_{\mu, \mysmall{A \phi_o}} &=& 
\Pi^{(1)}_{\mu \alpha,
\ssAA}\,\Pi^{(1)}_{\alpha, \mysmall{A \phi_o}}
                               \nonumber \\
\hat{\Pi}^{\mysmall{(2)R}}_{\mu, \mysmall{A \phi_o}} &=& 
\Pi^{(1)}_{\mu \alpha, \mysmall{A
Z}}\,\Pi^{(1)}_{\alpha, \mysmall{Z \phi_o}} +
\Pi^{(1)}_{\mu, \mysmall{A\phi_o}}\,\Pi^{(1)}_{\mysmall{\phi_o \phi_o}} \,. 
\eea
The reducible diagrams with an intermediate photon propagator satisfy the
WSTI by themselves. Indeed, 
\be
     p_{\mu}\,p_{\nu}\,\tilde{\Pi}^{\mysmall{(2)R}}_{\mu \nu, \ssAZ} 
     +i M_0 p_{\mu} \tilde{\Pi}^{\mysmall{(2)R}}_{\mu, \mysmall{A \phi_o}} 
     = 0 \,,
\label{aabv}
\ee
as it can be easily checked using \eqn{WSTIexp1} with $n=1$. On the
contrary, the remaining reducible diagrams must be added to the irreducible
$\mathcal{O}(g^4)$ contributions in order to satisfy the WSTI for the
photon--$Z$ mixing:
\be 
    p_{\mu} p_{\nu}\left[
      \frac{\hat{\Pi}^{\mysmall{(2)R}}_{\mu \nu, \ssAZ}}
	   {(2\pi)^4 i (p^2+M_0^2)}+
      \Pi^{\mysmall{(2)I}}_{\mu \nu,\ssAZ}\right] +
    i M_0 p_{\mu} \left[
      \frac{\hat{\Pi}^{\mysmall{(2)R}}_{\mu, \mysmall{A \phi_0}}}
	   {(2\pi)^4 i (p^2+M_0^2)}+
      \Pi^{\mysmall{(2)I}}_{\mu, \mysmall{A \phi_0}}\right] = 0.
\label{eq:WSTIRAZ}
\ee
%%%%%%%%%%%%%%%%%%%%%%%%%%%%%%%%%%%%%%%%%%%%%%%%%%%%%%%%%%%%%%%%%%%%%%%%%%%%%
\subsubsection{The $Z$ self-energy}

Also in the case of the WSTI for the ${\mathcal O}(g^4)$ $Z$ self-energy
it is convenient to separate the reducible contributions mediated by a
photon propagator from the rest of the reducible diagrams. In the
't~Hooft--Feynman gauge it is
\bea 
\pzz{\mysmall{(2)R}} &=& \frac{1}{(2\pi)^4 i} \left[ 
\frac{1}{p^2}\,\tilde{\Pi}^{\mysmall{(2)R}}_{
\mu \nu, \ssZZ}\, + \frac{1}{p^2 + M_0^2}\,
\hat{\Pi}^{\mysmall{(2)R}}_{\mu \nu, \ssZZ}\right]
                         \nonumber \\
\tilde{\Pi}^{\mysmall{(2)R}}_{\mu \nu, \ssZZ} &=& 
\Pi^{(1)}_{\mu \alpha, \ssZA}\,\Pi^{(1)}_{\alpha \nu, \ssAZ}
                         \nonumber \\
\hat{\Pi}^{\mysmall{(2)R}}_{\mu \nu, \ssZZ} &=& \Pi^{(1)}_{\mu
  \alpha, \ssZZ}\,\Pi^{(1)}_{\alpha \nu, \ssZZ} +
\Pi^{(1)}_{\mu, \mysmall{Z \phi_o}}\,\Pi^{(1)}_{\nu,\mysmall{\phi_o Z}}\,,
                                   \\
                         \nonumber \\
\pzp{\mysmall{(2)R}} &=& \frac{1}{(2\pi)^4 i} \left[ 
\frac{1}{p^2}\,\tilde{\Pi}^{\mysmall{(2)R}}_{\mu, \mysmall{Z \phi_o}}\, + 
\frac{1}{p^2 + 
M_0^2}\,\hat{\Pi}^{\mysmall{(2)R}}_{\mu, \mysmall{Z \phi_o}}\right] 
                         \nonumber \\
\tilde{\Pi}^{\mysmall{(2)R}}_{\mu, \mysmall{Z \phi_o}} &=& 
\Pi^{(1)}_{\mu \alpha,\ssZA}\,\Pi^{(1)}_{\alpha, \mysmall{A \phi_o}}
                         \nonumber \\
\hat{\Pi}^{\mysmall{(2)R}}_{\mu, \mysmall{Z \phi_o}} &=& 
\Pi^{(1)}_{\mu \alpha, \ssZZ}\,\Pi^{(1)}_{\alpha, \mysmall{Z
\phi_o}} +\Pi^{(1)}_{\mu, \mysmall{Z \phi_o}}\,
\Pi^{(1)}_{\mysmall{\phi_o \phi_o}} \,,
\eea
\bea 
\ppzpz{\mysmall{(2)R}} &=& \frac{1}{(2\pi)^4 i} \left[ \frac{1}{p^2}\,
\tilde{\Pi}^{\mysmall{(2)R}}_{\mysmall{\phi_o \phi_o}}\, + \frac{1}{p^2 + 
M_0^2}\,\hat{\Pi}^{\mysmall{(2)R}}_{\mysmall{\phi_o \phi_o}}\right]
                         \nonumber \\
\tilde{\Pi}^{\mysmall{(2)R}}_{\mysmall{\phi_o \phi_o}} &=& \Pi^{(1)}_{\alpha,
\mysmall{\phi_o A}}\,\Pi^{(1)}_{\alpha, \mysmall{A \phi_o}}
                         \nonumber \\
\hat{\Pi}^{\mysmall{(2)R}}_{\mysmall{\phi_o \phi_o}} &=& 
\Pi^{(1)}_{\alpha, \mysmall{\phi_o
Z}}\,\Pi^{(1)}_{\alpha, \mysmall{Z \phi_o}} +\Pi^{(1)}_{\mysmall{\phi_o
\phi_o}}\,\Pi^{(1)}_{\mysmall{\phi_o \phi_o}}, 
\eea
and, once again, the reducible diagrams mediated by a photon propagator
satisfy the WSTI by themselves, i.e.
\be 
      p_{\mu}\,p_{\nu}\,
      \tilde{\Pi}^{\mysmall{(2)R}}_{\mu \nu, \ssZZ} +
      M_0^2\,\tilde{\Pi}^{\mysmall{(2)R}}_{\mysmall{\phi_o \phi_o}} +
      2\,i\,p_{\mu}\,M_0\,
      \tilde{\Pi}^{\mysmall{(2)R}}_{\mu, \mysmall{Z \phi_o}} =0 \,,
\label{eq:WSTIRZZ}
\ee
as it can be easily checked using the one-loop WSTI for the photon--$Z$
mixing (\eqn{WSTIexp2} with $n=1$).
%%%%%%%%%%%%%%%%%%%%%%%%%%%%%%%%%%%%%%%%%%%%%%%%%%%%%%%%%%%%%%%%%%%%%%%%%%%%%
\subsubsection{The $W$ self-energy}

All the ${\mathcal O}(g^4)$ 1PR contributions to the WSTI for the $W$
self-energy are mediated, in the 't Hooft--Feynman gauge, by a charged
particle of mass $M$. A separate analysis of their contribution does not
lead, in this case, to particularly significant simplifications of the
structure of the WSTI.  However, some cancellations among the reducible
terms occur, allowing to obtain a relation that will be useful in the
discussion of the Dyson resummation of the $W$ propagator. The 1PR
quantities that contribute to the ${\mathcal O}(g^4)$ WSTI for the $W$
self-energy have the following form:
\bea 
\pww{\mysmall{(2)R}} &=&
\frac{1}{(2\pi)^4 i\,\left(p^2+M^2\right)}\,\left\{\left(\dd{W}{W}\right)^2 
\delta_{\mu \nu} \,+\,
p_{\mu}\,p_{\nu}\,\left[\phantom{\left(M^1\right)^M} 
\!\!\!\!\!\!\!\!\!\!\!\!\!\!\!\!\!\!\!
2\,\dd{W}{W}\,\pp{W}{W} \right. \right. 
%                           \nonumber \\
%&&+ \left. \left. p^2\,\left(\pp{W}{W}\right)^2 +
+ \left. \left. p^2\,\left(\pp{W}{W}\right)^2 +
M^2\,\left(\GG{W}{\phi}\right)^2 \right]\right\}
                           \nonumber \\ 
\pwp{\mysmall{(2)R}} &=& 
\frac{-i\,p_\mu\,M}{(2\pi)^4 i\,(p^2 + M^2)}\,\GG{W}{\phi}\left[\dd{W}{W}
+p^2\,\pp{W}{W} + \rr{\phi}{\phi}\right]
                           \nonumber \\ 
\ppp{\mysmall{(2)R}} &=&
\frac{1}{(2\pi)^4 i\,(p^2 +M^2)}\,\left[p^2\,M^2\, 
\left(\GG{W}{\phi}\right)^2 +\left(\rr{\phi}{\phi}\right)^2\right]\, .
\eea
Contracting the free indices with the corresponding external momenta,
summing the three contributions and employing \eqn{WSTIexp4} with
$n=1$, we obtain
\be
        (2\pi)^4 i \left[p_{\mu}\,p_{\nu}\,\pww{\mysmall{(2)R}} + 
	M^2\,\ppp{\mysmall{(2)R}} +
	2\,i\,p_{\mu}\,M\,\pwp{\mysmall{(2)R}} \right]
	=  
	p^2\,M^2 \,\left(\GG{W}{\phi}\right)^2-\rr{\phi}{\phi}\,
	\left[\dd{W}{W} + p^2\,\pp{W}{W}\right].
\label{aawwred} 
\ee
%--
%%%%%%%%%%%%%%%%%%%%%%%%%%%%%%%%%%%%%%%%%%%%%%%%%%%%%%%%%%%%%%%%%%%%%%%%%%%%%
\section{Dyson resummed propagators and their WST identities}
\label{sec-dyson}

We will now present the Dyson resummed propagators for the electroweak gauge
bosons. We will then employ the results of \sect{sec-WSTI} to show
explicitly, up to terms of ${\mathcal O}(g^4)$, that the resummed
propagators satisfy the WST identities.

Following definition \eqn{eq:Piij} for ${\Pi}_{ij}$, the function
${\Pi}^{\ssI}_{ij}$ represents the sum of all 1PI diagrams with two
external boson fields, $i$ and $j$, to all orders in perturbation theory (as
usual, the external Born propagators are not to be included in the
expression for ${\Pi}^{\ssI}_{ij}$).  As we did in
Eqs.~(\ref{eq:WSTIl}), we write explicitly its Lorentz structure,
\bq 
    {\Pi}^{\ssI}_{\mu \nu, \mysmall{V V}}  = 
    D^{\ssI}_{\mysmall{V V}}\,\delta_{\mu \nu} + 
    P^{\ssI}_{\mysmall{V V}}\,p_{\mu}\,p_{\nu}  
			 \quad
    {\Pi}^{\ssI}_{\mu, \mysmall{V S}} = 
    - i p_{\mu}\,M_{\ssS}\,G^{\ssI}_{\mysmall{V S}} 
			 \quad
    \Pi^{\ssI}_{\mysmall{S S}} = R^{\ssI}_{\mysmall{S S}}\,, 
\label{eq:PiLorentz}
\eq
where $V$ and $S$ indicate SM vector and scalar fields, and $p_{\mu}$ is the
incoming momentum of the vector boson (note: $ {\Pi}^{\ssI}_{\mu,
\mysmall{SV}} = -{\Pi}^{\ssI}_{\mu, \mysmall{VS}} $). We also
introduce the transverse and longitudinal projectors
\bea
     &t^{\mu \nu} = 
      \delta_{\mu \nu} - \frac{p_{\mu} p_{\nu}}{p^2}\, , \quad \quad
      l^{\mu \nu} = \frac{p_{\mu} p_{\nu}}{p^2}\,, &
                 \nonumber \\
     &t^{\mu \alpha}\,t^{\alpha \nu} =
      t^{\mu \nu} \, , \quad   l^{\mu \alpha}\,l^{\alpha \nu} =
      l^{\mu \nu} \, , \quad t^{\mu \alpha}\,l^{\alpha \nu} = 0 \,,
                 \nonumber \\
     & {\Pi}^{\ssI}_{\mu \nu, \mysmall{V V}} = 
    D^{\ssI}_{\mysmall{V V}}\,t_{\mu \nu} + 
    L^{\ssI}_{\mysmall{V V}}\,l_{\mu \nu}, \quad 
L^{\II}_{\mysmall{V V}} = D^{\ssI}_{\mysmall{V V}} + 
p^2\,P^{\ssI}_{\mysmall{V V}}\,. &
\label{LTLorentz}
\eea
%--
The full propagator for a field $i$ which mixes with a field $j$ via the
function ${\Pi}^{\ssI}_{ij}$ is given by the perturbative series
\bea
    \bar{\Delta}_{ii} 
    &=& 
    \Delta_{ii} \,+\, \Delta_{ii} \sum_{n=0}^{\infty} \,
      \prod_{l=1}^{n+1} \sum_{k_l} \Pi_{k_{l-1} k_{l}}^{\ssI} 
      \Delta_{k_{l} k_{l}}
\label{eq:aa32}
    \\ &=& 
    \Delta_{ii} \,+\, \Delta_{ii}\,{\Pi}^{\ssI}_{ii}\,\Delta_{ii} 
    \,+\, \Delta_{ii} \!\sum_{k_1=i,j} {\Pi}^{\ssI}_{i k_1}
    \Delta_{k_1 k_1} {\Pi}^{\ssI}_{k_1 i} 
    \Delta_{ii} \,+\, \cdots  \,, \nonumber
\eea
where $k_0=k_{n+1}=i$, while for $l\neq n+1$, $k_l$ can be $i$ or
$j$. ${\Delta}_{ii}$ is the Born propagator of the field $i$. We rewrite
\eqn{eq:aa32} as
\be
   \bar{\Delta}_{ii} =
   \Delta_{ii}\,\left[1 - \left({\Pi}\,\Delta\right)_{ii}\right]^{-1}, 
\label{eq:dysonR}
\ee
and refer to $\bar{\Delta}_{ii}$ as the resummed propagator.  The
quantity $({\Pi}\,\Delta)_{ii}$ is the sum of all the possible products of
Born propagators and self-energies, starting with a 1PI self-energy
${\Pi}^{\ssI}_{ii}$, or transition ${\Pi}^{\ssI}_{ij}$, and
ending with a propagator $\Delta_{ii}$, such that each element of the sum
cannot be obtained as a product of other elements in the sum.  A
diagrammatic representation of $({\Pi}\,\Delta)_{ii}$ is the following,
%----
\bqas \left({\Pi}\,\Delta\right)_{ii}\,\,\,=\,\,\,
  \vcenter{\hbox{
  \begin{picture}(60,0)(0,0)
  \SetScale{0.6}
  \SetWidth{2.}
  \GCirc(0,0){10}{1}
  \DashArrowLine(10,0)(60,0){2}
  \end{picture}}}
\hspace{-8mm} +\,\,  
  \vcenter{\hbox{
  \begin{picture}(100,0)(0,0)
  \SetScale{0.6}
  \SetWidth{2.}
  \GCirc(0,0){10}{0.7}
  \ArrowLine(10,0)(50,0)
  \GCirc(50,0){10}{0.7}
  \DashArrowLine(60,0)(100,0){2}
  \end{picture}}}
\hspace{-13mm} +\,\,
  \vcenter{\hbox{
  \begin{picture}(90,0)(0,0)
  \SetScale{0.6}
  \SetWidth{2.}
  \GCirc(0,0){10}{0.7}
  \ArrowLine(10,0)(50,0)
  \GCirc(50,0){10}{0}
  \ArrowLine(60,0)(100,0)
  \GCirc(100,0){10}{0.7}
  \DashArrowLine(110,0)(150,0){2}
  \end{picture}}}
 +\, \cdots
\eqas
%----
where the Born propagator of the field $i$ ($j$) is represented by a dotted
(solid) line, the white blob is the $i$ 1PI self-energy, and the dots at the
end indicate a sum running over an infinite number of 1PI $j$ self-energies
(black blobs) inserted between two 1PI $i$--$j$ transitions (gray blobs).

It is also useful to define, as an auxiliary quantity, the \emph{partially
resummed} propagator for the field $i$, $\hat{\Delta}_{ii}$, in which we
resum only the proper 1PI self-energy insertions ${\Pi}^{\ssI}_{ii}$,
namely,
\be 
    \hat{\Delta}_{ii} = \Delta_{ii}\,\left[1 - {\Pi}^{\ssI}_{ii}\,
    \Delta_{ii} \right]^{-1}. 
\label{eq:partialDyson}
\ee
If the particle $i$ were not mixing with $j$ through loops or two-leg vertex
insertions, $\hat{\Delta}_{ii}$ would coincide with the resummed propagator
$\bar{\Delta}_{ii}$. $\hat{\Delta}_{ii}$ can be graphically depicted as 
\bqas \hat{\Delta}_{ii} \, = \,
  \vcenter{\hbox{
  \begin{picture}(50,0)(0,0)
  \SetScale{0.6}
  \SetWidth{2.}
  \DashArrowLine(0,0)(50,0){2}
  \end{picture}}}
\hspace{-0.3cm} + \,\,\,
  \vcenter{\hbox{
  \begin{picture}(70,0)(0,0)
  \SetScale{0.6}
  \SetWidth{2.}
  \DashArrowLine(0,0)(50,0){2}
  \GCirc(50,0){10}{1}
  \DashArrowLine(60,0)(100,0){2}
  \end{picture}}}
\hspace{0cm} + \,\,\,
  \vcenter{\hbox{
  \begin{picture}(110,0)(0,0)
  \SetScale{0.6}
  \SetWidth{2.}
  \DashArrowLine(0,0)(50,0){2}
  \GCirc(50,0){10}{1}
  \DashArrowLine(60,0)(110,0){2}
  \GCirc(110,0){10}{1}
  \DashArrowLine(120,0)(160,0){2}
  \end{picture}}}
 +\,\,\,\, \cdots.
\eqas
Partially resummed propagators allow for a compact expression for
$({\Pi}\,\Delta)_{ii}$,
\be
    ({\Pi}\,\Delta)_{ii} = \Pi^{\ssI}_{ii} \Delta_{ii} +
    \Pi^{\ssI}_{ij} \hat{\Delta}_{jj} 
    \Pi^{\ssI}_{ji} \Delta_{ii},
\ee
so that the resummed propagator of the field $i$ can be cast in the form
\be
   \bar{\Delta}_{ii} =
   \Delta_{ii}\,\left[1 - \left(\Pi^{\ssI}_{ii} +
     \Pi^{\ssI}_{ij} \hat{\Delta}_{jj} 
     \Pi^{\ssI}_{ji} \right) \Delta_{ii}
     \right]^{-1}. 
\label{eq:dysonR2}
\ee
We can also define a resummed propagator for the $i$--$j$ transition. In
this case there is no corresponding Born propagator, and the resummed one is
given by the sum of all possible products of 1PI $i$ and $j$ self-energies,
transitions, and Born propagators starting with $\Delta_{ii}$ and ending
with $\Delta_{jj}$.  This sum can be simply expressed in the following
compact form,
\be \bar{\Delta}_{ij} = \bar{\Delta}_{ii} \,{\Pi}^{\ssI}_{ij}\,
    \hat{\Delta}_{jj}.
\label{eq:dysonRmix}
\ee
%
%%%%%%%%%%%%%%%%%%%%%%%%%%%%%%%%%%%%%%%%%%%%%%%%%%%%%%%%%%%%%%%%%%%%%%%%%%%%%
\subsection{The charged sector}
\label{subsec-dysoncharged}

We now apply Eqs.~(\ref{eq:partialDyson}, \ref{eq:dysonR2},
\ref{eq:dysonRmix}) to $W$ and charged Goldstone boson fields.
The {\em partially} resummed propagator of the charged Goldstone scalar
follows immediately from \eqn{eq:partialDyson}. The Born $W$ and $\phi$
propagators in the 't Hooft--Feynman gauge are
\be
    \Delta_{\mysmall{WW}}^{\mu \nu} = \frac{\delta_{\mu \nu}}
	  {p^2 +M^2}, \quad
    \Delta_{\phi \phi} = \frac{1}{p^2 +M^2},
\ee
where, for simplicity of notation, we have dropped the coefficients
$(2\pi)^4 i$. In the same gauge, the partially resummed $\phi$ and $W$
propagators are
\bea 
      \hat{\Delta}_{\phi \phi} &=& 
      \Delta_{\phi \phi}\,
      \left[1 -{\Pi}^{\ssI}_{\phi \phi}\,
	\Delta_{\phi \phi}\right]^{-1} = \,\,\,
      \left[p^2 + M^2 - R_{\phi \phi}^{\II} \right]^{-1}
\label{eq:partresPHI} \\
     \hat{\Delta}^{\mu \nu}_{\WW} &=& 
     \frac{1}{p^2 + M^2 - D^{\ssI}_{\WW}}\,
     \left(\delta_{\mu \nu}  +
     \frac{ p_{\mu} p_{\nu} P^{\ssI}_{\WW}}{p^2 + 
       M^2 - D_{\WW}^{\II} - p^2 P_{\WW}^{\II}  } \right).
\label{eq:partresW}
\eea
\eqn{eq:partresW} assumes a more compact form when expressed in
terms of the transverse and longitudinal projectors $t_{\mu \nu}$ and
$l_{\mu \nu}$,
\be
    \hat{\Delta}^{\mu \nu}_{\WW} = 
    \frac{t^{\mu \nu}}{p^2 +M^2 - D^{\ssI}_{\WW}} +
    \frac{l^{\mu \nu}}{p^2 +M^2 - L^{\ssI}_{\WW}}.
\ee
The resummed $W$ and $\phi$ propagators can be then derived from
\eqn{eq:dysonR2},
\bea 
      \bar{\Delta}_{\phi \phi} &\!\!\!=&\!\!\!
      \left[ p^2 +M^2 - R^{\ssI}_{\phi \phi} - 
      \frac{p^2\,M^2\,(G^{\ssI}_{{\ssW} \phi})^2}
	   {p^2 + M^2 - L_{\WW}^{\II}} \right]^{-1} \phantom{DD}
                 \\
    \bar{\Delta}^{\mu \nu}_{\WW} &\!\!\!=&\!\!\! 
    \frac{t^{\mu \nu}}{p^2 +M^2 - D^{\ssI}_{\WW}} +
    l^{\mu \nu} \!\!\left[p^2 \!+\! M^2 \!-\! 
    L_{\WW}^{\II} \!-\!
    \frac{p^2M^2(G^{\ssI}_{{\ssW} \phi})^2}{p^2 
      + M^2 - R^{\II}_{\phi \phi}} \right]^{-1}\!\!\!\!\!\!. \phantom{DD}
\eea
The resummed propagator for the $W$--$\phi$ transition is provided by
\eqn{eq:dysonRmix},
\be
    \bar{\Delta}^{\mu}_{{\ssW} \phi} = 
    \frac{-i p_{\mu} M G^{\ssI}_{{{\phi \ssW}}}}
    {p^2 + M^2 -R^{\ssI}_{\phi \phi}}\,
    \left[ p^2 + M^2 - L_{\WW}^{\II} - 
    \frac{p^2 M^2 (G^{\ssI}_{{\ssW} \phi})^2}{p^2 + M^2 - 
      R^{\ssI}_{\phi \phi}} \right]^{-1}\!\!\!\!\!\!. 
\label{aacv1}  
\ee
We will now show explicitly, up to terms of $\mathcal{O}(g^4)$, that the
resummed propagators defined above satisfy the following WST identity:
\be 
    p_{\mu}\,p_{\nu}\,\bar{\Delta}^{\mu \nu}_{\WW}\ +
    i\,p_{\mu}\,M\,\bar{\Delta}^{\mu}_{{\ssW} \phi} -
    i\,p_{\nu}\,M\,\bar{\Delta}^{\nu}_{\phi \ssW} + 
    M^2\,\bar{\Delta}_{\phi \phi} =1 \,,
\label{prp2}
\ee 
which, in turn, is satisfied if
\be
     p^2 M^2\left(G^{\ssI}_{{\ssW} \phi}\right)^2 + 
     M^2 R^{\ssI}_{\phi \phi} + p^2L_{\WW}^{\II} - 
     R^{\ssI}_{\phi \phi}L_{\WW}^{\II} + 
     2p^2 M^2 G^{\ssI}_{{\ssW} \phi} = 0 \, . 
\label{prp3}
\ee
This equation can be verified explicitly, up to terms of ${\mathcal
O}(g^4)$, using the WSTI for the $W$ self-energy: at ${\mathcal O}(g^2)$
\eqn{prp3} becomes simply 
\be
    M^2  R^{(1)}_{\phi \phi} + p^2 L^{(1)}_{\WW} +
    2p^2 M^2 G^{(1)}_{{\ssW} \phi} = 0 \,, 
\ee
which coincides with \eqn{WSTIexp4} for $n=1$.  To prove
\eqn{prp3} at ${\mathcal O}(g^4)$ we can combine the last of
Eqs.~(\ref{aWSTIl}) with $n=2$ and \eqn{aawwred} to get \footnote{For
simplicity of notation, in this section we dropped the coefficients
$(2\pi)^4 i$.}
\be 
     p^2 M^2 \left(G^{(1)}_{W \phi}\right)^2 +
     M^2 R^{(2)\II}_{\phi \phi} + 
     p^2 L^{(2)\II}_{\WW}- R^{(1)}_{\phi \phi}L^{(1)}_{\WW} + 
     2p^2 M^2 G^{(2)\II}_{W \phi}= 0\, .
\ee
%
%%%%%%%%%%%%%%%%%%%%%%%%%%%%%%%%%%%%%%%%%%%%%%%%%%%%%%%%%%%%%%%%%%%%%%%%%%%%%
\subsection{The neutral sector}
\label{subsec-dysonneutral}

The SM neutral sector involves the mixing of three boson fields, $A_{\mu}$,
$Z_{\mu}$ and $\phi_0$. As the definitions for the resummed propagators
presented at the beginning of \sect{sec-dyson} refer to the mixing of
only two boson fields, we will now discuss their generalization to the
three-field case.

Consider three boson fields $i$, $j$ and $k$ mixing up through radiative
corrections. For each of them we can define a {\em partially resummed}
propagator $\hat{\Delta}_{ll}$ ($l=i,j, ~\mbox{or}~ k$) according to
\eqn{eq:partialDyson}. For each pair of the three fields, say $(j,k)$, we can
also define the following {\em intermediate} propagators
\bea
   \tilde{\Delta}_{jj}(j,k) &=&
   \Delta_{jj}\,\left[1 - \left(\Pi^{\ssI}_{jj} +
     \Pi^{\ssI}_{jk} \hat{\Delta}_{kk} 
     \Pi^{\ssI}_{kj} \right) \Delta_{jj}
     \right]^{-1}      
\label{eq:intermediateDyson1}   \\
   \tilde{\Delta}_{jk}(j,k) &=& 
   \tilde{\Delta}_{jj}(j,k) \,{\Pi}^{\ssI}_{jk}\,
    \hat{\Delta}_{kk} \,,
\label{eq:intermediateDyson2}
\eea
where the parentheses on the l.h.s.~indicate the chosen pair of fields.
[$\tilde{\Delta}_{kk}(j,k)$ and $\tilde{\Delta}_{kj}(j,k)$ can be simply
derived from the above definitions by exchanging $j \leftrightarrow k$.] The
reader will immediately note that the r.h.s.~of the above
Eqs.~(\ref{eq:intermediateDyson1}, \ref{eq:intermediateDyson2}) are almost
identical to those of Eqs.~(\ref{eq:dysonR2}, \ref{eq:dysonRmix}), with the
appropriate renaming of the fields. Equations (\ref{eq:intermediateDyson1},
\ref{eq:intermediateDyson2}), introduced in the context of three-field
mixing, define however only {\em intermediate} propagators (denoted by the
tilde), while Eqs.~(\ref{eq:dysonR2}, \ref{eq:dysonRmix}), presented in the
analysis of the two-field mixing case, define the complete resummed
propagators (denoted by the bar). Indeed, the definition of full resummed
propagator in the three-field mixing scenario requires one further step: the
resummed propagator for a field $i$ mixing with the fields $j$ and $k$ via
the functions ${\Pi}^{\ssI}_{ij}$, ${\Pi}^{\ssI}_{ik}$ and
${\Pi}^{\ssI}_{jk}$ can be cast in the following form
\be
    \bar{\Delta}_{ii} =
    \Delta_{ii} \left[1 - \left(\Pi^{\ssI}_{ii} +
      \sum_{l,m} \, \Pi^{\ssI}_{il} \, \tilde{\Delta}_{lm}^{}\!(j,k) 
      \, \Pi^{\ssI}_{mi} \right) \Delta_{ii}
      \right]^{-1} \!\!\!\!\!\!\!,          
\label{eq:neutralDyson}
\ee
where $l$ and $m$ can be $j$ or $k$, while the resummed propagator for the
transition between the fields $i$ and $k$ is
\be
   \bar{\Delta}_{ik} = 
   \bar{\Delta}_{ii} \sum_{l=j,k} \Pi^{\ssI}_{il} 
   \tilde{\Delta}_{lk}^{} (j,k) \,. 
\label{eq:neutralMixDyson}
\ee
Armed with Eqs.~(\ref{eq:intermediateDyson1})--(\ref{eq:neutralMixDyson}),
we can now present the $A_{\mu}$, $Z_{\mu}$ and $A_{\mu}$--$Z_{\mu}$
propagators. First of all, the Born $A_{\mu}$, $Z_{\mu}$ and $\phi_0$
propagators in the 't~Hooft--Feynman gauge are
\be
    \Delta_{\ssAA}^{\mu \nu} = \frac{\delta_{\mu \nu}}
	  {p^2}, \quad
    \Delta_{\ssZZ}^{\mu \nu} = \frac{\delta_{\mu \nu}}
	  {p^2 +M_0^2}, \quad
    \Delta_{\phi_0 \phi_0} = \frac{1}{p^2 +M_0^2},
\ee
where, for simplicity of notation, we have dropped once again the
coefficients $(2\pi)^4 i$. The \emph{partially resummed} propagators (three)
can be immediately computed via \eqn{eq:partialDyson} and the {\em intermediate}
ones (twelve) via Eqs.~(\ref{eq:intermediateDyson1}) and
(\ref{eq:intermediateDyson2}). Finally, after some algebra,
Eqs.~(\ref{eq:neutralDyson}) and (\ref{eq:neutralMixDyson}) provide us with
the fully resummed propagators:
$    \bar{\Delta}_{\mysmall{VV}} = 
          t_{\mu \nu}  \bar{\Delta}_{\mysmall{VV}}^{\ssT} +
          l_{\mu \nu}  \bar{\Delta}_{\mysmall{VV}}^{\ssL},
$
with $V=A,Z$ and
\bea
     \bar{\Delta}_{\ssAA}^{\ssT} &=&
     \left[ p^2 -D^{\ssI}_{\ssAA} -
       \frac{(D^{\ssI}_{\ssAZ})^2}
	    {p^2+M_0^2-D^{\ssI}_{\ssZZ}} \right]^{-1}
\\
     \bar{\Delta}_{\ssZZ}^{\ssT} &=&
     \left[ p^2 +M_0^2-D^{\ssI}_{\ssZZ} -
       \frac{\left(D^{\ssI}_{\ssAZ}\right)^2}
	    {p^2-D^{\ssI}_{\ssAA}} \right]^{-1}
\\
     \bar{\Delta}_{\ssAZ}^{\ssT} &=& 
     D^{\ssI}_{\ssAZ}
	  \left[\left(p^2-D^{\ssI}_{\ssAA}\right)
	    \left(p^2+M_0^2-D^{\ssI}_{\ssZZ}\right)
	  -(D^{\ssI}_{\ssAZ})^2\right]^{-1}.
\eea
The expressions of the longitudinal components of these propagators are more
lengthy and we will only present them up to terms of ${\mathcal O}(g^4)$:
\bea
     \bar{\Delta}_{\ssAA}^{\ssL} &=&
     \left[ p^2 + {\mathcal O}(g^6) \right]^{-1}
\label{eq:fullpropAAL}
\\
     \bar{\Delta}_{\ssZZ}^{\ssL} &=&
     \left[ p^2 + M_0^2 -L_{\ssZZ}^{\ssI}
       -\frac{(L_{\ssAZ}^{\ssI})^2}{p^2}
       -\frac{p^2 M_0^2 (G_{\mysmall{Z \phi_o}}^{\ssI})^2}
       {p^2 +M_0^2} + {\mathcal O}(g^6) \right]^{-1}
\label{eq:fullpropZZL}
\\
     \bar{\Delta}_{\ssAZ}^{\ssL} &=&
     \frac{L_{\mysmall{AZ}}^{\ssI}}
	  {p^2\left(p^2 + M_0^2 -
	    L_{\ssZZ}^{\ssI}\right)} \,+\,
     \frac{M_0^2}{\left(p^2 + M_0^2\right)^2} 
       G_{\mysmall{A \phi_o}}^{\ssI}
       G_{\mysmall{Z \phi_o}}^{\ssI}\,+\, {\mathcal O}(g^6). 
       \phantom{MM}
\label{eq:fullpropAZL}
\eea
Equation (\ref{eq:fullpropAAL}) achieves its compact form due to the use of
the WSTI (\ref{WSTIexp1}) and (\ref{WSTIexp2}) with $n=1,2$. Also
\eqn{eq:fullpropAZL} has been simplified using
$L^{\mysmall{(1)}}_{\ssAA}=0$ (i.e. \eqn{WSTIexp1} with $n=1$).
We point out that if we use the one-loop WSTI for the photon self-energy,
\eqn{WSTIexp1}, the transverse part of the resummed $A$--$Z$
propagator becomes, up to terms of ${\mathcal O}(g^4)$,
\be
     \bar{\Delta}_{\ssAZ}^{\ssT} = 
     D^{\ssI}_{\ssAZ}
     \left[p^2 \left(1+P^{\ssI}_{\ssAA}\right)
       \left(p^2+M_0^2-D^{\ssI}_{\ssZZ}\right)
       \right]^{-1} + {\mathcal O}(g^6),
\ee
thus showing a pole at $p^2 = 0$ if $D^{\ssI}_{\ssAZ}(p^2 =
0)$ were not vanishing because of the rediagonalization of the neutral
sector.
 
In order to show explicitly, up to terms of ${\mathcal O}(g^4)$, that the
above resummed propagators satisfy their WSTI, we also present the resummed
propagators involving the neutral Higgs-Kibble scalar $\phi_0$:
\bea 
      \bar{\Delta}^{\mu}_{\mysmall{A \phi_o}} \!\!\!\!&=&\!\!\!\!
      -i p_\mu \frac{M_0}{p^2} \left[
	\frac{G_{\mysmall{Z \phi_o}}^{\ssI}
	L_{\ssAZ}^{\ssI}}{\left(p^2+M_0^2\right)^2} +
	\frac{G_{\mysmall{A \phi_o}}^{\ssI}}
	{p^2 + M_0^2 - R_{\mysmall{\phi_o \phi_o}}^{\ssI}} \right] +
       {\mathcal O}(g^6)
\label{eq:fullpropAP}               \\
      \bar{\Delta}^{\mu}_{\mysmall{Z \phi_o}} \!\!\!\!&=&\!\!\!\!
      \frac{-i p_\mu M_0}{p^2+ M_0^2-L_{\ssZZ}^{\ssI}} 
	\left[
	\frac{G_{\mysmall{A \phi_o}}^{\ssI}
	L_{\ssAZ}^{\ssI}}{p^2\left(p^2+M_0^2\right)} +
	\frac{G_{\mysmall{Z \phi_o}}^{\ssI}}
	{p^2 + M_0^2 - R_{\mysmall{\phi_o \phi_o}}^{\ssI}} 
	\right] \!+\!
       {\mathcal O}(g^6) \phantom{MM}
\label{eq:fullpropZP}               \\
       \bar{\Delta}_{\mysmall{\phi_o \phi_o}} \!\!\!\!&=&\!\!\!\!
       \left[p^2 \!+\! M_0^2 -\! R_{\mysmall{\phi_o \phi_o}}^{\ssI} 
       \!-M_0^2 \!\left(G_{\mysmall{A \phi_o}}^{\ssI}\right)^2
       \!\!-\! \frac{p^2 M_0^2 }{p^2+M_0^2} 
       \!\left(G_{\mysmall{Z \phi_o}}^{\ssI}\right)^2
	 \right]^{-1}  \!\!\!\!+ {\mathcal O}(g^6).
\label{eq:fullpropPP}
\eea
With these results, and with the WSTI (\ref{WSTIexp1})--(\ref{WSTIexp3}),
(\eqn{eq:WSTIRAZ}) and (\eqn{eq:WSTIRZZ}), we can finally prove, up to
${\mathcal O}(g^4)$, the following WSTI for the resummed $A$, $Z$ and
$A$--$Z$ propagators,
\bea
      && p_{\mu}\,p_{\nu}\, \bar{\Delta}_{\ssAA}^{\mu \nu} \,=\, 1 
\\
      && p_{\mu}\,p_{\nu}\, \bar{\Delta}_{\ssAZ}^{\mu \nu} \,+\, 
      ip_{\mu} M_0 \,\bar{\Delta}^{\mu}_{\mysmall{A \phi_o}}  \,=\, 0 
\\
      && p_{\mu}\,p_{\nu}\, \bar{\Delta}_{\ssZZ}^{\mu \nu} \,+\, 
      M_0^2 \,\bar{\Delta}_{\mysmall{\phi_o \phi_o}} \,+\, 2ip_{\mu} M_0 \,
      \bar{\Delta}^{\mu}_{\mysmall{Z \phi_o}}  \,=\, 1 \,. 
\eea
%
%%%%%%%%%%%%%%%%%%%%%%%%%%%%%%%%%%%%%%%%%%%%%%%%%%%%%%%%%%%%%%%%%%%%%%%%%%%%%
\section{The LQ basis}
\label{sec-LQ}

For the purpose of the renormalization, it is more convenient to extract
from the quantities defined in the previous sections the factors involving
the weak mixing angle $\theta$. To achieve this goal, we employ the LQ basis
\cite{Bardin:1999ak}, which relates the photon and $Z$ fields to a new 
pair of fields, $L$ and $Q$:
\bea
     \left(
     \begin{array}{c}
       Z_{\mu} \\ A_{\mu}
     \end{array}
     \right)
     =
     \left(
     \begin{array}{cc}
       \ctw &\; 0
       \\ \stw & 1/\stw
     \end{array}
     \right)
     \left(
     \begin{array}{c}
       L_{\mu} \\ Q_{\mu}
     \end{array}
     \right).
\label{eq:ZALQ}
\eea
Consider the fermion currents $j^{\mu}_{\ssA}$ and
$j^{\mu}_{\ssZ}$ coupling to the photon and to the $Z$. As the
Lagrangian must be left unchanged under this transformation, namely
$j^{\mu}_{\ssZ}\,Z_{\mu} + j^{\mu}_{\ssA}\,A_{\mu} =
j^{\mu}_{\ssL}\,L_{\mu} + j^{\mu}_{\mysmall{Q}}\,Q_{\mu}$, 
the currents transform as
\bea
        \left(
	\begin{array}{c}
	  j^{\mu}_{\ssZ} \\ j^{\mu}_{\ssA}
	\end{array}
	\right)
	=
	\left(
	\begin{array}{cc}
	  1/\ctw &\; -\stws/\ctw
	  \\ 0 & \stw
	\end{array}
	\right)
	\left(
	\begin{array}{c}
	  j^{\mu}_{\ssL} \\ j^{\mu}_{\mysmall{Q}}
	\end{array}
	\right).
\eea
If we rewrite the SM Lagrangian in terms of the fields $L$ and $Q$, and
perform the same transformation (\eqn{eq:ZALQ}) on the FP ghosts fields
(from ($X_{\ssA}$,$X_{\ssZ}$) to
($X_{\ssL}$,$X_{\mysmall{Q}}$)), then all the interaction terms of
the SM Lagrangian are independent of $\theta$. Note that this is true only
if the relation $M/\ctw=M_0$ is employed, wherever necessary, to remove the
remaining dependence on $\theta$. In this way the dependence on the weak
mixing angle is moved to the kinetic terms of the $L$ and $Q$ fields which,
clearly, are not mass eigenstates.

The relevant fact for our discussion is that the couplings of $Z$,
photon, $X_{\ssZ}$ and $X_{\ssA}$ are related to those of the
fields $L$ and $Q$, $X_{\ssL}$ and $X_{\mysmall{Q}}$ by identities
like the ones described, in a diagrammatic way, in the following figure:
\vspace{1mm}
\bqas
  \vcenter{\hbox{
  \begin{picture}(70,0)(0,0)
  \SetScale{0.6}
  \SetWidth{2.}
  \Line(0,0)(40,0)
  \ArrowLine(40,0)(85,35)
  \ArrowLine(85,-35)(40,0) 
  \Text(-2,4)[cb]{$Z$}
  \Text(57,15)[cb]{$f$}
  \Text(57,-20)[cb]{$\overline{f}$}
  \end{picture}}}
\hspace{-0.3cm}  =  \frac{1}{\ctw}\,\hspace{0.3cm}
  \vcenter{\hbox{
  \begin{picture}(70,0)(0,0)
  \SetScale{0.6}
  \SetWidth{2.}
  \Line(0,0)(40,0)
  \ArrowLine(40,0)(85,35)
  \ArrowLine(85,-35)(40,0) 
  \Text(-2,4)[cb]{$L$}
  \Text(57,15)[cb]{$f$}
  \Text(57,-20)[cb]{$\overline{f}$}
  \end{picture}}}
\hspace{-0.3cm} - \frac{\stws}{\ctw}\, \hspace{0.3cm}
   \vcenter{\hbox{
  \begin{picture}(70,0)(0,0)
  \SetScale{0.6}
  \SetWidth{2.}
  \Line(0,0)(40,0)
  \ArrowLine(40,0)(85,35)
  \ArrowLine(85,-35)(40,0) 
  \Text(-2,4)[cb]{$Q$}
  \Text(57,15)[cb]{$f$}
  \Text(57,-20)[cb]{$\overline{f}$}
  \end{picture}}}
\eqas
\vspace{5mm}
\bqas
  \vcenter{\hbox{
  \begin{picture}(70,0)(0,0)
  \SetScale{0.6}
  \SetWidth{2.}
  \Line(0,0)(100,0)
  \ArrowLine(50,0)(85,35)
  \ArrowLine(15,35)(50,0) 
  \Text(-2,4)[cb]{$A$}
  \Text(60,4)[cb]{$Z$}
  \Text(57,20)[cb]{$W$}
  \end{picture}}}
\hspace{-0.3cm}  =  \frac{\stw}{\ctw}\,\hspace{0.3cm}
  \vcenter{\hbox{
  \begin{picture}(70,0)(0,0)
  \SetScale{0.6}
  \SetWidth{2.}
  \Line(0,0)(100,0)
  \ArrowLine(50,0)(85,35)
  \ArrowLine(15,35)(50,0) 
  \Text(-2,4)[cb]{$Q$}
  \Text(60,4)[cb]{$L$}
  \Text(57,20)[cb]{$W$}
  \end{picture}}}
\hspace{-0.3cm} - \frac{\stwc}{\ctw}\, \hspace{0.3cm}
   \vcenter{\hbox{
  \begin{picture}(70,0)(0,0)
  \SetScale{0.6}
  \SetWidth{2.}
  \Line(0,0)(100,0)
  \ArrowLine(50,0)(85,35)
  \ArrowLine(15,35)(50,0) 
  \Text(-2,4)[cb]{$Q$}
  \Text(60,4)[cb]{$Q$}
  \Text(57,20)[cb]{$W$}
  \end{picture}}} 
 \,.
\eqas
As the couplings of the fields $L$, $Q$, $X_{\ssL}$ and
$X_{\mysmall{Q}}$ do not depend on $\theta$, all the dependence on this
parameter is factored out in the coefficients in the r.h.s.~of these
identities. 

Since $\theta$ appears only in the couplings of the fields $A$, $Z$,
$X_{\ssA}$ and $X_{\ssZ}$ (once again, the relation $M/\ctw=M_0$
must also be employed, wherever necessary), it is possible to single out
this parameter in the two-loop self-energies of the vector bosons.
Consider, for example, the transverse part of the photon two-loop
self-energy $D^{\mysmall{(2)}}_{\ssAA}$ (which includes the
contribution of both irreducible and reducible diagrams).  All
diagrams contributing to $D^{\mysmall{(2)}}_{\ssAA}$ can be
classified in two classes: those including $(i)$ one internal $A$, $Z$,
$X_{\ssA}$ or $X_{\ssZ}$ field, and $(ii)$ those not
containing any of these fields.  The complete dependence on $\theta$ can be
factored out by expressing the external photon couplings and the internal
$A$, $Z$ $X_{\ssA}$ or $X_{\ssZ}$ couplings of the diagrams
of class $(i)$ in terms of the couplings of the fields $L$, $Q$,
$X_{\ssL}$ and $X_{\mysmall{Q}}$, namely
\be
       D^{\mysmall{(2)}}_{\ssAA} =
       \stws \left[ 
	 \frac{1}{\ctws} f^{\ssAA}_{1} +
	 f^{\ssAA}_{2} +
	 \stws f^{\ssAA}_{3}
       \right]\!,
\label{fI}
\ee 
where the functions $f^{\ssAA}_i$ $(i=1,2,3)$ are
$\theta$-independent. Similarly, we can factor out the $\theta$ dependence
of the transverse part of the two-loop photon--$Z$ mixing and $Z$
self-energy,
%--
\bea
      D^{\mysmall{(2)}}_{\ssAZ} &=& 
      \frac{\stw}{\ctw} \left[
	\frac{1}{\ctws} f^{\ssAZ}_{1} +
	 f^{\ssAZ}_{2} +
	 \stws f^{\ssAZ}_{3} +
	 \stwq f^{\ssAZ}_{4}
	 \right],
                               \\
      D^{\mysmall{(2)}}_{\ssZZ} &=& 
      \frac{1}{\ctws} \left[
	\frac{1}{\ctws} f^{\ssZZ}_{1} +
	     f^{\ssZZ}_{2} +
	     \stws f^{\ssZZ}_{3} +
	     \stwq f^{\ssZZ}_{4} +
	     \stwx f^{\ssZZ}_{5}
	     \right],
\label{fII}
\eea
where, once again, the functions $f^{\ssAZ}_i$ and
$f^{\ssZZ}_i$ $(i=1,\ldots,5)$ do not depend on $\theta$. Analogous
relations hold for the longitudinal components of the two-loop
self-energies.

We note that $D^{\mysmall{(2)}}_{\ssAZ}$ and
$D^{\mysmall{(2)}}_{\ssZZ}$ also contain a third class of diagrams
containing more than one internal $Z$ (or $X_{\ssZ}$) field (up to
three, in $D^{\mysmall{(2)}}_{\ssZZ}$). However, the diagrams of
this class involve the trilinear vertex $Z H Z$ (or $\Xb_{\ssZ}
H X_{\ssZ}$), which does not induce any new $\theta$ dependence.

However, from the point of view of renormalization it is more convenient to
distinguish between the $\theta$ dependence originating from external legs and
the one introduced by external legs. We define, to all orders,
%--
\bqa
D_{\ssAA} &=& \stws\,\Pi_{\ssQ\ssQ\,;\,{\rm ext}}\,p^2 =
\stws\,\sum_{n=1}^{\infty}\,\lpar\frac{g^2}{16\,\pi^2}\rpar^n\,
\Pi^{(n)}_{\ssQ\ssQ\,;\,{\rm ext}}\,p^2,
\nl
D_{\ssAZ} &=& \frac{\stw}{\ctw}\,\Sigma_{\ssAZ\,;\,{\rm ext}} =
\frac{\stw}{\ctw}\,\sum_{n=1}^{\infty}\,
\lpar\frac{g^2}{16\,\pi^2}\rpar^n\,\Sigma^{(n)}_{\ssAZ\,;\,{\rm ext}},
\nl
D_{\ssZZ} &=& \frac{1}{\ctws}\,\Sigma_{\ssZZ\,;\,{\rm ext}} =
\frac{1}{\ctws}\,\sum_{n=1}^{\infty}\,
\lpar\frac{g^2}{16\,\pi^2}\rpar^n\,\Sigma^{(n)}_{\ssZZ\,;\,{\rm ext}},
\eqa
%--
\bq
\Sigma^{(n)}_{\ssAZ\,;\,{\rm ext}} =
\Sigma^{(n)}_{3\ssQ\,;\,{\rm ext}} - 
\stws\,\Pi^{(n)}_{\ssQ\ssQ\,;\,{\rm ext}}\,p^2,
\quad
\Sigma^{(n)}_{\ssZZ\,;\,{\rm ext}} =
\Sigma^{(n)}_{33\,;\,{\rm ext}} - 
2\,\stws\,\Sigma^{(n)}_{3\ssQ\,;\,{\rm ext}} +
\stwq\,\Pi^{(n)}_{\ssQ\ssQ\,;\,{\rm ext}}\,p^2.
\eq
%--
Furthermore, our procedure is such that
%--
\bq
\Sigma^{(n)}_{3\ssQ\,;\,{\rm ext}} =
\Pi^{(n)}_{3\ssQ\,;\,{\rm ext}}\,p^2,
\eq
%--
with $\Pi^{(n)}_{3\ssQ\,;\,{\rm ext}}$ regular at $p^2 = 0$.
At $\ord{g^2}$ the external quantities are $\theta$-independent while,
at $\ord{g^4}$ the relation with the coefficients of \eqns{fI}{fII} is
%--
\bqa
\Pi^{(2)}_{\ssQ\ssQ\,;\,{\rm ext}}\,p^2 &=&
\frac{1}{\ctws}\,f^{\ssAA}_1 + f^{\ssAA}_2 + f^{\ssAA}_3\,\stws,
\nl
\Sigma^{(2)}_{3\ssQ\,;\,{\rm ext}} &=&
        \frac{1}{\ctws}\,(
           f^{\ssAA}_1
          + f^{\ssAZ}_1
          )
          - f^{\ssAA}_1
          + f^{\ssAZ}_2
       + \stws\,(
           f^{\ssAA}_2
          + f^{\ssAZ}_3
          )
       + \stwq\,(
           f^{\ssAA}_3
          + f^{\ssAZ}_4
          )
\nl
\Sigma^{(2)}_{33\,;\,{\rm ext}} &=&
        \frac{1}{\ctws}\,(
           f^{\ssAA}_1
          + 2\,f^{\ssAZ}_1
          + f^{\ssZZ}_1
          )
          - f^{\ssAA}_1
          - 2\,f^{\ssAZ}_1
          + f^{\ssZZ}_2
       + \stws\,(
          - f^{\ssAA}_1
          + 2\,f^{\ssAZ}_2
          + f^{\ssZZ}_3
          )
\nl
{}&+& \stwq\,(
          f^{\ssAA}_2
          + 2\,f^{\ssAZ}_3
          + f^{\ssZZ}_4
          )
       + \stwx\,(
          f^{\ssAA}_3
          + 2\,f^{\ssAZ}_4
          + f^{\ssZZ}_5
          ),
\label{atOg4}
\eqa
%--
and $\stw, \ctw$ in \eqn{atOg4} should be evaluated at $\ord{g^0}$, in any
renormalization scheme, for two-loop accuracy. 

Consider the process $\barf f \to \barh h$; taking into account Dyson 
resummed propagators and neglecting, for the moment, vertices and boxes we 
write
%--
\bqa
{\cal M}(\barf f \to \barh h) &=& -\,\tpfi\,\Bigl[
e^2\,Q_f Q_h\,\gapu{\mu}\,\otimes\,\gapu{\mu}\,\bar{\Delta}_{\ssAA}^{\ssT}
+ \frac{e g}{2\,\ctw}\,Q_f\,
\gapu{\mu}\,\otimes\,\gapu{\mu}\,(v_h + a_h\,\gfd)\,
\bar{\Delta}_{\ssZ\ssA}^{\ssT}
\nl
{}&+& \frac{e g}{2\,\ctw}\,Q_h
\gapu{\mu}\,(v_f + a_f\,\gfd)\,\otimes\,\gapu{\mu}\,
\bar{\Delta}_{\ssZ\ssA}^{\ssT}
+ \frac{g^2}{4 \ctws}
\gapu{\mu} (v_f + a_f\,\gfd)\otimes\,\gapu{\mu}(v_h + a_h \gfd)
\bar{\Delta}_{\ssZZ}^{\ssT} \Bigr]
\label{amp}
\eqa
%--
where $f$ and $h$ are fermions with quantum numbers $Q_I, I_{3i},\,i=f,h$;
furthermore we have introduced
%--
\bq
v_f = I_{3f} - 2\,Q_f\,\stws, \qquad
a_f = I_{3f},
\eq
%--
with $e^2 = g^2\,\stws$. Always neglecting terms proportional to fermion
masses it is useful to introduce an effective weak-mixing angle as follows:
%--
\bq
s^2_{\rm eff} = \stws\,\Bigl[1 - \frac{\Pi_{\ssAZ\,;\,{\rm ext}}}{1 - 
\stws\,\Pi_{\ssAA\,;\,{\rm ext}}} \Bigr],
\quad V_f = I_{3f} - 2\,Q_f\,s^2_{\rm eff}.
\label{sineff}
\eq
%--
The amplitude of \eqn{amp} can be cast into the following form:
%--
\bqa
{\cal M}(\barf f \to \barh h) &=& -\,\tpfi\,\Bigl[
\gapu{\mu}\,\otimes\,\gapu{\mu}\,
\frac{1}{1 - \stws\,\Pi_{\ssAA\,;\,{\rm ext}}}\,
\frac{e^2\,Q_f Q_h}{p^2}
\nl
{}&& \frac{g^2}{4\,\ctws}\,
\gapu{\mu}\,(V_f + a_f\,\gfd)\,\otimes\,\gapu{\mu}\,(V_h + a_h\,\gfd)\,
\bar{\Delta}_{\ssZZ}^{\ssT}
\Bigr].
\label{eff}
\eqa
%--
The functions $\Pi_{\ssAA\,;\,{\rm ext}}, \Pi_{\ssAZ\,;\,{\rm ext}}$ and
$\Sigma_{\ssZZ\,;\,{\rm ext}}$ start at $\ord{g^2}$ in perturbation
theory. Equation (\ref{eff}) shows the nice effect of absorbing -- to all
orders -- non-diagonal transitions into a redefinition of $\stws$ and forms
the basis for introducing renormalization equations in the neutral sector,
e.g.\ the one associated with the fine-structure constant
$\alpha$. Questions related to gauge-parameter independence of Dyson
resummation, e.g.\ in \eqn{sineff}, are not addressed here, but we will
present a detailed discussion in Part III, where their relevance will be
investigated.

%--
\section{Conclusions}
%--
In this paper we prepared the ground to perform a comprehensive
renormalization procedure of the Standard Model at the two-loop level; with
minor changes our results can be extended to an arbitrary gauge theory with
spontaneously broken symmetry.

The same set of problems that we encountered in this paper may receive
different answers; for instance, one could decide to work in the
background-field method and treat differently the problem of diagonalization
of the neutral sector in the SM. Our solution has been extended
beyond one-loop and it is an integral part of a renormalization procedure
which goes from fundamentals to applications. The whole set of {\em new}
Feynman rules of our Appendices has been coded in $\GS$ and has proven its
value in several applications, including the proof of the WST identities.

In this paper we outlined peculiar aspects of tadpoles in a spontaneously
broken gauge theory and extended beyond one-loop a strategy to diagonalize
the neutral sector of the SM, order-by-order in perturbation
theory. The obtained results have been used as the starting point in the
construction of the renormalized Lagrangian of the SM and in the
computation of (pseudo-)observables up to two loops.
%--
\Acknowledgments
%--
We gratefully acknowledge several important discussions with Dima Bardin,
Ansgar Denner, Stefan Dittmaier and Sandro Uccirati.  The work of A.~F. was
supported in part by the Swiss National Science Foundation (SNF) under
contract 200020-109162.
%--
\newpage 
%--
\appendix
%--
\section{Appendix: Feynman rules for \bm $\beta_h$  \ubm vertices}
%--
In this appendix we present the new set of diagrammatic rules induced by
our approach. The Feynman rules for the $\beta_h$ vertices are extremely 
simple and can be immediately derived from \eqn{eq:LSI}:
$$
\begin{array}{lcll}
  H & \put(60,3){\circle*{5}}\put(0,3){\line(1,0){60}}
  ~~~~~~~~~~~~~~~~~~&           &~~~~   -2M\beta_h/g
  \\ \\
  H & \put(30,3){\circle*{5}}\put(0,3){\line(1,0){60}} 
  ~~~~~~~~~~~~~~~~~~& H         &~~~~   -\beta_h
  \\ \\
  \phi_0  & \put(30,3){\circle*{5}}\put(0,3){\line(1,0){60}}
  ~~~~~~~~~~~~~~~~~~& \phi_0    &~~~~   -\beta_h
  \\ \\
  \phi_+  & \put(30,3){\circle*{5}}\put(0,3){\line(1,0){60}}
  ~~~~~~~~~~~~~~~~~~& \phi_-    &~~~~   -\beta_h, 
\end{array}
$$
where $\beta_h = \beta_{h_1} g^2 + \beta_{h_2} g^4 + \cdots ~$ and $M$ is
the bare $W$ mass. If working with the rediagonalized neutral sector, simply
replace $g$ by $\Mgb$. Multiply each vertex by a factor $(2\pi)^4 i$. As
usual, we have included the combinatorial factors for identical fields (see
Appendix D of ref.~\cite{Veltman:1994wz}).

%%%%%%%%%%%%%%%%%%%%%%%%%%%%%%%%%%%%%%%%%%%%%%%%%%%%%%%%%%%%%%%%%%%%%%%%%%%%%
\section{Appendix: Feynman rules for \bm $\bb$  \ubm vertices}
%--
In this appendix we present the $\bb$ vertices.  They can be read off the
Lagrangian terms of Eqs.~(\ref{eq:LSIt}), (\ref{eq:LSDt}), (\ref{eq:LFP})
and (\ref{eq:Lfmass}), including the combinatorial factors for identical
fields. Also, $\bb =\beta_{t_1} g^2 +\beta_{t_2} g^4 + \cdots$. Simply
replace $g$ by $\Mgb$ if working with the rediagonalized neutral sector. The
two-leg $\bb$ vertices are:
\vspace{0.5cm}
\renewcommand{\arraystretch}{0.60}
$$
\begin{array}{lcllll}
  H & \put(30,3){\circle*{5}}\put(0,3){\line(1,0){60}} 
  ~~~~~~~~~~~~~~~~~~& H 
  & & \;\;  -\left(3\mh^2/2\right)\bb\left(\bb+2\right) &
  \\~\\
  \phi_0  & \put(30,3){\circle*{5}}\put(0,3){\line(1,0){60}}
  ~~~~~~~~~~~~~~~~~~& \phi_0 
  & & \;\;  -\left(\mh^2/2\right)\bb\left(\bb+2\right) &
  \\~\\
  \phi_+  & \put(30,3){\circle*{5}}\put(0,3){\line(1,0){60}}
  ~~~~~~~~~~~~~~~~~~& \phi_- 
  & & \;\;  -\left(\mh^2/2\right)\bb\left(\bb+2\right) &
  \\~\\
  Z_\mu & \put(30,3){\circle*{5}}\put(0,3){\line(1,0){60}} 
  ~~~~~~~~~~~~~~~~~~& Z_\nu  
  & & \;\;  -M_0^2\bb\left(\bb+2\right) \,\delta_{\mu \nu}&
  \\~\\
  W^+_\mu  & \put(30,3){\circle*{5}}\put(0,3){\line(1,0){60}}
  ~~~~~~~~~~~~~~~~~~& W^-_\nu    
  & & \;\;  -M^2\bb\left(\bb+2\right) \,\delta_{\mu \nu}&
  \\~\\
  Z_\mu  & \put(30,3){\circle*{5}}\put(0,3){\line(1,0){60}}
  \put(42,-5){$\leftarrow$}\put(55,-5){$p$}
  ~~~~~~~~~~~~~~~~~~& \phi_0 
  & & \;\;   i p_\mu M_0 \bb &
  \\~\\
  W^+_\mu & \put(30,3){\circle*{5}}\put(0,3){\line(1,0){60}} 
  \put(42,-5){$\leftarrow$}\put(55,-5){$p$}
  ~~~~~~~~~~~~~~~~~~& \phi_-  
  & & \;\;   i p_\mu M \bb &
  \\~\\
  W^-_\mu  & \put(30,3){\circle*{5}}\put(0,3){\line(1,0){60}}
  \put(42,-5){$\leftarrow$}\put(55,-5){$p$}
  ~~~~~~~~~~~~~~~~~~& \phi_+
  & & \;\;   i p_\mu M \bb &
  \\~\\
  \bar{f} & \put(30,3){\circle*{5}}\put(0,3){\line(1,0){60}} 
  ~~~~~~~~~~~~~~~~~~& f
  & & \;\;  -m_f \bb                               &
  \\~\\
  \Xb^+  & \put(30,3){\circle*{5}}\put(0,3){\line(1,0){60}}
  ~~~~~~~~~~~~~~~~~~& X^+ 
  & & \;\;   -\xi_{\ssW} M^2 \bb            &
  \\~\\
  \Xb^-  & \put(30,3){\circle*{5}}\put(0,3){\line(1,0){60}}
  ~~~~~~~~~~~~~~~~~~& X^-
  & & \;\;   -\xi_{\ssW} M^2 \bb            &
  \\~\\
  \Xb_{\ssZ} & \put(30,3){\circle*{5}}\put(0,3){\line(1,0){60}} 
  ~~~~~~~~~~~~~~~~~~& X_{\ssZ}  
  & & \;\;   -\xi_{\ssZ} M^2_0 \bb          &
\end{array}
$$
\renewcommand{\arraystretch}{1}

\vspace{0.2cm}

\noindent
where $M_0=M/\cos_\theta$ and $\bb\left(\bb+2\right)$ $= 2 \beta_{t_1}
g^2+\left(\beta_{t_1}^2+2\beta_{t_2}\right)g^4+O\left(g^6\right)$.  Each
vertex must be multiplied by the usual factor $(2\pi)^4 i$. 

$-$ The three-leg $\bb$ vertices are:
\vspace{0.5cm}
$$
\begin{array}{lcllll}
 H &
 \put(10,3){\line(1,0){20}} \put(30,3){\circle*{5}}
 \put(30,3){\line(1,1){15}} \put(30,3){\line(1,-1){15}}
 \put(60,15){$H$} \put(60,-15){$H$}  ~~~~~~~~~~~~~~~~~~~~~&
  & & -g\bb \left(3\mh^2/2M\right)&
  \\~\\
 H &
 \put(10,3){\line(1,0){20}} \put(30,3){\circle*{5}}
 \put(30,3){\line(1,1){15}} \put(30,3){\line(1,-1){15}}
 \put(60,15){$\phi_0$} \put(60,-15){$\phi_0$}  ~~~~~~~~~~~~~~~~~~~~~&
  & & -g\bb \left(\mh^2/2M\right)&
  \\~\\
 H &
 \put(10,3){\line(1,0){20}} \put(30,3){\circle*{5}}
 \put(30,3){\line(1,1){15}} \put(30,3){\line(1,-1){15}}
 \put(60,15){$\phi_+$} \put(60,-15){$\phi_-$}  ~~~~~~~~~~~~~~~~~~~~~&
  & & -g\bb \left(\mh^2/2M\right)&
  \\~\\
 A_\mu &
 \put(10,3){\line(1,0){20}} \put(30,3){\circle*{5}}
 \put(30,3){\line(1,1){15}} \put(30,3){\line(1,-1){15}}
 \put(60,15){$W_\nu^+$} \put(60,-15){$\phi_-$}  ~~~~~~~~~~~~~~~~~~~~~&
  & & +ig\bb \stw M \,\delta_{\mu \nu}  &
  \\~\\
 A_\mu &
 \put(10,3){\line(1,0){20}} \put(30,3){\circle*{5}}
 \put(30,3){\line(1,1){15}} \put(30,3){\line(1,-1){15}}
 \put(60,15){$W_\nu^-$} \put(60,-15){$\phi_+$}  ~~~~~~~~~~~~~~~~~~~~~&
  & & -ig\bb \stw M \,\delta_{\mu \nu}  &
\\~\\
 Z_\mu &
 \put(10,3){\line(1,0){20}} \put(30,3){\circle*{5}}
 \put(30,3){\line(1,1){15}} \put(30,3){\line(1,-1){15}}
 \put(60,15){$W_\nu^+$} \put(60,-15){$\phi_-$}  ~~~~~~~~~~~~~~~~~~~~~&
  & & -ig\bb \stws M_0 \, \delta_{\mu \nu}  &
  \\~\\
 Z_\mu &
 \put(10,3){\line(1,0){20}} \put(30,3){\circle*{5}}
 \put(30,3){\line(1,1){15}} \put(30,3){\line(1,-1){15}}
 \put(60,15){$W_\nu^-$} \put(60,-15){$\phi_+$}  ~~~~~~~~~~~~~~~~~~~~~&
   & & +ig\bb \stws M_0 \, \delta_{\mu \nu}  &
  \\~\\
 H &
 \put(10,3){\line(1,0){20}} \put(30,3){\circle*{5}}
 \put(30,3){\line(1,1){15}} \put(30,3){\line(1,-1){15}}
 \put(60,15){$W_\mu^+$} \put(60,-15){$W_\nu^-$}  ~~~~~~~~~~~~~~~~~~~~~&
  & & -g\bb M\, \delta_{\mu \nu}  &
  \\~\\
 H &
 \put(10,3){\line(1,0){20}} \put(30,3){\circle*{5}}
 \put(30,3){\line(1,1){15}} \put(30,3){\line(1,-1){15}}
 \put(60,15){$Z_\mu$} \put(60,-15){$Z_\nu$}  ~~~~~~~~~~~~~~~~~~~~~&
  & & -g\bb \left(M/\ctws\right)\delta_{\mu \nu}  &
\end{array}
$$
where $\stw=\sin \theta$, $\ctw=\cos \theta$ and, once again, each vertex 
must be multiplied by the factor $(2\pi)^4 i$.  The $\bb$ tadpole $H$~~
\put(50,3){\circle*{5}}\put(0,3){\line(1,0){50}} ~~~~~~~~~~~~~~~~~~ is:
\bea
  &&(2\pi)^4 i \left(\mh^2 M\right)
  \left[-\frac{1}{g}\bb\left(\bb+1\right)\left(\bb+2\right)\right]~=
  \nonumber \nonumber \\ &&(2\pi)^4 i\left(\mh^2 M\right)\left[-2\beta_{t_1}
  g - \left(3\beta_{t_1}^2 +2 \beta_{t_2}\right) g^3
  +O\left(g^5\right)\right].\nonumber
\eea
\renewcommand{\arraystretch}{1}
%%%%%%%%%%%%%%%%%%%%%%%%%%%%%%%%%%%%%%%%%%%%%%%%%%%%%%%%%%%%%%%%%%%%%%%%%%%%%
\section{Appendix: Feynman rules for \bm $\Gamma$  \ubm vertices}
%--
In this appendix we present the $\Gamma$ vertices.  The new $\Gamma$
vertices introduced by the replacement $g\rightarrow\Mgb(1+\Gamma)$ in the
SM Lagrangian are listed here up to terms of ${\mathcal{O}}(\Mgb^4)$ in the
$R_\xi$ gauges. All primes and bars over $A_{\mu}$, $Z_{\mu}$, $M$, $\mh$
and $\theta$ have been dropped, except over $\Mgb$. Also, $\Gamma = \Gamma_1
\, \Mgb^2 + \Gamma_2 \,\Mgb^4 + \cdots$. As usual, each vertex must be
multiplied by the factor $(2\pi)^4 i$.  The following two-leg $\Gamma$
vertices are in the $\beta_t$ scheme. For the $\beta_h$ scheme, just set
$\beta_t=0$.
\vspace{0.6cm}
\renewcommand{\arraystretch}{0.8}
$$
\begin{array}{lcllll}
  A_\mu & \put(0,3){\line(1,0){60}}\put(30,3){\redcircle{4}}
  ~~~~~~~~~~~~~~~~~~& A_\nu  
  & & \!\!\! \;\;\;\;-\delta_{\mu \nu} [\,\Mgb^4 \stws M^2 \Gamma_1^2]
  &
  \\ \\ \\
   Z_\mu & \put(0,3){\line(1,0){60}} \put(30,3){\redcircle{4}}
  ~~~~~~~~~~~~~~~~~~& Z_\nu  
  & & \!\!\!\!\!\! \;\;\;\;-2\delta_{\mu \nu} [\,\Mgb^2 M^2 \Gamma_1 
     +\Mgb^4 M^2 (\Gamma_2+2\Gamma_1 \beta_{t_1} +\ctws\Gamma_1^2/2)]
  &
  \\ \\ \\
  A_\mu & \put(0,3){\line(1,0){60}} \put(30,3){\redcircle{4}}
  ~~~~~~~~~~~~~~~~~~& Z_\nu  
  & & \!\!\!\!\!\! \;\;\;\;-\delta_{\mu \nu} (\stw/\ctw)[\,\Mgb^2 M^2 \Gamma_1 
    +\Mgb^4 M^2 (\Gamma_2 +2\Gamma_1 \beta_{t_1} +\ctws \Gamma_1^2)]
  &
  \\ \\ \\
  W^+_\mu  & \put(0,3){\line(1,0){60}} \put(30,3){\redcircle{4}}
  ~~~~~~~~~~~~~~~~~~& W^-_\nu    
  & & \!\!\!\!\!\! \;\;\;\;-2\delta_{\mu \nu} [\,\Mgb^2 M^2 \Gamma_1 
    +\Mgb^4 M^2(\Gamma_2 +2\Gamma_1 \beta_{t_1} +\Gamma_1^2/2)]
  &
  \\ \\ \\
  A_\mu  & \put(0,3){\line(1,0){60}} \put(30,3){\redcircle{4}}
  \put(42,-5){$\leftarrow$}\put(55,-5){$p$}
  ~~~~~~~~~~~~~~~~~~& \phi_0 
  & & \!\!\!\!\!\! \;\;\;\;i p_\mu \stw M [\Mgb^2 \Gamma_1 +
    \Mgb^4 (\Gamma_2 + \Gamma_1 \beta_{t_1})]
  &
  \\ \\ \\
  Z_\mu  & \put(0,3){\line(1,0){60}} \put(30,3){\redcircle{4}}
  \put(42,-5){$\leftarrow$}\put(55,-5){$p$}
  ~~~~~~~~~~~~~~~~~~& \phi_0 
  & & \!\!\!\!\!\! \;\;\;\;i p_\mu \ctw M [\Mgb^2 \Gamma_1 +
    \Mgb^4 (\Gamma_2 + \Gamma_1 \beta_{t_1})]
  &
  \\ \\ \\
  W^+_\mu & \put(0,3){\line(1,0){60}} \put(30,3){\redcircle{4}}
  \put(42,-5){$\leftarrow$}\put(55,-5){$p$}
  ~~~~~~~~~~~~~~~~~~& \phi_-  
  & &  \!\!\!\!\!\! \;\;\;\;i p_\mu M [\Mgb^2 \Gamma_1 +
    \Mgb^4 (\Gamma_2 + \Gamma_1 \beta_{t_1})]
  &
  \\ \\ \\
  W^-_\mu  & \put(0,3){\line(1,0){60}} \put(30,3){\redcircle{4}}
  \put(42,-5){$\leftarrow$}\put(55,-5){$p$}
  ~~~~~~~~~~~~~~~~~~& \phi_+
  & &  \!\!\!\!\!\! \;\;\;\;i p_\mu M [\Mgb^2 \Gamma_1 +
    \Mgb^4 (\Gamma_2 + \Gamma_1 \beta_{t_1})]
  &
  \\ \\ \\
  \Xb^+  & \put(0,3){\line(1,0){60}} \put(30,3){\redcircle{4}}
  ~~~~~~~~~~~~~~~~~& X^+ 
  & &  \!\!\!\!\!\! \;\;\;\;   -\xi_{\ssW} M^2  [\Mgb^2 \Gamma_1 +
    \Mgb^4 (\Gamma_2 + \Gamma_1 \beta_{t_1})]
                                  &
  \\ \\ \\
  \Xb^-  & \put(0,3){\line(1,0){60}} \put(30,3){\redcircle{4}}
  ~~~~~~~~~~~~~~~~~~& X^-
  & &  \!\!\!\!\!\! \;\;\;\;-\xi_{\ssW} M^2  [\Mgb^2 \Gamma_1 +
    \Mgb^4 (\Gamma_2 + \Gamma_1 \beta_{t_1})]
                                       &
  \\ \\ \\
  \Xb_{\ssZ}&\put(0,3){\line(1,0){60}}\put(30,3){\redcircle{4}}
  ~~~~~~~~~~~~~~~~~~& X_{\ssZ}  
  & &  \!\!\!\!\!\! \;\;\;\;-\xi_{\ssZ} M^2  [\Mgb^2 \Gamma_1 +
    \Mgb^4 (\Gamma_2 + \Gamma_1 \beta_{t_1})]
  &
  \\ \\ \\
  \Xb_{\ssZ}&\put(0,3){\line(1,0){60}}\put(30,3){\redcircle{4}}
  ~~~~~~~~~~~~~~~~~~& X_{\ssA}  
  & &  \!\!\!\!\!\! \;\;\;\;-\xi_{\ssZ} (\stw/\ctw) M^2  [\Mgb^2 \Gamma_1 +
    \Mgb^4 (\Gamma_2 + \Gamma_1 \beta_{t_1})]
                                         &
\end{array}
$$
\renewcommand{\arraystretch}{1.3}

\newpage

\noindent 
$-$ The three-leg $\Gamma$ vertices are (all momenta are flowing inwards):
\vspace{0.4cm}
%--
% 3-LEG PAGE 1
%
$$
\begin{array}{lcllll}
 H &
 \put(10,3){\line(1,0){20}} \put(30,3){\redcircle{4}}
 \put(30,3){\line(1,1){15}} \put(30,3){\line(1,-1){15}}
 \put(60,15){$Z_\mu$} \put(60,-15){$Z_\nu$}  ~~~~~~~~~~~~~~~&
  & & \;\;\;\;\;\;  \Mgb^3\Gamma_1\, [-2M\delta_{\mu\nu}]
  &
  \\~\\
 H &
 \put(10,3){\line(1,0){20}} \put(30,3){\redcircle{4}}
 \put(30,3){\line(1,1){15}} \put(30,3){\line(1,-1){15}}
 \put(60,15){$A_\mu$} \put(60,-15){$Z_\nu$}  ~~~~~~~~~~~~~~~&
  & & \;\;\;\;\;\;  \Mgb^3\Gamma_1\, [-(\stw/\ctw)M\delta_{\mu\nu}]
  &
  \\~\\
 H &
 \put(10,3){\line(1,0){20}} \put(30,3){\redcircle{4}}
 \put(30,3){\line(1,1){15}} \put(30,3){\line(1,-1){15}}
 \put(60,15){$W^+_\mu$} \put(60,-15){$W^-_\nu$}  ~~~~~~~~~~~~~~~&
  & & \;\;\;\;\;\;  \Mgb^3\Gamma_1\, [-2M\delta_{\mu\nu}]
  \\~\\
 A_\mu &
 \put(10,3){\line(1,0){20}} \put(30,3){\redcircle{4}}
 \put(42,7){$k$} \put(42,-6){$q$} 
 \put(30,3){\line(1,1){15}} \put(30,3){\line(1,-1){15}}
 \put(60,15){$H$} \put(60,-15){$\phi_0$}  ~~~~~~~~~~~~~~~&
  & & \;\;\;\;\;\;  \Mgb^3\Gamma_1\, (i\stw/2)(q_\mu-k_\mu)
  &
  \\~\\
 A_\mu &
 \put(10,3){\line(1,0){20}} \put(30,3){\redcircle{4}}
 \put(42,7){$k$} \put(42,-6){$q$} 
 \put(30,3){\line(1,1){15}} \put(30,3){\line(1,-1){15}}
 \put(60,15){$\phi_-$} \put(60,-15){$\phi_+$}  ~~~~~~~~~~~~~~~&
  & & \;\;\;\;\;\;  \Mgb^3\Gamma_1\, (\stw/2)(q_\mu-k_\mu)
  &
  \\~\\
 Z_\mu &
 \put(10,3){\line(1,0){20}} \put(30,3){\redcircle{4}}
 \put(42,7){$k$} \put(42,-6){$q$} 
 \put(30,3){\line(1,1){15}} \put(30,3){\line(1,-1){15}}
 \put(60,15){$H$} \put(60,-15){$\phi_0$}  ~~~~~~~~~~~~~~~&
  & & \;\;\;\;\;\;  \Mgb^3\Gamma_1\, (i\ctw/2)(q_\mu-k_\mu)
  &
  \\~\\
 Z_\mu &
 \put(10,3){\line(1,0){20}} \put(30,3){\redcircle{4}}
 \put(42,7){$k$} \put(42,-6){$q$} 
 \put(30,3){\line(1,1){15}} \put(30,3){\line(1,-1){15}}
 \put(60,15){$\phi_-$} \put(60,-15){$\phi_+$}  ~~~~~~~~~~~~~~~&
   & & \;\;\;\;\;\;  \Mgb^3\Gamma_1\, (\ctw/2)(q_\mu-k_\mu)
  &
  \\~\\
 A_\mu&
 \put(10,3){\line(1,0){20}} \put(30,3){\redcircle{4}}
 \put(30,3){\line(1,1){15}} \put(30,3){\line(1,-1){15}}
 \put(60,15){$\phi_-$} \put(60,-15){$W^+_\nu$}  ~~~~~~~~~~~~~~~&
  & & \;\;\;\;\;\;  \Mgb^3\Gamma_1\, [i\stw M \delta_{\mu\nu}]
  &
  \\~\\
 A_\mu &
 \put(10,3){\line(1,0){20}} \put(30,3){\redcircle{4}}
 \put(30,3){\line(1,1){15}} \put(30,3){\line(1,-1){15}}
 \put(60,15){$W^-_\nu$} \put(60,-15){$\phi_+$}  ~~~~~~~~~~~~~~~&
  & & \;\;\;\;\;\;  \Mgb^3\Gamma_1\, [-i\stw M \delta_{\mu\nu}]
  &
\end{array}
$$
%

% 3-LEG PAGE 2
%
$$
\begin{array}{lcllll}
 Z_\mu &
 \put(10,3){\line(1,0){20}} \put(30,3){\redcircle{4}}
 \put(30,3){\line(1,1){15}} \put(30,3){\line(1,-1){15}}
 \put(60,15){$\phi_-$} \put(60,-15){$W^+_\nu$}  ~~~~~~~~~~~~~~~~~~&
  & & \;\;\;\;  \Mgb^3\Gamma_1\, (-i\stws M/c) \,\delta_{\mu\nu}
  &
  \\~\\
 Z_\mu &
 \put(10,3){\line(1,0){20}} \put(30,3){\redcircle{4}}
 \put(30,3){\line(1,1){15}} \put(30,3){\line(1,-1){15}}
 \put(60,15){$W^-_\nu$} \put(60,-15){$\phi_+$}  ~~~~~~~~~~~~~~~~~~&
  & & \;\;\;\;  \Mgb^3\Gamma_1\, (i\stws M/c) \,\delta_{\mu\nu}
  &
  \\~\\
 W^+_\mu &
 \put(10,3){\line(1,0){20}} \put(30,3){\redcircle{4}}
 \put(42,7){$k$} \put(42,-6){$q$} 
 \put(30,3){\line(1,1){15}} \put(30,3){\line(1,-1){15}}
 \put(60,15){$\phi_0$} \put(60,-15){$\phi_-$}  ~~~~~~~~~~~~~~~~~~&
  & & \;\;\;\;  \Mgb^3\Gamma_1\, (q_\mu-k_\mu)/2
  \\~\\
 W^-_\mu &
 \put(10,3){\line(1,0){20}} \put(30,3){\redcircle{4}}
 \put(42,7){$k$} \put(42,-6){$q$} 
 \put(30,3){\line(1,1){15}} \put(30,3){\line(1,-1){15}}
 \put(60,15){$\phi_+$} \put(60,-15){$\phi_0$}  ~~~~~~~~~~~~~~~~~~&
  & & \;\;\;\;  \Mgb^3\Gamma_1\, (q_\mu-k_\mu)/2
  &
  \\~\\
 W^+_\mu &
 \put(10,3){\line(1,0){20}} \put(30,3){\redcircle{4}}
 \put(42,7){$k$} \put(42,-6){$q$} 
 \put(30,3){\line(1,1){15}} \put(30,3){\line(1,-1){15}}
 \put(60,15){$H$} \put(60,-15){$\phi_-$}  ~~~~~~~~~~~~~~~~~~&
  & & \;\;\;\;  \Mgb^3\Gamma_1\, i(q_\mu-k_\mu)/2
  &
  \\~\\
 W^-_\mu &
 \put(10,3){\line(1,0){20}} \put(30,3){\redcircle{4}}
 \put(42,7){$k$} \put(42,-6){$q$} 
 \put(30,3){\line(1,1){15}} \put(30,3){\line(1,-1){15}}
 \put(60,15){$H$} \put(60,-15){$\phi_+$}  ~~~~~~~~~~~~~~~~~~&
  & & \;\;\;\;  \Mgb^3\Gamma_1\, i(q_\mu-k_\mu)/2
  &
\end{array}
$$

\vspace{15mm} $-$ The trilinear $\Gamma$ vertices with FP ghosts are:
\vspace{2mm}
$$
\begin{array}{lcllll}
 \Xb^- &
 \put(10,3){\line(1,0){20}} \put(30,3){\redcircle{4}}
 \put(12,7){$p$}
 \put(30,3){\line(1,1){15}} \put(30,3){\line(1,-1){15}}
 \put(60,15){$A_\nu$} \put(60,-15){$X^-$}  ~~~~~~~~~~~~~~~~~~&
   & & \;\;\;\;  \Mgb^3\Gamma_1\, \stw p_\nu/\xi_{\ssW}
  &
  \\~\\
 \Xb^+ &
 \put(10,3){\line(1,0){20}} \put(30,3){\redcircle{4}}
 \put(12,7){$p$}
 \put(30,3){\line(1,1){15}} \put(30,3){\line(1,-1){15}}
 \put(60,15){$A_\nu$} \put(60,-15){$X^+$}  ~~~~~~~~~~~~~~~~~~&
  & & \;\;\;\;  \Mgb^3\Gamma_1\, (-\stw p_\nu/\xi_{\ssW})
  &
  \\~\\
 \Xb^- &
 \put(10,3){\line(1,0){20}} \put(30,3){\redcircle{4}}
 \put(12,7){$p$}
 \put(30,3){\line(1,1){15}} \put(30,3){\line(1,-1){15}}
 \put(60,15){$Z_\nu$} \put(60,-15){$X^-$}  ~~~~~~~~~~~~~~~~~~&
  & & \;\;\;\;  \Mgb^3\Gamma_1\, \ctw p_\nu/\xi_{\ssW}
  &
\end{array}
$$
%
% 3-LEG PAGE 3
%
$$
\begin{array}{lcllll}
 \Xb^+ &
 \put(10,3){\line(1,0){20}} \put(30,3){\redcircle{4}}
 \put(12,7){$p$}
 \put(30,3){\line(1,1){15}} \put(30,3){\line(1,-1){15}}
 \put(60,15){$Z_\nu$} \put(60,-15){$X^+$}  ~~~~~~~~~~~~~~~~~~&
  & & \;\;\;\;  \Mgb^3\Gamma_1\, (-\ctw p_\nu/\xi_{\ssW})
  &
  \\~\\
 \Xb^- &
 \put(10,3){\line(1,0){20}} \put(30,3){\redcircle{4}}
 \put(12,7){$p$}
 \put(30,3){\line(1,1){15}} \put(30,3){\line(1,-1){15}}
 \put(60,15){$W^-_\nu$} \put(60,-15){$X_{\ssZ}$}  ~~~~~~~~~~~~~~~~~~&
  & & \;\;\;\;  \Mgb^3\Gamma_1\, (-\ctw p_\nu/\xi_{\ssW})
  &
  \\~\\
 \Xb^- &
 \put(10,3){\line(1,0){20}} \put(30,3){\redcircle{4}}
 \put(12,7){$p$}
 \put(30,3){\line(1,1){15}} \put(30,3){\line(1,-1){15}}
 \put(60,15){$W^-_\nu$} \put(60,-15){$X_{\ssA}$}  ~~~~~~~~~~~~~~~~~~&
  & & \;\;\;\;  \Mgb^3\Gamma_1\, (-\stw p_\nu/\xi_{\ssW})
  \\~\\
 \Xb_{\ssZ} &
 \put(10,3){\line(1,0){20}} \put(30,3){\redcircle{4}}
 \put(12,7){$p$}
 \put(30,3){\line(1,1){15}} \put(30,3){\line(1,-1){15}}
 \put(60,15){$W^+_\nu$} \put(60,-15){$X^-$}  ~~~~~~~~~~~~~~~~~~&
  & & \;\;\;\;  \Mgb^3\Gamma_1\, (-\ctw p_\nu/\xi_{\ssZ})
  &
  \\~\\
 \Xb_{\ssA} &
 \put(10,3){\line(1,0){20}} \put(30,3){\redcircle{4}}
 \put(12,7){$p$}
 \put(30,3){\line(1,1){15}} \put(30,3){\line(1,-1){15}}
 \put(60,15){$W^+_\nu$} \put(60,-15){$X^-$}  ~~~~~~~~~~~~~~~~~~&
  & & \;\;\;\;  \Mgb^3\Gamma_1\, (-\stw p_\nu/\xi_{\ssA})
  &
  \\~\\
 \Xb_{\ssZ} &
 \put(10,3){\line(1,0){20}} \put(30,3){\redcircle{4}}
 \put(12,7){$p$}
 \put(30,3){\line(1,1){15}} \put(30,3){\line(1,-1){15}}
 \put(60,15){$W^-_\nu$} \put(60,-15){$X^+$}  ~~~~~~~~~~~~~~~~~~&
  & & \;\;\;\;  \Mgb^3\Gamma_1\, \ctw p_\nu/\xi_{\ssZ}
  &
  \\~\\
 \Xb^+ &
 \put(10,3){\line(1,0){20}} \put(30,3){\redcircle{4}}
 \put(12,7){$p$}
 \put(30,3){\line(1,1){15}} \put(30,3){\line(1,-1){15}}
 \put(60,15){$W^+_\nu$} \put(60,-15){$X_{\ssZ}$}  ~~~~~~~~~~~~~~~~~~&
   & & \;\;\;\;  \Mgb^3\Gamma_1\, \ctw p_\nu/\xi_{\ssW}
  &
  \\~\\
 \Xb_{\ssA} &
 \put(10,3){\line(1,0){20}} \put(30,3){\redcircle{4}}
 \put(12,7){$p$}
 \put(30,3){\line(1,1){15}} \put(30,3){\line(1,-1){15}}
 \put(60,15){$W^-_\nu$} \put(60,-15){$X^+$}  ~~~~~~~~~~~~~~~~~~&
  & & \;\;\;\;  \Mgb^3\Gamma_1\, \stw p_\nu/\xi_{\ssA} 
  &
  \\~\\
 \Xb^+ &
 \put(10,3){\line(1,0){20}} \put(30,3){\redcircle{4}}
 \put(12,7){$p$}
 \put(30,3){\line(1,1){15}} \put(30,3){\line(1,-1){15}}
 \put(60,15){$W^+_\nu$} \put(60,-15){$X_{\ssA}$}  ~~~~~~~~~~~~~~~~~~&
  & & \;\;\;\;  \Mgb^3\Gamma_1\, \stw p_\nu/\xi_{\ssW}
  &
\end{array}
$$
%
% 3-LEG PAGE 4
%
$$
\begin{array}{lcllll}
 \Xb^+ & 
 \put(10,3){\line(1,0){20}} \put(30,3){\redcircle{4}}
 \put(30,3){\line(1,1){15}} \put(30,3){\line(1,-1){15}}
 \put(60,15){$\phi_0$} \put(60,-15){$X^+$}  ~~~~~~~~~~~~~~~~~~&
  & & \;\;\;\;  \Mgb^3\Gamma_1\, (i M\xi_{\ssW}/2)
  &
  \\~\\
 \Xb^- &
 \put(10,3){\line(1,0){20}} \put(30,3){\redcircle{4}}
 \put(30,3){\line(1,1){15}} \put(30,3){\line(1,-1){15}}
 \put(60,15){$\phi_0$} \put(60,-15){$X^-$}  ~~~~~~~~~~~~~~~~~~&
  & & \;\;\;\;  \Mgb^3\Gamma_1\, (-i M\xi_{\ssW}/2)
  &
  \\~\\
 \Xb^+ &
 \put(10,3){\line(1,0){20}} \put(30,3){\redcircle{4}}
 \put(30,3){\line(1,1){15}} \put(30,3){\line(1,-1){15}}
 \put(60,15){$H$} \put(60,-15){$X^+$}  ~~~~~~~~~~~~~~~~~~&
  & & \;\;\;\;  \Mgb^3\Gamma_1\, (-M\xi_{\ssW}/2)
  \\~\\
 \Xb^- &
 \put(10,3){\line(1,0){20}} \put(30,3){\redcircle{4}}
 \put(30,3){\line(1,1){15}} \put(30,3){\line(1,-1){15}}
 \put(60,15){$H$} \put(60,-15){$X^-$}  ~~~~~~~~~~~~~~~~~~&
  & & \;\;\;\;  \Mgb^3\Gamma_1\, (-M\xi_{\ssW}/2)
  &
  \\~\\
 \Xb_{\ssZ} &
 \put(10,3){\line(1,0){20}} \put(30,3){\redcircle{4}}
 \put(30,3){\line(1,1){15}} \put(30,3){\line(1,-1){15}}
 \put(60,15){$\phi_+$} \put(60,-15){$X^-$}  ~~~~~~~~~~~~~~~~~~&
  & & \;\;\;\;  \Mgb^3\Gamma_1\, (i M\xi_{\ssZ}/2 \ctw)
  &
  \\~\\
 \Xb_{\ssZ} &
 \put(10,3){\line(1,0){20}} \put(30,3){\redcircle{4}}
 \put(30,3){\line(1,1){15}} \put(30,3){\line(1,-1){15}}
 \put(60,15){$\phi_-$} \put(60,-15){$X^+$}  ~~~~~~~~~~~~~~~~~~&
  & & \;\;\;\;  \Mgb^3\Gamma_1\, (-i M\xi_{\ssZ}/2 \ctw)
  &
  \\~\\
 \Xb_{\ssZ} &
 \put(10,3){\line(1,0){20}} \put(30,3){\redcircle{4}}
 \put(30,3){\line(1,1){15}} \put(30,3){\line(1,-1){15}}
 \put(60,15){$H$} \put(60,-15){$X_{\ssA}$}  ~~~~~~~~~~~~~~~~~~&
   & & \;\;\;\;  \Mgb^3\Gamma_1\, (-\stw M\xi_{\ssZ}/2 \ctw)
  &
  \\~\\
 \Xb_{\ssZ} &
 \put(10,3){\line(1,0){20}} \put(30,3){\redcircle{4}}
 \put(30,3){\line(1,1){15}} \put(30,3){\line(1,-1){15}}
 \put(60,15){$H$} \put(60,-15){$X_{\ssZ}$}  ~~~~~~~~~~~~~~~~~~&
  & & \;\;\;\;  \Mgb^3\Gamma_1\, (- M\xi_{\ssZ}/2)
  &
  \\~\\
 \Xb^- &
 \put(10,3){\line(1,0){20}} \put(30,3){\redcircle{4}}
 \put(30,3){\line(1,1){15}} \put(30,3){\line(1,-1){15}}
 \put(60,15){$\phi_-$} \put(60,-15){$X_{\ssZ}$}  ~~~~~~~~~~~~~~~~~~&
  & & \;\;\;\;  \Mgb^3\Gamma_1\, (i \ctw M\xi_{\ssW}/2)
  &
\end{array}
$$
%
% 3-LEG PAGE 5
%
$$
\begin{array}{lcllll}
 \Xb^- &
 \put(10,3){\line(1,0){20}} \put(30,3){\redcircle{4}}
 \put(30,3){\line(1,1){15}} \put(30,3){\line(1,-1){15}}
 \put(60,15){$\phi_-$} \put(60,-15){$X_{\ssA}$}  ~~~~~~~~~~~~~~~~~~&
  & & \;\;\;\;  \Mgb^3\Gamma_1\, (i \stw M\xi_{\ssW}/2)
  &
  \\~\\
 \Xb^+ &
 \put(10,3){\line(1,0){20}} \put(30,3){\redcircle{4}}
 \put(30,3){\line(1,1){15}} \put(30,3){\line(1,-1){15}}
 \put(60,15){$\phi_+$} \put(60,-15){$X_{\ssZ}$}  ~~~~~~~~~~~~~~~~~~&
  & & \;\;\;\;  \Mgb^3\Gamma_1\, (-i \ctw M\xi_{\ssW}/2)
  &
  \\~\\
 \Xb^+ &
 \put(10,3){\line(1,0){20}} \put(30,3){\redcircle{4}}
 \put(30,3){\line(1,1){15}} \put(30,3){\line(1,-1){15}}
 \put(60,15){$\phi_+$} \put(60,-15){$X_{\ssA}$}  ~~~~~~~~~~~~~~~~~~&
  & & \;\;\;\;  \Mgb^3\Gamma_1\, (-i \stw M\xi_{\ssW}/2)
  &
\end{array}
$$
\vspace{1cm}

\noindent The three-leg $\Gamma$ vertices introduced by the pure Yang--Mills
Lagrangian are not listed here as they can be immediately derived from the
usual Yang--Mills vertices (see, e.g., the appendix D of
ref.~\cite{Veltman:1994wz}) by simply replacing $g \rightarrow \Mgb \Gamma$.

\vspace{1cm}

\noindent $-$ The trilinear $\Gamma$ vertices with fermions are:
\vspace{5mm}
$$
\begin{array}{lcllll}
 A_\mu &
 \put(10,3){\line(1,0){20}} \put(30,3){\redcircle{4}}
 \put(30,3){\line(1,1){15}} \put(30,3){\line(1,-1){15}}
 \put(60,15){$\bar{f}$} \put(60,-15){$f$}  ~~~~~~~~~~~~~~~~~~&
  & & \;\;\;\;  \Mgb^3\Gamma_1\, (i \stw I_3/2) \,\gamma_\mu (1+\gamma_5)
  &
  \\~\\
 Z_\mu &
 \put(10,3){\line(1,0){20}} \put(30,3){\redcircle{4}}
 \put(30,3){\line(1,1){15}} \put(30,3){\line(1,-1){15}}
 \put(60,15){$\bar{f}$} \put(60,-15){$f$}  ~~~~~~~~~~~~~~~~~~&
  & & \;\;\;\;  \Mgb^3\Gamma_1\, (i \ctw I_3/2) \,\gamma_\mu (1+\gamma_5)
  &
  \\~\\
 W^+_\mu &
 \put(10,3){\line(1,0){20}} \put(30,3){\redcircle{4}}
 \put(30,3){\line(1,1){15}} \put(30,3){\line(1,-1){15}}
 \put(60,15){$\bar{u}$} \put(60,-15){$d$}  ~~~~~~~~~~~~~~~~~~&
  & & \;\;\;\;  \Mgb^3\Gamma_1\, (i /2\sqrt2) \,\gamma_\mu (1+\gamma_5)
  &
  \\~\\
 W^-_\mu &
 \put(10,3){\line(1,0){20}} \put(30,3){\redcircle{4}}
 \put(30,3){\line(1,1){15}} \put(30,3){\line(1,-1){15}}
 \put(60,15){$\bar{d}$} \put(60,-15){$u$}  ~~~~~~~~~~~~~~~~~~&
  & & \;\;\;\;  \Mgb^3\Gamma_1\, (i /2\sqrt2) \,\gamma_\mu (1+\gamma_5)
  &
\end{array}
$$
\newpage

\noindent $-$ The four-leg $\Gamma$ vertices are:
\vspace{0.6cm}
\renewcommand{\arraystretch}{1.5}
$$
\begin{array}{lcllll}
  &
 \put(-15, +7){$H$}
 \put(-15,-15){$H$}
 \put(10,15){\line(1,-1){30}} 
 \put(10,-15){\line(1,+1){30}} 
 \put(25,0){\redcircle{4}}
 \put(+50,  7){$Z_\mu$}
 \put(+50,-15){$Z_\nu$} &
   & & ~~~~~~~~~~~~~~~~~~ \;\; - \Mgb^4\Gamma_1\, \delta_{\mu\nu}
  &
  \\~\\
  &
 \put(-15, +7){$H$}
 \put(-15,-15){$H$}
 \put(10,15){\line(1,-1){30}} 
 \put(10,-15){\line(1,+1){30}} 
 \put(25,0){\redcircle{4}}
 \put(+50,  7){$A_\mu$}
 \put(+50,-15){$Z_\nu$} &
   & & ~~~~~~~~~~~~~~~~~~ \;\;\Mgb^4\Gamma_1\, (-\stw/2 \ctw)\, \delta_{\mu\nu}
  &
  \\~\\
  &
 \put(-15, +7){$H$}
 \put(-15,-15){$H$}
 \put(10,15){\line(1,-1){30}} 
 \put(10,-15){\line(1,+1){30}} 
 \put(25,0){\redcircle{4}}
 \put(+50,  7){$W^+_\mu$}
 \put(+50,-15){$W^-_\nu$} &
   & & ~~~~~~~~~~~~~~~~~~ \;\;-\Mgb^4\Gamma_1\, \delta_{\mu\nu}
  &
  \\~\\
  &
 \put(-15, +7){$\phi_0$}
 \put(-15,-15){$\phi_0$}
 \put(10,15){\line(1,-1){30}} 
 \put(10,-15){\line(1,+1){30}} 
 \put(25,0){\redcircle{4}}
 \put(+50,  7){$Z_\mu$}
 \put(+50,-15){$Z_\nu$} &
   & & ~~~~~~~~~~~~~~~~~~ \;\;-\Mgb^4\Gamma_1\, \delta_{\mu\nu}
  &
  \\~\\
  &
 \put(-15, +7){$\phi_0$}
 \put(-15,-15){$\phi_0$}
 \put(10,15){\line(1,-1){30}} 
 \put(10,-15){\line(1,+1){30}} 
 \put(25,0){\redcircle{4}}
 \put(+50,  7){$A_\mu$}
 \put(+50,-15){$Z_\nu$} &
   & & ~~~~~~~~~~~~~~~~~~ \;\;\Mgb^4\Gamma_1\, (-\stw/2 \ctw)\, \delta_{\mu\nu}
  &
  \\~\\
  &
 \put(-15, +7){$\phi_0$}
 \put(-15,-15){$\phi_0$}
 \put(10,15){\line(1,-1){30}} 
 \put(10,-15){\line(1,+1){30}} 
 \put(25,0){\redcircle{4}}
 \put(+50,  7){$W^+_\mu$}
 \put(+50,-15){$W^-_\nu$} &
   & & ~~~~~~~~~~~~~~~~~~ \;\;-\Mgb^4\Gamma_1\, \delta_{\mu\nu}
  &
  \\~\\
  &
 \put(-15, +7){$\phi_+$}
 \put(-15,-15){$\phi_-$}
 \put(10,15){\line(1,-1){30}} 
 \put(10,-15){\line(1,+1){30}} 
 \put(25,0){\redcircle{4}}
 \put(+50,  7){$A_\mu$}
 \put(+50,-15){$A_\nu$} &
   & & ~~~~~~~~~~~~~~~~~~ \;\;\Mgb^4\Gamma_1\, (-2\stws )\, \delta_{\mu\nu}
  &
  \\~\\
  &
 \put(-15, +7){$\phi_+$}
 \put(-15,-15){$\phi_-$}
 \put(10,15){\line(1,-1){30}} 
 \put(10,-15){\line(1,+1){30}} 
 \put(25,0){\redcircle{4}}
 \put(+50,  7){$Z_\mu$}
 \put(+50,-15){$Z_\nu$} &
   & & ~~~~~~~~~~~~~~~~~~ \;\;\Mgb^4\Gamma_1\,(1-2\ctws)\,\delta_{\mu\nu}
  &
  \\~\\
  &
 \put(-15, +7){$\phi_+$}
 \put(-15,-15){$\phi_-$}
 \put(10,15){\line(1,-1){30}} 
 \put(10,-15){\line(1,+1){30}} 
 \put(25,0){\redcircle{4}}
 \put(+50,  7){$A_\mu$}
 \put(+50,-15){$Z_\nu$} &
   & & ~~~~~~~~~~~~~~~~~~ 
\;\;\Mgb^4\Gamma_1\,(\stw/2 \ctw -2\stw\ctw)\,\delta_{\mu\nu}
  &
  \\~\\
  &
 \put(-15, +7){$\phi_+$}
 \put(-15,-15){$\phi_-$}
 \put(10,15){\line(1,-1){30}} 
 \put(10,-15){\line(1,+1){30}} 
 \put(25,0){\redcircle{4}}
 \put(+50,  7){$W^+_\mu$}
 \put(+50,-15){$W^-_\nu$} &
   & & ~~~~~~~~~~~~~~~~~~ \;\;-\Mgb^4\Gamma_1\, \delta_{\mu\nu}
  &
\end{array}
$$
$$
\begin{array}{lcllll}
  &
 \put(-15, +7){$A_\mu$}
 \put(-15,-15){$\phi_0$}
 \put(10,15){\line(1,-1){30}} 
 \put(10,-15){\line(1,+1){30}} 
 \put(25,0){\redcircle{4}}
 \put(+50,  7){$\phi_+$}
 \put(+50,-15){$W^-_\nu$} &
   & & ~~~~~~~~~~~~~~~~~~  \;\;\Mgb^4\Gamma_1\, (\stw/2)\, \delta_{\mu\nu}
  &
  \\~\\
  &
 \put(-15, +7){$A_\mu$}
 \put(-15,-15){$\phi_0$}
 \put(10,15){\line(1,-1){30}} 
 \put(10,-15){\line(1,+1){30}} 
 \put(25,0){\redcircle{4}}
 \put(+50,  7){$\phi_-$}
 \put(+50,-15){$W^+_\nu$} &
   & & ~~~~~~~~~~~~~~~~~~ \;\;\Mgb^4\Gamma_1\, (\stw/2)\, \delta_{\mu\nu}
  &
  \\~\\
  &
 \put(-15, +7){$A_\mu$}
 \put(-15,-15){$H$}
 \put(10,15){\line(1,-1){30}} 
 \put(10,-15){\line(1,+1){30}} 
 \put(25,0){\redcircle{4}}
 \put(+50,  7){$\phi_+$}
 \put(+50,-15){$W^-_\nu$} & 
   & & ~~~~~~~~~~~~~~~~~~ \;\;\Mgb^4\Gamma_1\, (-i\stw/2)\, \delta_{\mu\nu}
  &
  \\~\\
  &
 \put(-15, +7){$A_\mu$}
 \put(-15,-15){$H$}
 \put(10,15){\line(1,-1){30}} 
 \put(10,-15){\line(1,+1){30}} 
 \put(25,0){\redcircle{4}}
 \put(+50,  7){$\phi_-$}
 \put(+50,-15){$W^+_\nu$} &
   & & ~~~~~~~~~~~~~~~~~~ \;\;\Mgb^4\Gamma_1\, (i\stw/2)\, \delta_{\mu\nu}
  &
  \\~\\
  &
 \put(-15, +7){$Z_\mu$}
 \put(-15,-15){$\phi_0$}
 \put(10,15){\line(1,-1){30}} 
 \put(10,-15){\line(1,+1){30}} 
 \put(25,0){\redcircle{4}}
 \put(+50,  7){$\phi_+$}
 \put(+50,-15){$W^-_\nu$} &
   & & ~~~~~~~~~~~~~~~~~~ \;\;\Mgb^4\Gamma_1\,(-\stws /2 \ctw)\,\delta_{\mu\nu}
  &
  \\~\\
  &
 \put(-15, +7){$Z_\mu$}
 \put(-15,-15){$\phi_0$}
 \put(10,15){\line(1,-1){30}} 
 \put(10,-15){\line(1,+1){30}} 
 \put(25,0){\redcircle{4}}
 \put(+50,  7){$\phi_-$}
 \put(+50,-15){$W^+_\nu$} &
   & & ~~~~~~~~~~~~~~~~~~ \;\;\Mgb^4\Gamma_1\,(-\stws /2 \ctw)\,\delta_{\mu\nu}
  &
  \\~\\
  &
 \put(-15, +7){$Z_\mu$}
 \put(-15,-15){$H$}
 \put(10,15){\line(1,-1){30}} 
 \put(10,-15){\line(1,+1){30}} 
 \put(25,0){\redcircle{4}}
 \put(+50,  7){$\phi_+$}
 \put(+50,-15){$W^-_\nu$} &
   & & ~~~~~~~~~~~~~~~~~~ \;\;\Mgb^4\Gamma_1\,(i\stws /2 \ctw)\,\delta_{\mu\nu}
  &
  \\~\\
  &
 \put(-15, +7){$Z_\mu$}
 \put(-15,-15){$H$}
 \put(10,15){\line(1,-1){30}} 
 \put(10,-15){\line(1,+1){30}} 
 \put(25,0){\redcircle{4}}
 \put(+50,  7){$\phi_-$}
 \put(+50,-15){$W^+_\nu$} &
   & &~~~~~~~~~~~~~~~~~~ \;\;\Mgb^4\Gamma_1\,(-i\stws /2 \ctw)\,\delta_{\mu\nu}
  &
\end{array}
$$
\vspace{1cm}

\noindent The four-leg $\Gamma$ vertices introduced by the pure Yang--Mills
Lagrangian are not listed here as they can be immediately derived from the
usual Yang--Mills vertices (see, e.g., the Appendix D of
ref.~\cite{Veltman:1994wz}) by simply replacing $g^2 \rightarrow
\Mgb^2 \Gamma(2+\Gamma)$.
%--
\clearpage
%--

\end{document}